\documentclass[times,english,sort&compress]{elsarticle}
\usepackage{color}
\usepackage[colorlinks=true] {hyperref}
\usepackage{float}
\usepackage{stmaryrd}
\usepackage{babel}
\usepackage{units}
\usepackage{textcomp}
\usepackage{textcomp}
\usepackage{amsmath}
\usepackage{commath}
\usepackage{amssymb}
\usepackage{graphicx}
\usepackage{esint}

\usepackage{geometry}
\makeatletter

\usepackage{epstopdf}

\journal{Condensed Matter}

\begin{document}
\begin{frontmatter}

\title{The variational reduction for low-dimensional Fermi gases and Bose - Fermi mixtures: A brief review}

\author[UFRO]{Pablo D\'{\i}az}
\author[UTA]{David Laroze}
\author[TELAVIV]{Boris A. Malomed\corref{mycorrespondingauthor}}
\cortext[mycorrespondingauthor]{Corresponding author}
\ead{malomed@post.tau.ac.il}

\address[UFRO]{Departamento de Ciencias F\'{i}sicas, Universidad de La Frontera, Casilla 54-D, Temuco, Chile.}
\address[UTA]{Instituto de Alta Investigaci\'{o}n, CEDENNA, Universidad de Tarapac\'{a}, Casilla 7D, Arica, Chile.}
\address[TELAVIV]{Department of Physical Electronics, School of Electrical Engineering, Faculty of Engineering, Tel Aviv University, Tel Aviv IL-69978, Israel.}

\begin{abstract}
We present a summary of some recent theoretical results for matter-wave patterns in Fermi and Bose-Fermi degenerate gases, obtained in the framework of the quasi-mean-field approximation. We perform a dimensional reduction from the  three-dimensional (3D) equations of motion to 2D and 1D effective equations. In both cases, comparison of the low-dimensional reductions to the full model is performed, showing very good agreement for ground-state solutions. Some complex dynamical regimes are reported too for the corresponding 1D systems.
\end{abstract}


\end{frontmatter}

\section{Introduction}

\label{S1}

Ultracold atomic gases have been widely explored from both experimental and
theoretical point of view due to their ability to emulate many effects from
condensed-matter physics and create novel states of quantum matter. Various
results obtained in this area have been reviewed in many publications --
see, in particular, Refs. \cite%
{Bongs04,Jaksch05,Giorgini08,Bloch08,Spielman12,Goldman14,Zhai15,Malomed18}.
Important experimental tools, the application of which opens ways to the
observation diverse novel phenomena in the quantum gases, are, \textit{inter
alia}, optical-lattice (OL) potentials, the use of the Feshbach resonance
(FR)\ to control the strength of interactions between atoms, and the
implementation of the effective spin-orbit coupling \cite%
{Snoek2011a,Frohlich2011,BoseSOC1,BoseSOC2,FermiSOC}.

The effective spatial dimension of the setting in which quantum gases are
created strongly affects the ensuing physics. The use of confining
potentials makes it possible to reduce the dimension from 3D to 2D and 1D.
In particular, the dimensional reduction of confined Bose gases can be
approximated by means of the variational method \cite%
{Salasnich02a,Salasnich02b,Salasnich2D,Salasnich10}. Recently, similar
approaches for ultracold Fermi gases in confining potentials have been
elaborated in Refs. \cite{Adhikari2006,Adhikari2007,Malomed2009,Diaz12}.
These reductions make it possible to study the complex dynamics and pattern
formation in ultracold gases in 2D and 1D settings. In this context, the
study of dark solitons in ultracold gases was reported in Bose-Einstein
condensates (BECs) \cite{Burger99}, and further developed later \cite%
{Becker08,Weller08}. For dark solitons in Fermi gases, several works have
reported theoretical and experimental results \cite%
{Antezza07,Scott11,Liao11,Yefsah13,Ku16,Syrwid18,Alphen18}. The reduced 1D
equation for Fermi gases was used for studies of interactions between dark
solitons \cite{Diaz12}.

The earliest experimental studies of Bose-Fermi mixtures (BFMs) were
performed with lithium isotopes \cite{Partridge01,Schreck01}, as well as in $%
^{174}$Yb-$^{6}$Li \cite{Hansen11} and $^{87}$Rb-$^{40}$K \cite{Heinze11}
settings. Much interest has been also drawn to heavy-atom mixtures, such as $%
^{87}$Sr-$^{84}$Sr \cite{Tey10}. These isotopes, which are characterized by
a large nuclear spin, have been proposed for the design of prototype
quantum-information processing devices. The use of FRs in the mixtures plays
a major role, as it allows one to control nonlinear interactions between the
species. For the $^{87}$Rb-$^{40}$K mixture, the FR has been observed in
Ref. \cite{Best09,Cumby13}, and a giant FR effect was reported in the $^{85}$%
Rb-$^{6}$Li system \cite{Deh10}. Further, in the $^{6} $Li-$^{133}$Cs
mixture five FRs have been observed \cite{Tung13}, and over 30 resonances
are expected to be available in the $^{23}$Na-$^{40}$K system \cite{Park12}.
Multiple heteronuclear FRs were reported in the triply quantum-degenerate
mixture of bosonic $^{41}$K and two fermionic species, $^{40}$K and $^{6}$Li
\cite{Wu11}. In a recently elaborated framework, the BFM is transformed into
a strongly interacting isotopic mixture immersed into a Fermi sea, with the
help of a wide $s$-wave resonance for the $^{41}$K-$^{40}$K combination.
Many theoretical works have addressed the dynamics of BFMs under various
conditions \cite%
{Lelas09,Watanabe08,Kain11,Mering11,Jun11,Ludwig11,Bertaina13}. To describe
the ground state (GS) of the mixture, the quasi-mean-field theory may be a
useful approach \cite{Adhikari08,Maruyama09,Iskin09,Snoek11,Nishida06}. In
this framework, the use of FRs was studied in $^{23}$Na-$^{6}$Li, $^{87}$Rb-$%
^{40}$K, $^{87}$Rb-$^{6}$Li, $^{3}$He-$^{4}$He, $^{173}$Yb-$^{174}$Yb, and $%
^{87}$Sr-$^{84}$Sr mixtures \cite{Salasnich07a,Gautam11}. Recently,
effective 1D and 2D nonlinear Schr\"{o}dinger equations have been derived
for BFMs in cigar-shaped and disc-shaped configurations \cite{Diaz15}, using
the variational approximation (VA) along the lines previously developed in
Refs. \cite{Salasnich02a,Diaz12}. In addition, dark solitons in BFMs have
been analyzed in Ref. \cite{Tylutki16}. Here, we address, in particular,
dark solitons in the $^{7}$Li-$^{6}$Li BFM, using the effective
low-dimensional equations derived in Ref. \cite{Diaz15}.

The general aim of the present article is to present a brief review of the
spatial reduction for Fermi gases and BFMs, based on the VA. In particular,
we outline the procedure for implementing the 2D and 1D reduction, starting
from the full 3D equations of motions. To test the accuracy of the
approximations, we present a comparison of the results with full 3D
numerical simulations. Using the corresponding effective equations, we
address various dynamical settings, such as dark solitons and their
interactions. In the case of BFMs, we consider the construction of GSs,
varying the interaction strength. Finally, for the 1D situation, we address
the formation of dark solitons in the mixture, and compare the corresponding
1D solution to results of the full numerical simulations, observing good
agreement between them. The presentation is arranged as follows: the Fermi
gases and BFMs are considered, severally,in Secs. \ref{S2} and \ref{S3}, and
the paper is concluded in Sec \ref{S4}.

\section{The Fermi Gas}

\label{S2}

We consider a dilute superfluid formed by $N$ fermionic atoms of mass $m_{%
\mathrm{F}}$ and spin $s_{\mathrm{F}}$, loaded into an optical trap at zero
temeprature. We apply the local density approximation \cite{Bloch08} to the
description of this setting. The corresponding dynamical equations can be
derived from the action functional%%
\begin{equation}
\mathcal{S}=\int {dtd{\mathbf{r}}\mathcal{L}},  \label{Eq1}
\end{equation}%
where the Lagrangian density is
\begin{equation}
{\mathcal{L}}=\frac{{i\hbar }}{2\lambda _{1}}\left( {\Psi ^{\ast }\frac{{%
\partial {\Psi }}}{{\partial t}}-{\Psi }\frac{{\partial \Psi ^{\ast }}}{{%
\partial t}}}\right) -\frac{{\hbar ^{2}}}{{2\lambda _{2}{m_{\mathrm{F}}}}}{%
\left\vert {\nabla {\Psi }}\right\vert ^{2}}-{U(\mathbf{r})}{\left\vert {%
\Psi }\right\vert ^{2}}-\frac{3}{5}\beta \frac{{\hbar ^{2}}}{{2{m_{\mathrm{F}%
}}}}{C_{\mathrm{F}}}{\left\vert {\Psi }\right\vert ^{10/3}}-\frac{{{g_{%
\mathrm{F}}}}}{2}{\left\vert {\Psi }\right\vert ^{4}},  \label{Eq2}
\end{equation}%
$\Psi \left( \mathbf{r},t\right) $ being a complex order parameter, whose
norm is equal to the number of particles. Here ${C_{\mathrm{F}}}={\left[ {6{%
\pi ^{2}}/\left( {2{s_{\mathrm{F}}}+1}\right) }\right] ^{2/3}}$ is a
constant that depends on spin $s_{\mathrm{F}}$, ${g_{\mathrm{F}}}=4\pi {%
\hbar ^{2}}({a_{\mathrm{F}}}/{m_{\mathrm{F}}})[2s_{\mathrm{F}}/(2s_{\mathrm{F%
}}+1)]$ with scattering length $a_{\mathrm{F}}$, which determines
interactions of fermions belonging to different spin states
(the interactions which are not forbidden by the Pauli principle) \cite%
{Giorgini08}, and ${U(\mathbf{r})}$ is to an external potential applied to
fermions.
\begin{table}[tbh]
\begin{center}
\begin{tabular}{|c|c|c|c|c|}
\hline
Regime & $\lambda_1$ & $\lambda_2$ & $\beta$ & $s_\mathrm{F}$ \\ \hline
Polarized & 1 & 1 & 1 & 0 \\
BCS & 2 & 4 & 1 & 1/2 \\
Unitary & 2 & 4 & 0.44 & 1/2 \\ \hline
\end{tabular}%
\end{center}
\caption{$\protect\lambda _{1}$, $\protect\lambda _{2}$, $\protect\beta $,
and $s_{\mathrm{F}}$ for three different regimes in the Fermi-gas dynamics
\protect\cite{Manini05,Salasnich09,Ancilotto09,Ancilotto12}.}
\label{TT1}
\end{table}
Parameters $\lambda _{1}$, $\lambda _{2}$, $\beta $, and $s_{\mathrm{F}}$ in
Eq. (\ref{Eq2}) correspond to three different regimes addressed in this
article, which are listed in Table \ref{TT1}. It is relevant
to mention that the spin polarization may affect some parameters, such as
coefficient ${C_{\mathrm{F}}}${\ \cite{Andreev18}}.

Lagrangian density (\ref{Eq2}) gives rise to the following The
Euler-Lagrange equation,
\begin{equation}
\frac{i\hbar }{\lambda _{1}}\frac{{\partial {\Psi }}}{{\partial t}}=\left[ -%
\frac{{\hbar ^{2}}}{{2{\lambda _{2}}{m_{\mathrm{F}}}}}{\nabla ^{2}}+{U}+g_{%
\mathrm{F}}{{\left\vert {\Psi }\right\vert }^{2}}\right. \left. +\frac{{%
\hbar ^{2}}}{{2{m_{\mathrm{F}}}}}\beta {C_{\mathrm{F}}}{{\left\vert {\Psi }%
\right\vert }^{4/3}}\right] {\Psi },  \label{Eq3}
\end{equation}%
which as an effective quasi-mean-field equation for the fermi gas under the
consideration; note that it may be rewritten in the form of
hydrodynamic equations \cite{Kim2004,Adhikari2006b}. More details on the
derivation of this equation are given in Appendix A. Below, we focus on the
BCS (Bardeen-Cooper-Schrieffer) setting, referring to atoms of $^{6}$Li with
mass $6$ a.u.

In numerical simulations we use the fourth-order Runge-Kutta
method in time, and the centered second-order finite-difference method for
handling the spatial discretization. In the next two subsections we reduce
the full 3D equation to the corresponding 2D and 1D effective equations,
using the VA proposed in Ref. \cite{Diaz12}.

\subsection{The two-dimensional reduction}

We derive effective 2D equations, applying the VA to the Fermi gas in the
disk-shaped trap. For this purpose, we consider an external potential
composed of two terms: the parabolic (harmonic-oscillator) one accounting
for the confinement in the $z$ direction, transverse to the disk's plane,
and the in-plane potential, $U_{\mathrm{2D}}$: %%
\begin{equation}
U\left( {\mathbf{r}}\right) =\frac{1}{2}m_{\mathrm{F}}\omega _{z}^{2}{z^{2}+}%
U_{\mathrm{2D}}\left( {\mathbf{r}_{\bot },t}\right) .  \label{Eq4}
\end{equation}%
The initial ansatz assumes, as usual, the factorization of the 3D wave
function into a product of functions of $z$ and $\mathbf{r}_{\bot }$, the
former one being the Gaussian ground state of the harmonic-oscillator
potential \cite{Salasnich02a}:
\begin{equation}
\Psi \left( {{\mathbf{r}},t}\right) =\frac{1}{{{\pi ^{1/4}}}\sqrt{{\xi (%
\mathbf{r}_{\bot },t)}}}{\exp }\left( -\frac{z^{2}}{2(\xi (\mathbf{r}_{\bot
},t))^{2}}\right) \phi \left( {\mathbf{r}_{\bot },t}\right) .  \label{Eq5}
\end{equation}%
The Gaussian is subject to the unitary normalization, with transverse width $%
\xi $ considered as a variational parameter, while the 2D wave function, $%
\phi $, is normalized to the number of atoms. Therefore, the reduction from
3D to 2D implies that the system of equations should be derived for the pair
of functions $\phi \left( \mathbf{r}_{\bot },t\right) $ and $\xi \left(
\mathbf{r}_{\bot },t\right) $, using the reduced action functional, which is
obtained by integrating the 3D action over the $z$-coordinate: %%%
\begin{equation}
\mathcal{S}{_{\mathrm{2D}}}=\int {dtdxdy{\mathcal{L}}_{\mathrm{2D}}},
\label{Eq6}
\end{equation}%
where the respective Lagrangian density is %%%
\begin{eqnarray}
{\mathcal{L}_{{\mathrm{2D}}}} &=&i\frac{\hbar }{2\lambda _{1}}\left( {{\phi
^{\ast }}{\partial _{t}}\phi -\phi {\partial _{t}}{\phi ^{\ast }}}\right) -%
\frac{{\hbar ^{2}}}{{2\lambda _{2}m_{\mathrm{F}}}}{\left\vert {{\nabla
_{\bot }}\phi }\right\vert ^{2}}-{U_{{\mathrm{2D}}}}{\left\vert {\phi }%
\right\vert ^{2}}-\frac{{\hbar ^{2}}}{{2m_{\mathrm{F}}}}\frac{{3\beta C_{2D}}%
}{{5{\xi ^{2/3}}}}{\left\vert {\phi }\right\vert ^{10/3}}-\frac{g_{\mathrm{F}%
}}{{2{{\left( {2\pi }\right) }^{1/2}}\xi }}{\left\vert {\phi }\right\vert
^{4}}  \notag \\
&&-\frac{{\hbar ^{2}}}{{4m_{\mathrm{F}}\lambda _{2}{\xi ^{2}}}}{\left\vert {%
\phi }\right\vert ^{2}}-\frac{1}{4}{{m_{\mathrm{F}}\omega _{z}^{2}{\xi ^{2}}}%
}{\left\vert {\phi }\right\vert ^{2}},  \label{Eq7}
\end{eqnarray}%
$C_{\mathrm{2D}}\equiv {(3/5)^{1/2}}{(6/(2s_{\mathrm{F}}+1))^{2/3}}\pi $,
the last two terms being produced by the reduction to 2D, the penultimate
term corresponding to the spread in the confined dimension. Hence, the
Euler-Lagrange equations, derived by varying the 2D action, which is
generated by Lagrangian (\ref{Eq7}), with respect to $\phi $ and $\xi $ take
the form of
\begin{eqnarray}
i\frac{\hbar }{\lambda _{1}}{\partial _{t}}{\phi } &=&\left[ {-\frac{\hbar
^{2}}{{2{\lambda _{2}}{m_{\mathrm{F}}}}}\nabla _{\bot }^{2}+{U_{2\mathrm{D}}}%
+\frac{g_{\mathrm{F}}}{{\sqrt{2\pi }{\xi }}}{\left\vert {\phi }\right\vert
^{2}}}\right. +\frac{{\hbar ^{2}}}{2m_{\mathrm{F}}}\frac{\beta }{{\xi ^{2/3}}%
}{C_{2\mathrm{D}}}{\left\vert {\phi }\right\vert ^{4/3}}+\frac{{{\hbar ^{2}}}%
}{{4\lambda _{2}{m_{\mathrm{F}}}\xi ^{2}}}  \notag \\
&&+\left. {\frac{1}{4}}{m_{\mathrm{F}}}\omega _{z}^{2}\xi ^{2}\right] {\phi }%
,  \label{Eq8}
\end{eqnarray}%
\begin{equation}
m_{\mathrm{F}}\omega _{z}^{2}\xi ^{4}-\frac{{2{\hbar ^{2}}}}{5m_{\mathrm{F}}}%
\beta {C_{2\mathrm{D}}}{\left\vert {\phi }\right\vert ^{4/3}}\xi ^{4/3}-%
\frac{g_{\mathrm{F}}}{\sqrt{2\pi }}{\left\vert {\phi }\right\vert ^{2}}{\xi }%
-\frac{{{\hbar ^{2}}}}{\lambda _{2}m_{\mathrm{F}}}=0.  \label{Eq9}
\end{equation}%
Algebraic equation (\ref{Eq9}) for $\xi $ cannot be solved analytically,
therefore we used the Newton's method to solve it numerically. The necessity
to find $\xi $ at each step of the integration is a numerical complication
of a minimal cost compared to the 3D integration of the underlying equation (%
\ref{Eq3}). Note that a further simplifications can be achieved
by assuming in Eq. (\ref{Eq5}) that the Gaussian width is a constant $\xi (%
\mathbf{r}_{\bot },t)=\xi _{0}$. In this case $\xi $, naturally, does not
depend on $\phi $. Then, the solution of Eq. (\ref{Eq9}) with the density
tending to zero can be calculated analytically and it is given by $\xi
_{0}=\lambda _{2}^{-1/4}\sqrt{\hbar /m_{\mathrm{F}}\omega _{z}}$.

\begin{figure}[tbp]
\begin{centering}
\begin{tabular}{lll}
(a) & (b) &  \\
\includegraphics[width=0.45\textwidth]{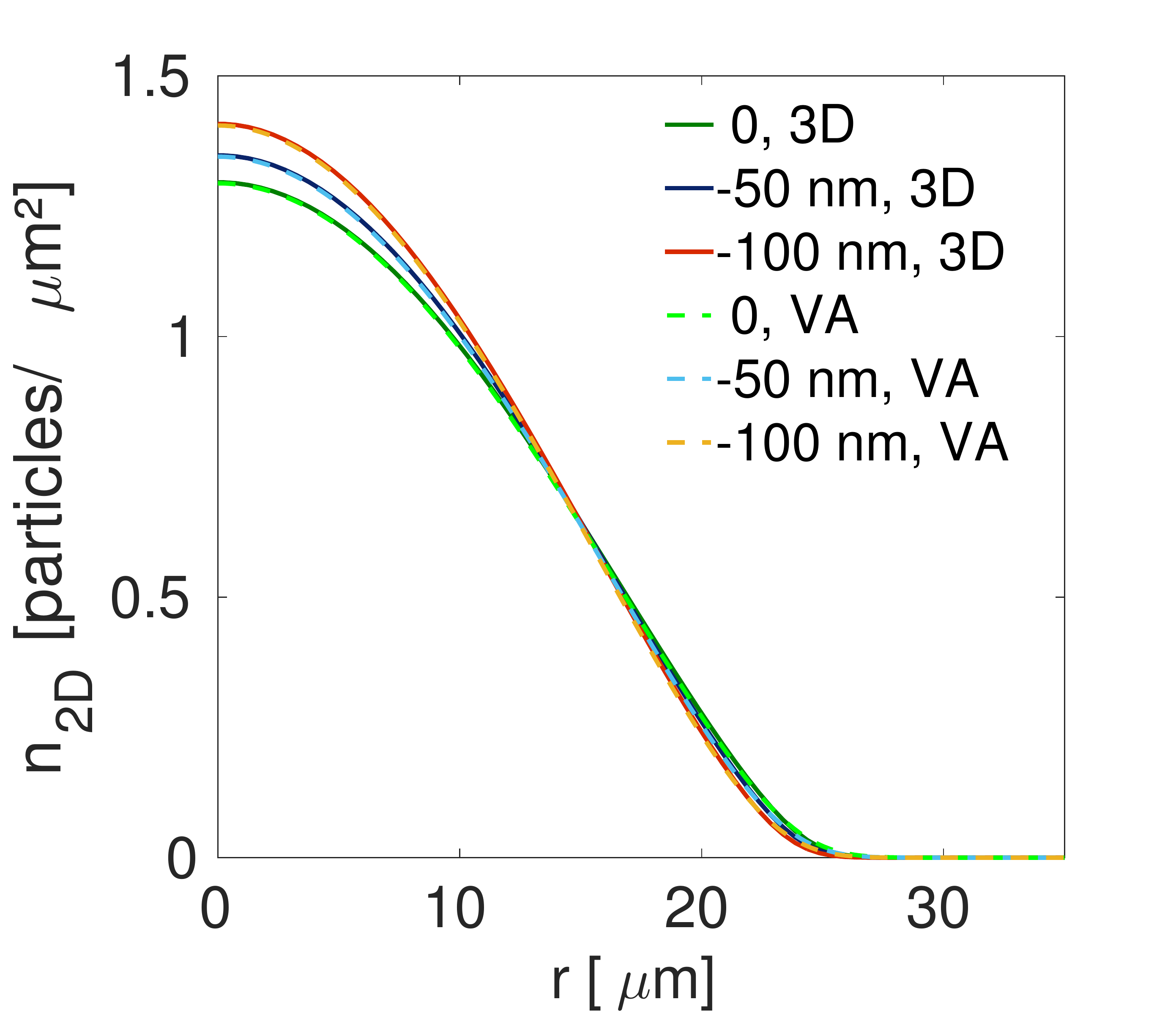} &
\includegraphics[width=0.45\textwidth]{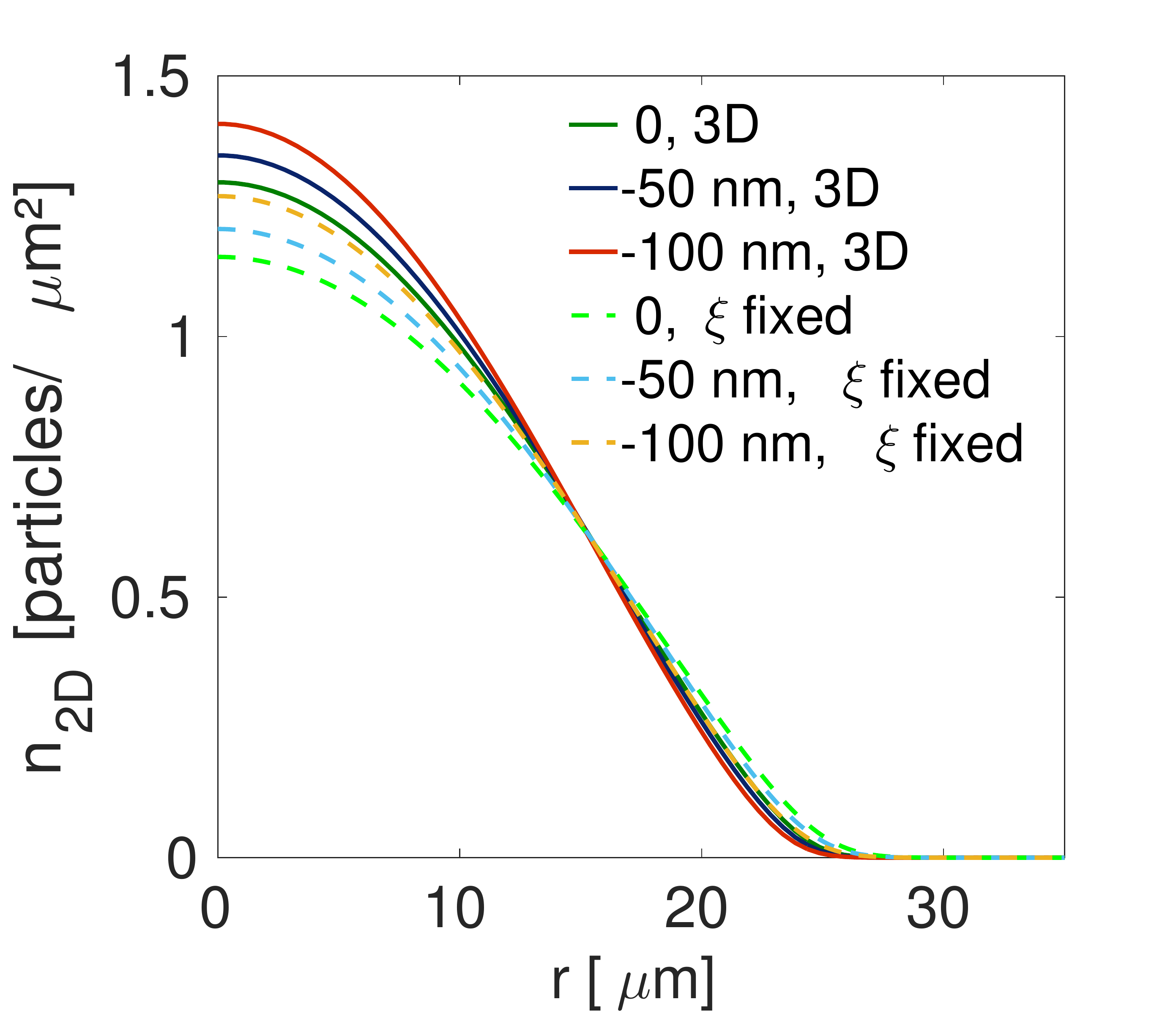} &
\end{tabular}%
\caption{(a) The 2D radial density, $\protect n_{\mathrm{2D}}(r)$, as obtained from the full 3D equation, and the 2D reduction derived with the help of the VA. (b) The 2D radial density, $\protect n_{\mathrm{2D}}(r)$, as obtained from the full 3D equation, and the 2D reduction derived with the help of the VA, assuming that the Gaussian width is a constant: $\xi_0 = \sqrt{\hbar/2m_{\mathrm{F}}\omega _{z} }$. Different curves correspond to the indicated values of $a_{s}=(0,-50,-100)\mathrm{nm}$. The other parameters are $N=1000$, $\omega_{x}=\omega_{y}=1050\mathrm{Hz}$, $\omega_{z}=21\mathrm{kHz}$ and $A=0$. The panel (a) is taken from Ref. \cite{Diaz12}.}
\label{Fig1}
\end{centering}
\end{figure}

We consider a 2D potential consisting of the axisymmetric parabolic
potential and the superposition of two triangular OLs:
\begin{equation}
{U_{\mathrm{2D}}}=A\sum\limits_{b=1}^{2}{\sum\limits_{a=1}^{3}{{{\sin }^{2}}%
\left( \frac{2\pi }{\lambda }\;({{\mathbf{k}_{a,b}}\cdot {\mathbf{r}_{\bot }}%
})\right) }}+\frac{1}{2}\omega _{x}^{2}{x^{2}}+\frac{1}{2}\omega _{y}^{2}{%
y^{2}},  \label{Eq10}
\end{equation}%
where $\{\mathbf{k}_{a,1}\}$ and $\{\mathbf{k}_{a,2}\}$ are triplets of
unitary vectors of both triangular lattices, which are separated by a
specific angle $\theta $. Here $A$ denotes the lattice's amplitude, and $%
(\omega _{x},\omega _{y})$ are frequencies of the magnetic-optical trapping
potential. In the absence of the OLs ($A=0$), we have verified the accuracy
of the 2D reduction by comparing results generated by this approximation to
those obtained by integrating the underlying 3D equation (\ref{Eq3}). The
respective GS was found by means of the imaginary-time integration based on
the fourth-order Runge-Kutta algorithm with $\Delta t=0.5$ $\mu $s. The
spatial discretization for the simulations was performed with $\Delta x=0.25$
$\mu $m and $\Delta y=0.25$ $\mu $m. The comparison is displayed in 
panel (a) of Figure \ref{Fig1}, where the radial-density
profiles are plotted. We can observe excellent agreement between the reduced
2D and full 3D descriptions. This result suggests one to use the Eqs. (\ref%
{Eq8}) and (\ref{Eq9}) for studying 2D patterns. Panel (b)
of Figure \ref{Fig1} shows a comparison of 3D full numerical simulations versus 
the VA, assuming a constant width $\xi _{0}$. One can observe that
the latter approximation produces less accurate results, which is at least
ten times worse than the VA with a density-dependent width.

\begin{figure}[tbp]
\begin{centering}
\begin{tabular}{lll}
(a) & (b) &  \\
\includegraphics[width=0.3\textwidth]{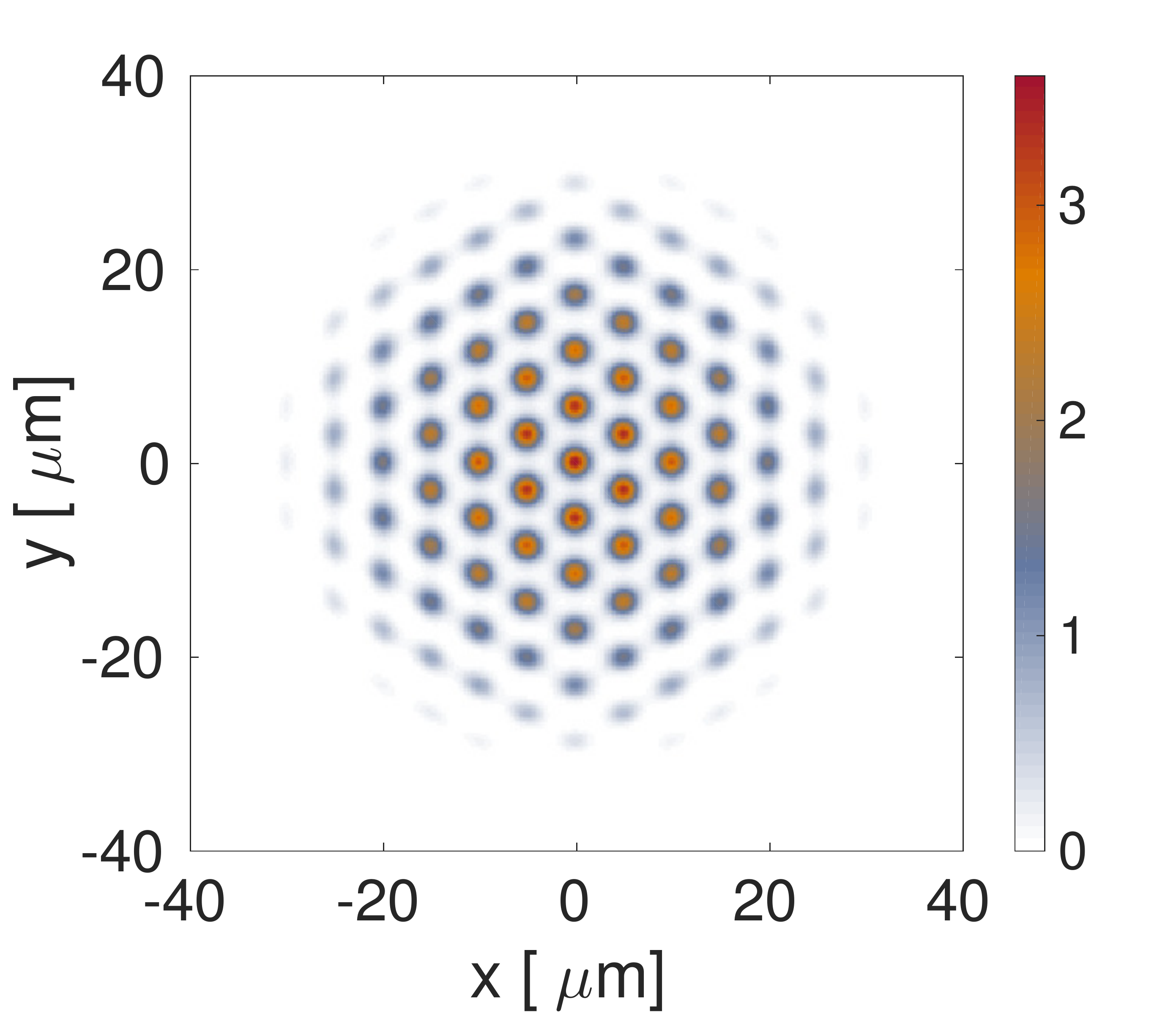} & %
\includegraphics[width=0.3\textwidth]{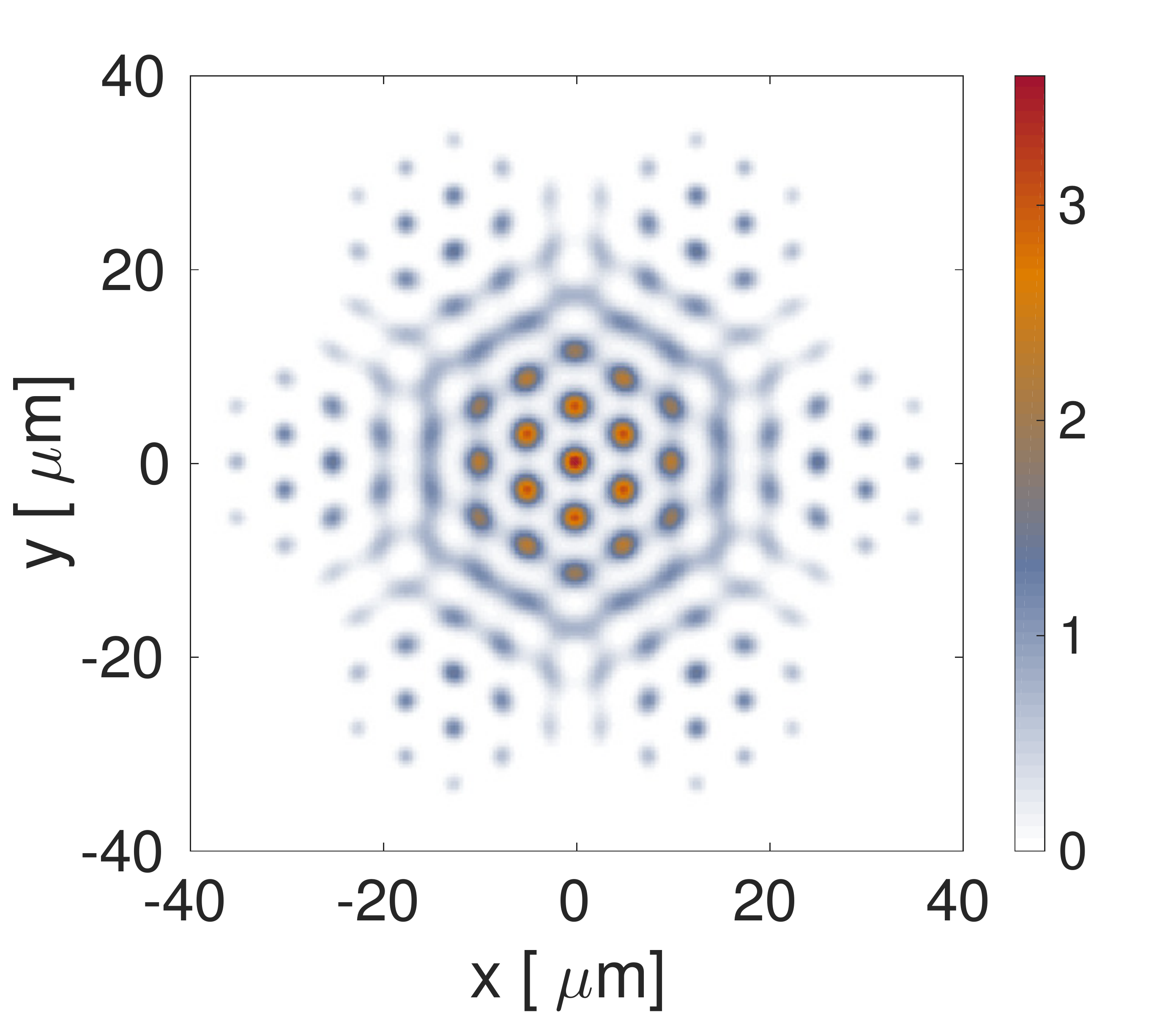} &  \\
&  &  \\
(c) & (d) &  \\
\includegraphics[width=0.3\textwidth]{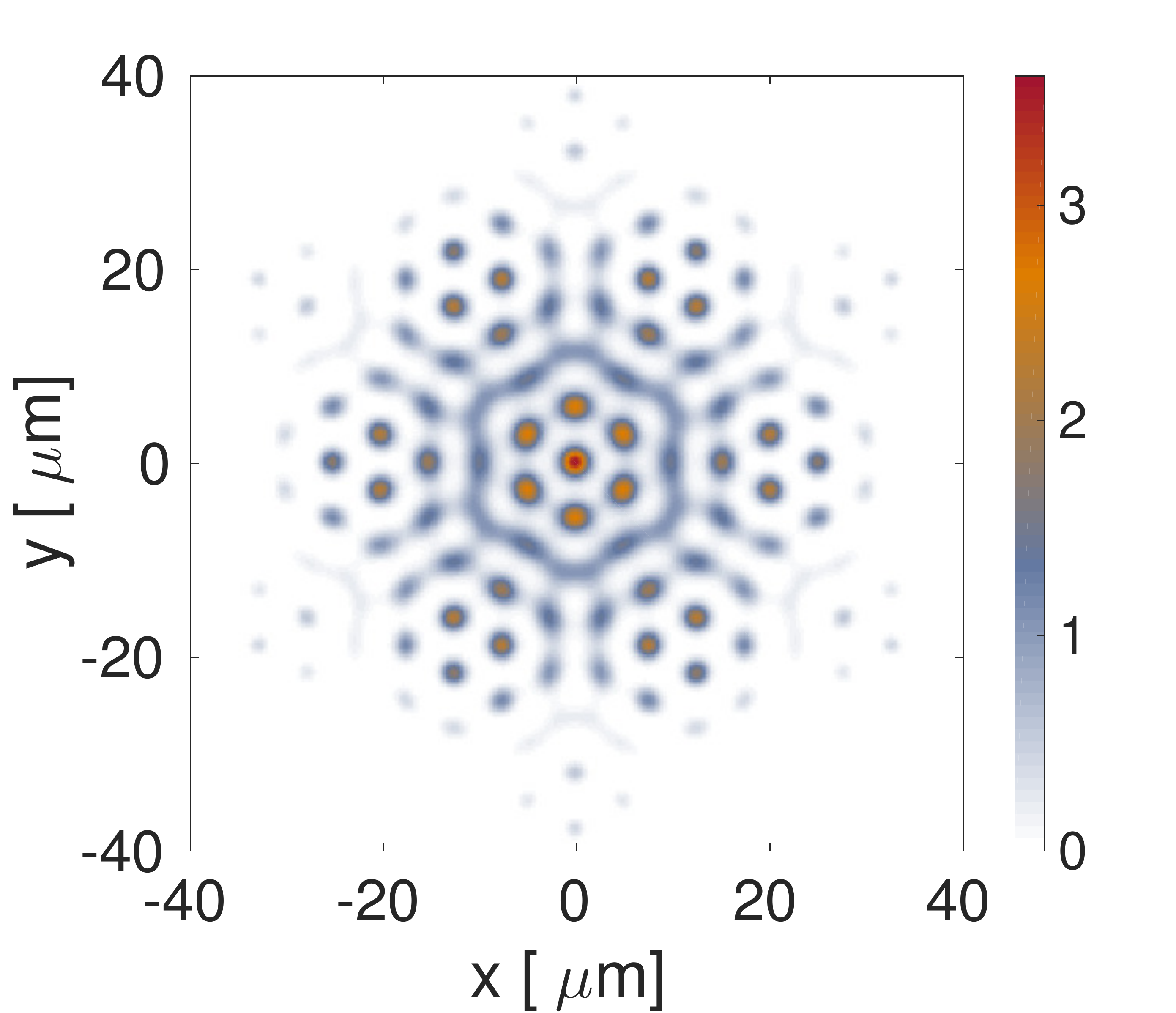} & %
\includegraphics[width=0.3\textwidth]{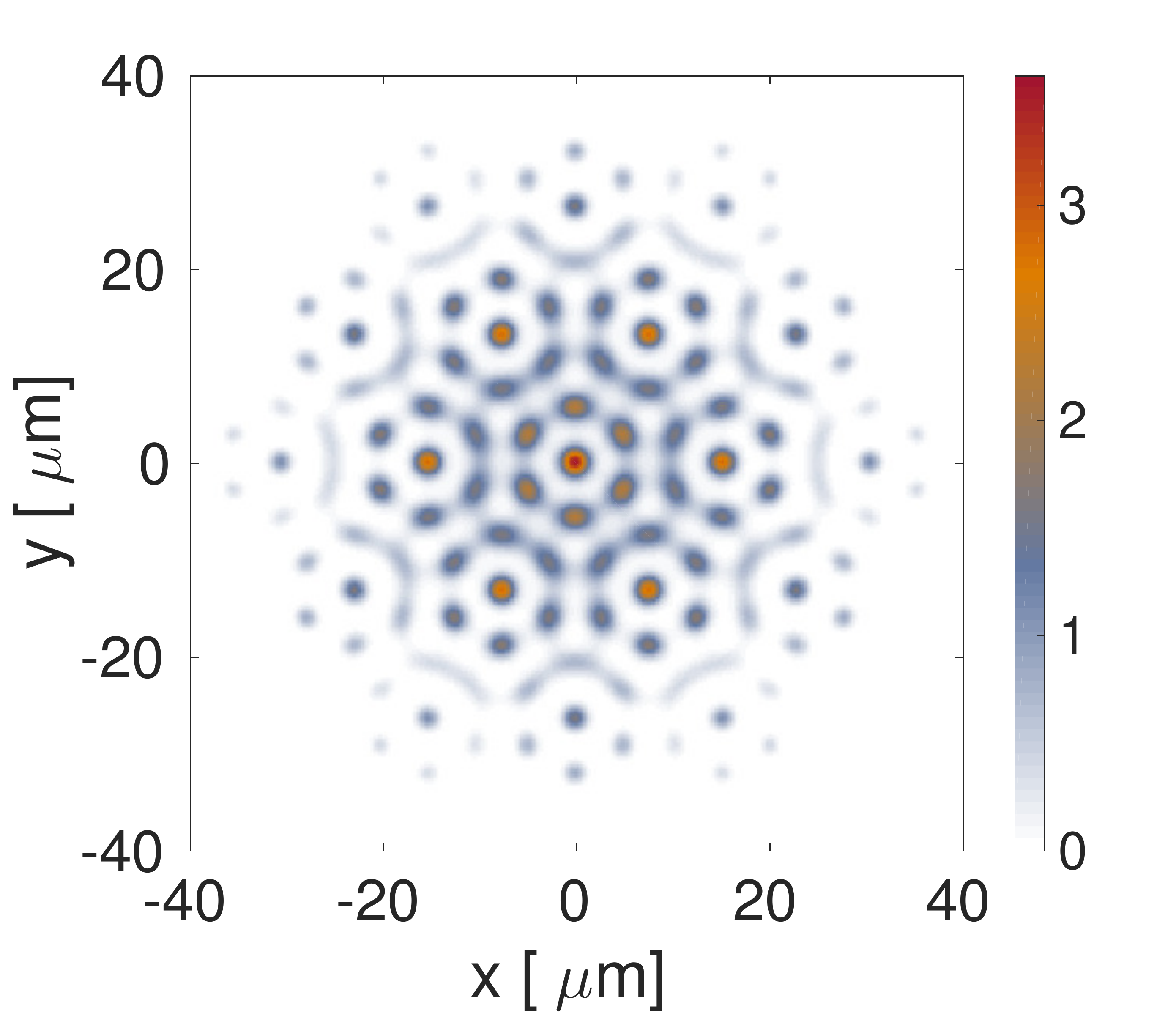} &  \\
&  &  \\
\end{tabular}%
\caption{Density $\protect n_{2D}$ as a function of coordinates $x$ and $y$ for four different angles between the triangular OLs $\theta=(5º,10º,15º,20º)$. The fixed parameters are $N=1000$, $\omega_x=\omega_y=1050\mathrm{Hz}$, $\omega_z=52.5\mathrm{kHz}$, $A=1.74\times10^{-29}\mathrm{J}$, $\lambda=10\mu\mathrm{m}$ and $a_s=200\mathrm{nm}$.}
\label{Fig2R}
\end{centering}
\end{figure}

Figure \ref{Fig2R} shows the density as a function of coordinates $x$ and $y$
when the OLs are taken into account. We observe that this particular
combination of the OLs (a superlattice) produces a pattern in the form of
the superstructure, with the number of density peaks varying when the angle
between the unitary vectors increases. Note that the multitude of different
coexisting robust multi-peak patterns suggests that this setting has a
potential for the use as a data-storage system.

\subsection{The one-dimensional reduction}

Next, we consider the system confined in one dimension, which implies a
cigar-shaped configuration, elongated in the $z$ direction. In this case,
the potential trap acting in transeverse plane is the harmonic oscillator in
the transverse plane:
\begin{equation}
U\left( {{\mathbf{r}}}\right) =U_{1\mathrm{D}}\left( {z}\right) +\frac{1}{2}%
m\omega _{t}^{2}r^{2},  \label{Eq11}
\end{equation}%
where $r^{2}={x^{2}}+{y^{2}}$. It is assumed that the potential in the
transverse direction is much stronger than the axial one. The simplest
option is to adopt a Gaussian shape in the transverse plane, which
represents the ground state of a the 2D harmonic oscillator , similar to
what is adopted above in the case of the 2D reduction. As a result, the
variable-separation assumption can be applied, defining the 3D wave function
as \cite{Salasnich02a,Salasnich02b,Salasnich2D}
\begin{equation}
\Psi \left( {{\mathbf{r}},t}\right) =\frac{1}{{{\pi ^{1/2}}{\sigma }\left( {%
z,t}\right) }}{\exp {\ \left( -\frac{r^{2}}{{2{\sigma }{{\left( {z,t}\right)
}^{2}}}}\right) }}f\left( {z,t}\right) ,  \label{Eq12}
\end{equation}%
where $f$ is normalized to $N$, such that the 1D density is ${n_{1\mathrm{D}}%
}={\left\vert f\right\vert ^{2}}$. Here $\sigma $ is the Gaussian width,
which is a function of $z$ and time. After some algebra, similar to that
performed above, one derives the Euler-Lagrange equations:
\begin{equation}
i\frac{\hbar }{\lambda _{1}}{\partial _{t}}{f}=\left[ {\ -\frac{\hbar ^{2}}{{%
2{\lambda _{2}}{m_{\mathrm{F}}}}}\partial _{z}^{2}+U_{1\mathrm{D}}+\frac{g_{%
\mathrm{F}}}{{2\pi \sigma ^{2}}}{{\left\vert {f}\right\vert }^{2}}}+\frac{%
\hbar ^{2}}{{2{m_{\mathrm{F}}}}}\beta \frac{C_{1\mathrm{D}}}{{\sigma ^{4/3}}}%
{{\left\vert {f}\right\vert }^{4/3}}+\frac{\hbar ^{2}}{2{m_{\mathrm{F}}}%
\lambda _{2}\sigma ^{2}}\frac{1}{2}{m_{\mathrm{F}}}\omega _{t}^{2}\sigma ^{2}%
\right] {f,}  \label{Eq13}
\end{equation}%
\begin{equation}
m_{\mathrm{F}}\omega _{t}^{2}\sigma ^{4}-\frac{2}{5}\frac{\hbar ^{2}}{m_{%
\mathrm{F}}}\beta C_{1\mathrm{D}}|f|^{4/3}\sigma ^{2/3}-\frac{\hbar ^{2}}{%
\lambda _{2}m_{\mathrm{F}}}-\frac{g_{\mathrm{F}}}{{2\pi }}|f|^{2}=0,
\label{Eq14}
\end{equation}%
where $C_{1\mathrm{D}}=(3/5){(6\pi (2{s_{\mathrm{F}}}+1))^{2/3}}$. Similar
to the 2D case, algebraic equation (\ref{Eq14}) is solved using the Newton's
method, and here too the quasi-BCS regime is addressed. We set $U_{1\mathrm{D%
}}=0$, $a_{\mathrm{F}}=-5$ nm, and $\omega _{t}=1000$ Hz, these parameters
being in the range of experimental values \cite{Yefsah13}. Since Eqs. (\ref%
{Eq13}) and (\ref{Eq14}) produce results which agree well with the full 3D
simulations \cite{Diaz12}, one can use the effective 1D equations to study
more complex dynamical behavior, such as that of dark solitons \cite%
{Yefsah13,Alphen18}.

\begin{figure}[tbp]
\begin{centering}
\begin{tabular}{lll}
(a) & (b) &  \\
\includegraphics[width=0.3\textwidth]{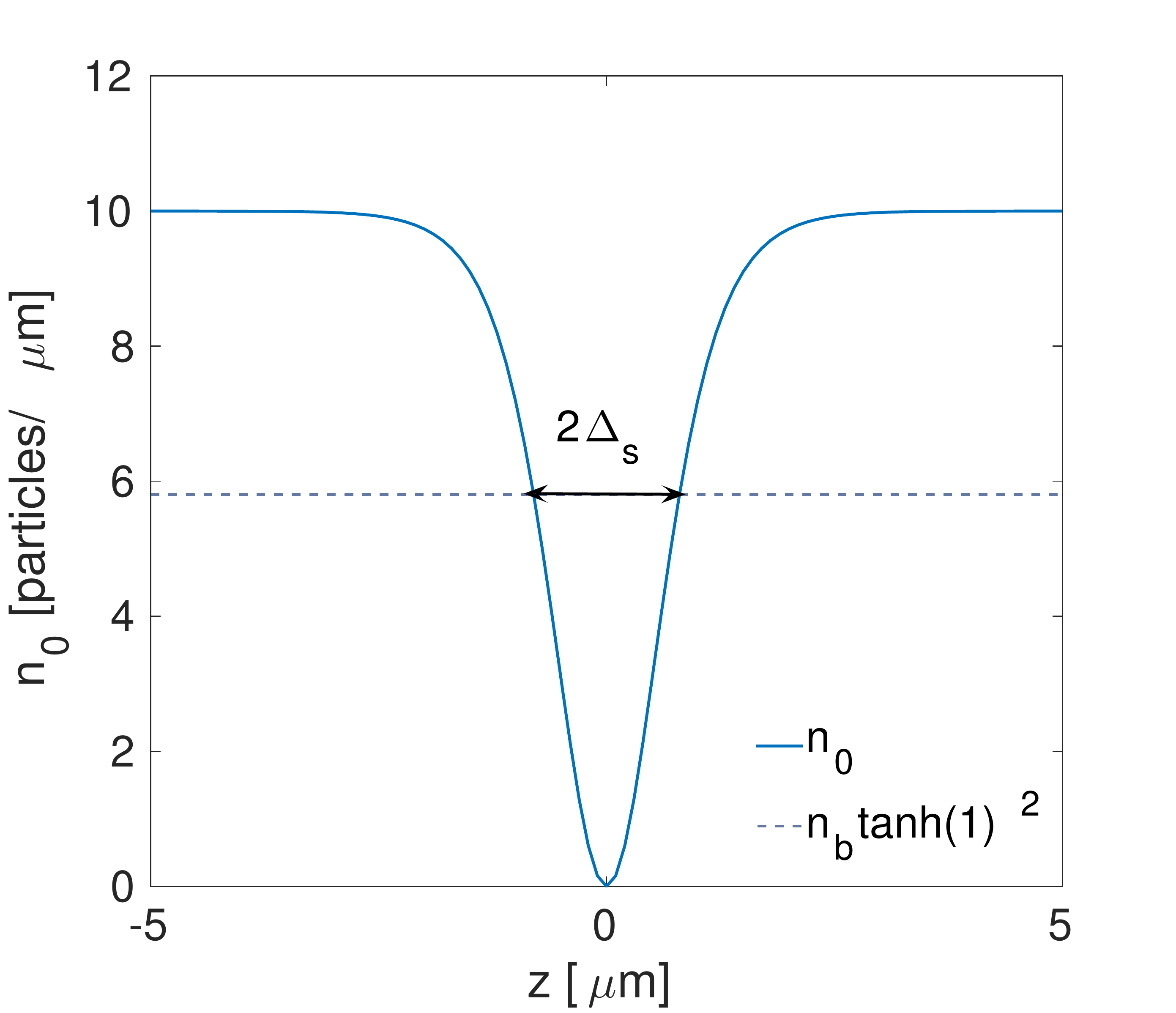} & %
\includegraphics[width=0.3\textwidth]{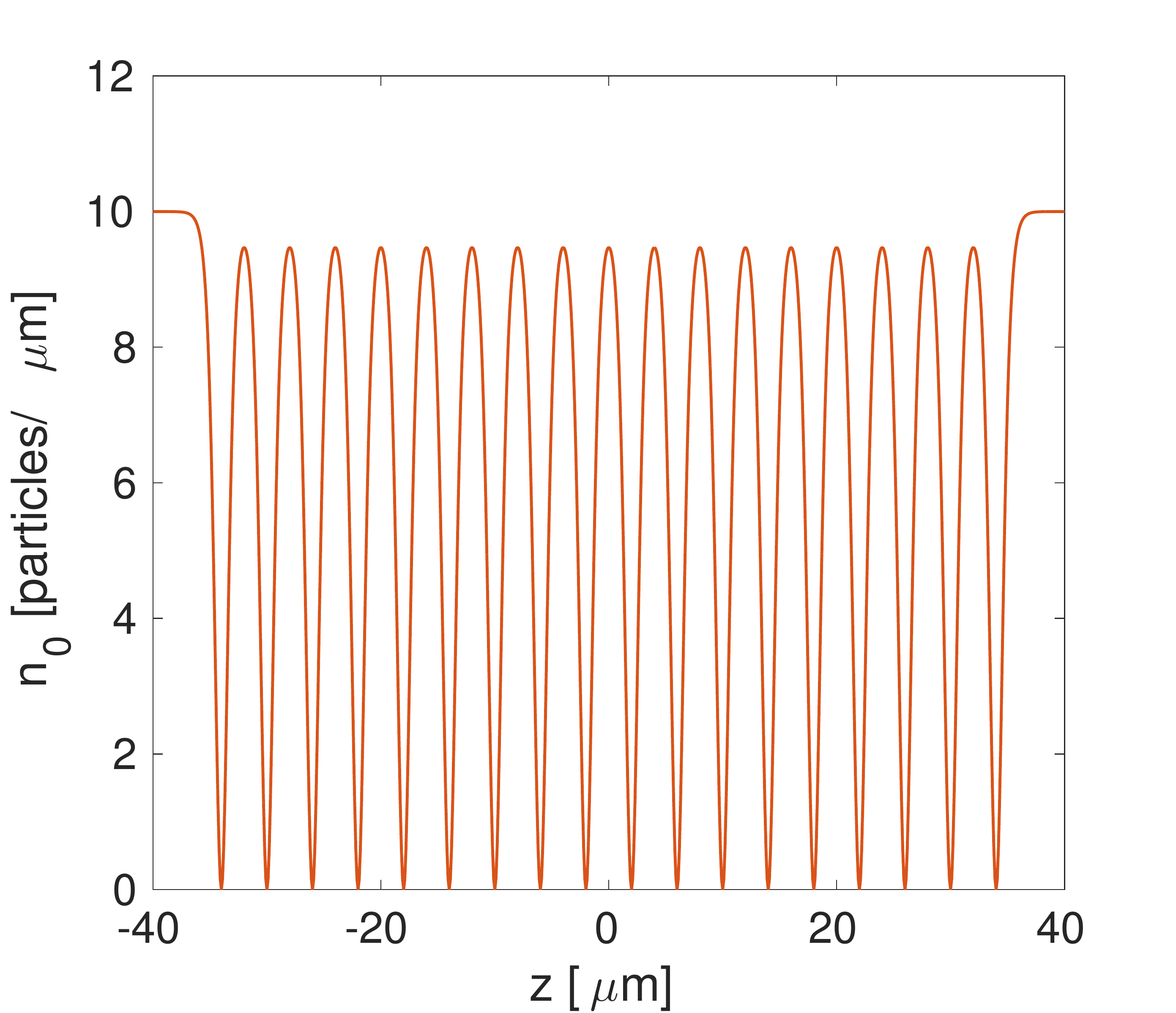} &
\end{tabular}%
\caption{The initial 1D density for one soliton (a) and eighteen dark solitons (b).
In both cases, $\Delta_s=0.8 \mu$m and $n_b=10$ are used. The other fixed parameters
are $a_\mathrm{F}=-5$ nm and $\omega_t=1000$ Hz.}
\label{FIG3R}
\end{centering}
\end{figure}

To generate a dark soliton, it is posible to consider the initial condition
with zero imaginary part, $f_{I}(z,t=0)=0$, while the real part is given by $%
f_{R}(z,t=0)=f_{b}\tanh ((z-z_{s})/\Delta _{s})$, where $f_{b}$ and
$\Delta _{s}$ are the soliton's amplitude and width,
respectively. We have found that values of the square amplitude and width, $%
n_{b}=10$ particles/$\mu $m and $\Delta _{s}=0.8~\mu $m, respectively, can
be chosen to minimize the background noise. If we consider a set of $N_{s}$
dark solitons, the initial condition for the imaginary part is again zero, $%
f_{I}=0$, while the real part can be cast in the form of
\begin{equation}
f_{R}\left( z,t_{0}\right) =\frac{1}{2}+\frac{1}{2}\sum%
\limits_{j=1}^{N_{s}/2}{\left[ {\left( {-1}\right) }^{1+j}\left( {\tanh
\left( {\frac{{z-{z}}_{j}}{\Delta _{s}}}\right) +1}\right) +{{\left( {-1}%
\right) }^{j}}\left( {\tanh \left( {\frac{{z-{z_{-j}}}}{\Delta _{s}}}\right)
+1}\right) \right] },
\end{equation}%
where the positions of the solitons are $z_{j}$ and $z_{-j}$ on the positive
and negative $z$ half-axes, receptively. Moreover, the widths of the
solitons ($\Delta _{s}$) are considered the same, and that the number of
initial solitons $N_{s}$ is even. This initial ansatz was normalized to
secure the correct density of the wave function, $n_{b}=|f_{b}|^{2}$.
Then, the system was simulated with the help of the standard
fourth-order Runge-Kutta method with $\Delta t=0.095$ $\mu $s. The spatial
discretization for the simulations was performed with $\Delta z=0.100$ $\mu $%
m. Figure \ref{FIG3R} shows the shape of the initial conditions for the case
of one and $N_{s}=18$ dark solitons.

\begin{figure}[tbp]
\begin{centering}
\begin{tabular}{lll}
(a) & (b) &  \\
\includegraphics[width=0.3\textwidth]{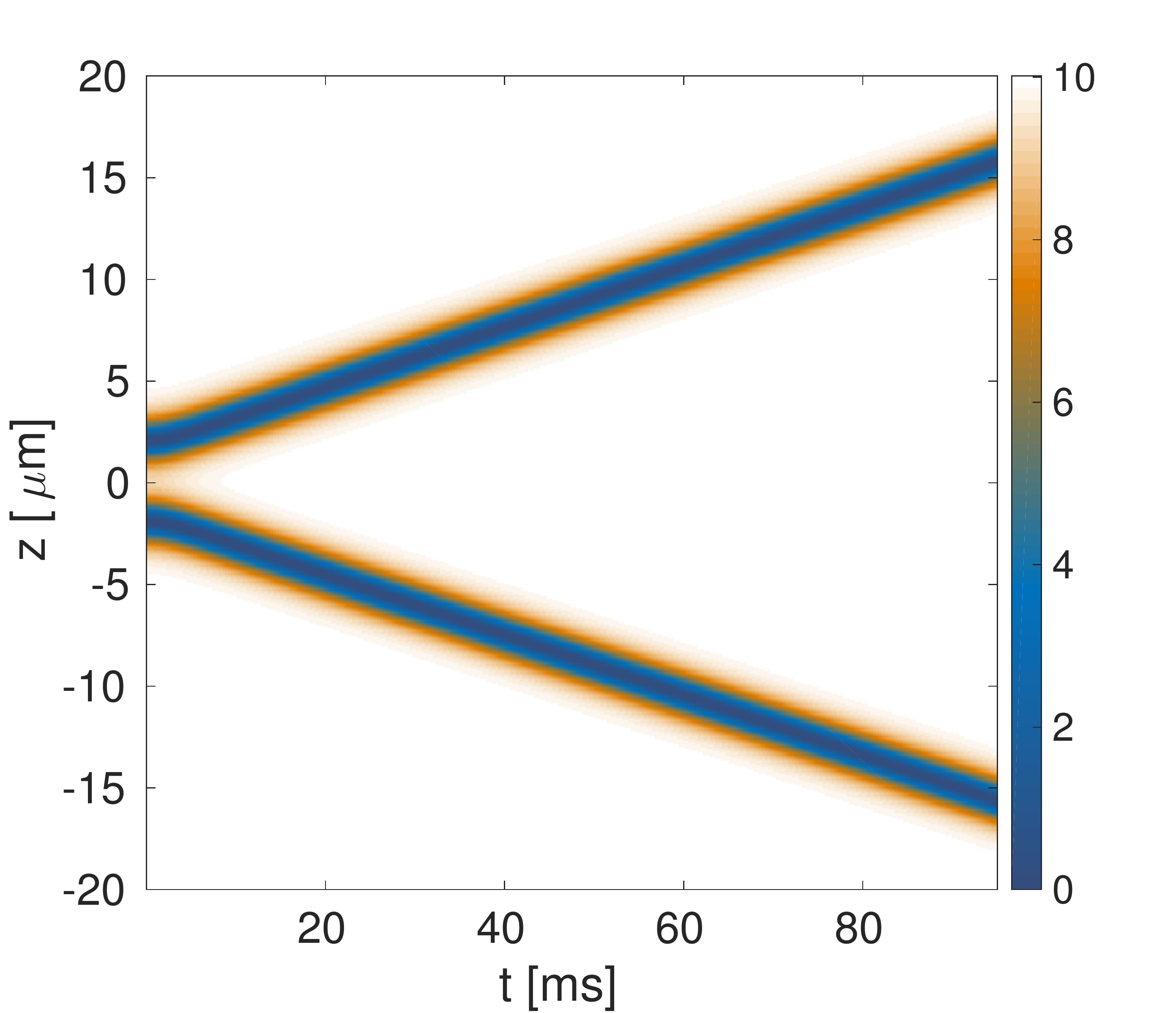} & %
\includegraphics[width=0.3\textwidth]{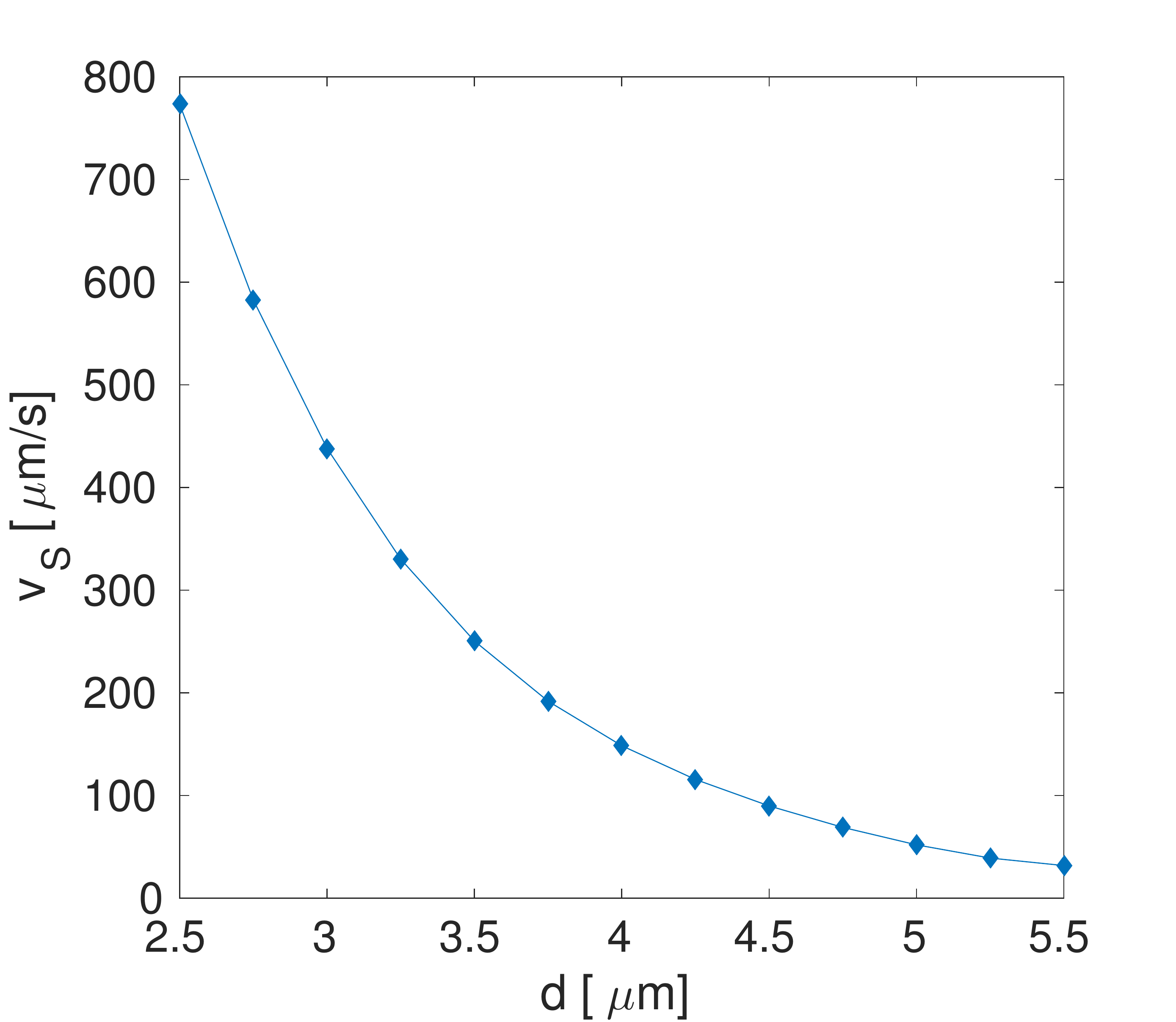} &  \\
&  &  \\
\end{tabular}%
\caption{(a) The spatiotemporal diagram for the density, $n_{1D}$, when the initial
core-core separation between the two dark solitons is
$d=4$ $\mu$m. (b) The speed of the solitons at $t=90$ms as a function of $d$. The
other fixed parameters are the same as in Fig. \ref{FIG3R}.}
\label{FIG4R}
\end{centering}
\end{figure}

\begin{figure}[tbp]
\begin{centering}
\begin{tabular}{lll}
(a) & (b) &  \\
\includegraphics[width=0.3\textwidth]{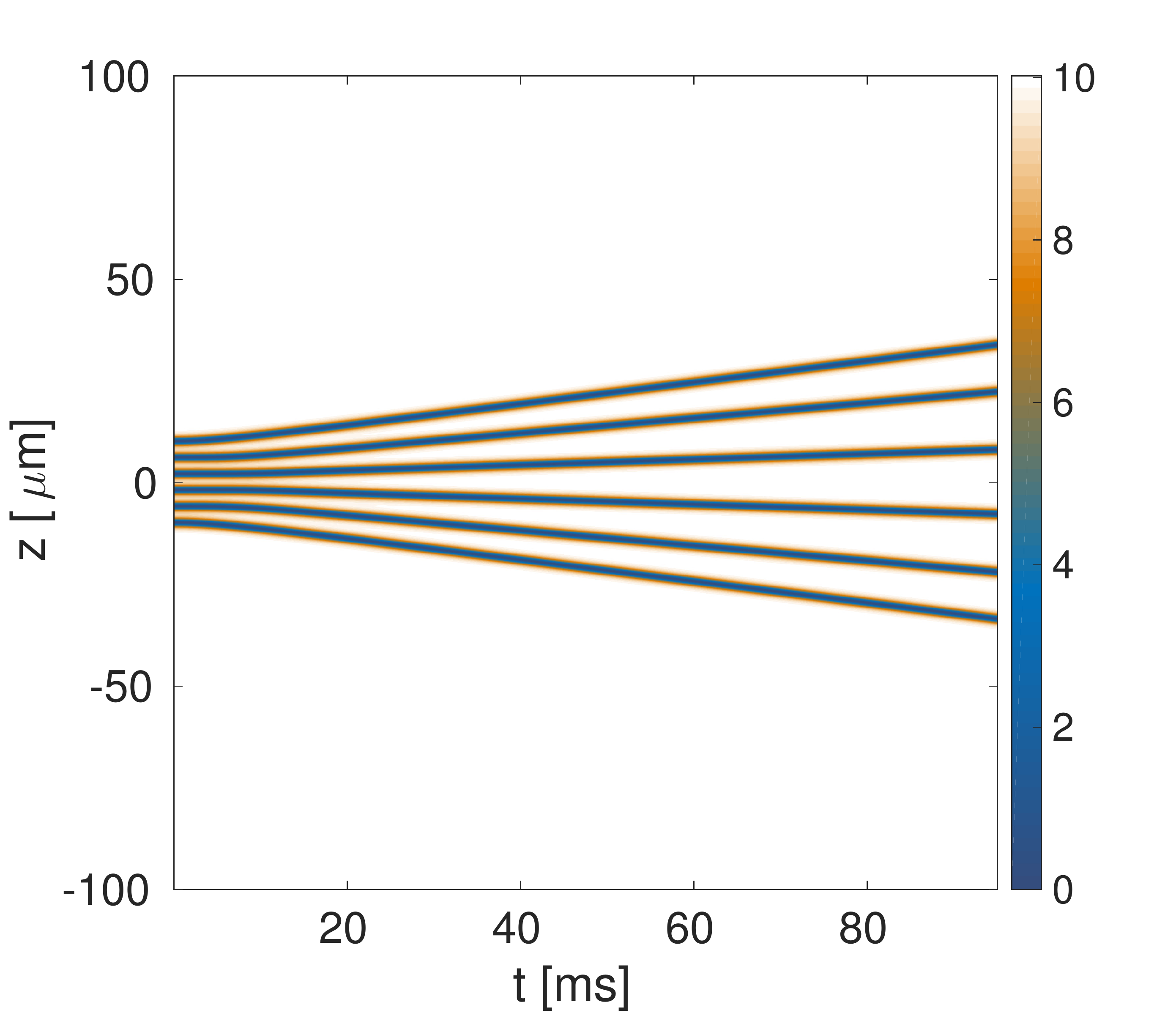} & %
\includegraphics[width=0.3\textwidth]{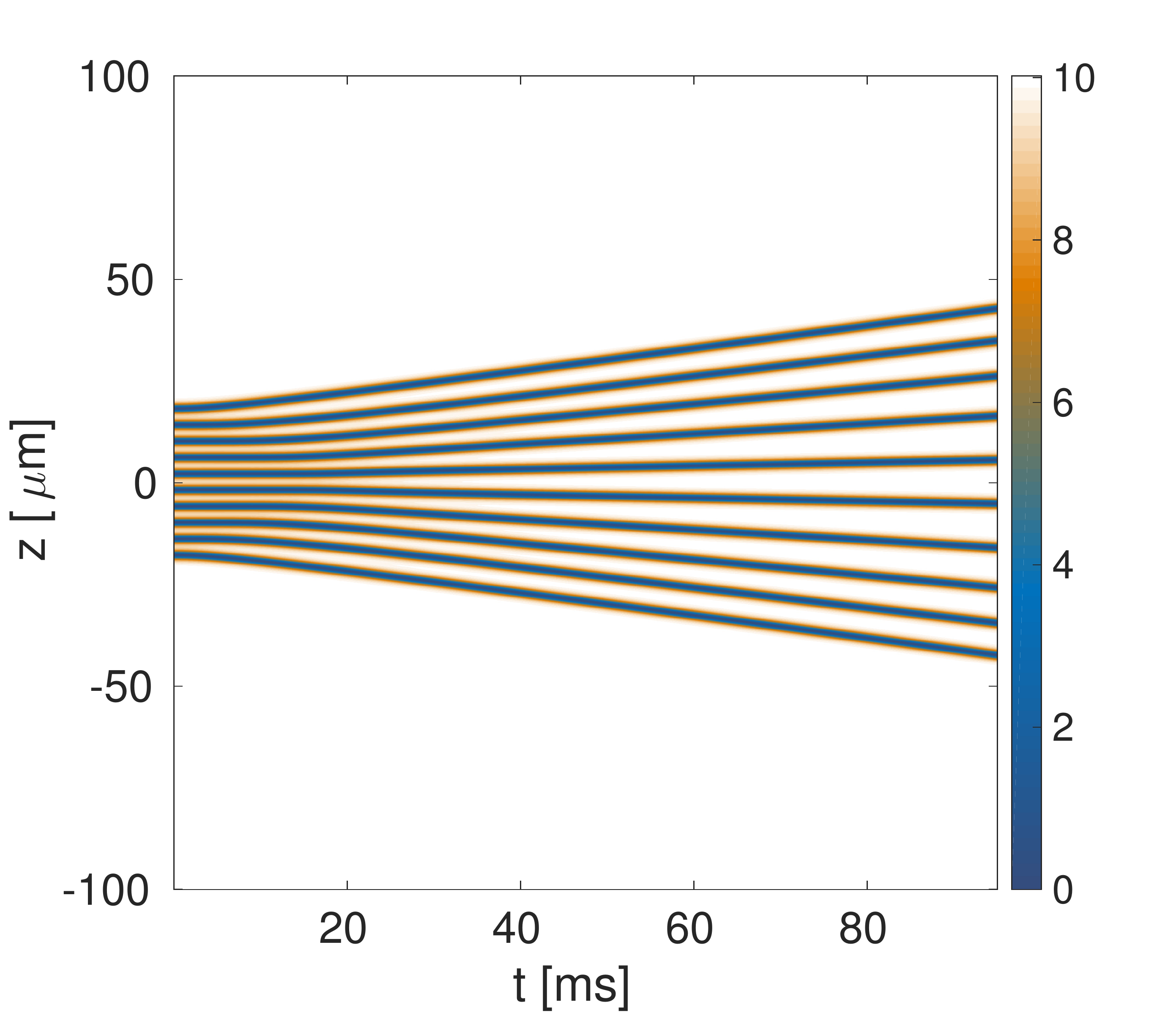} &  \\
&  &  \\
(c) & (d) &  \\
\includegraphics[width=0.3\textwidth]{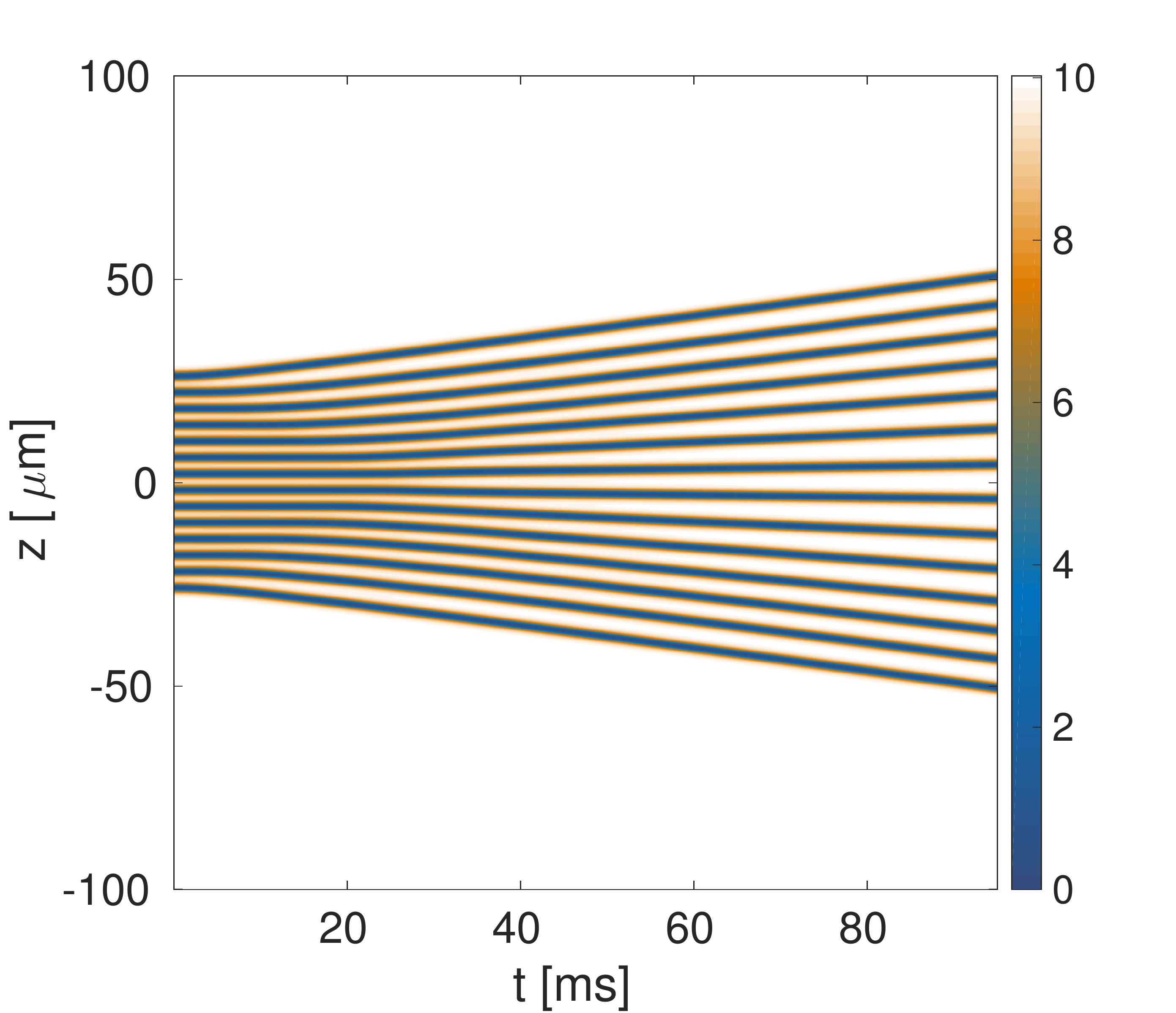} & %
\includegraphics[width=0.3\textwidth]{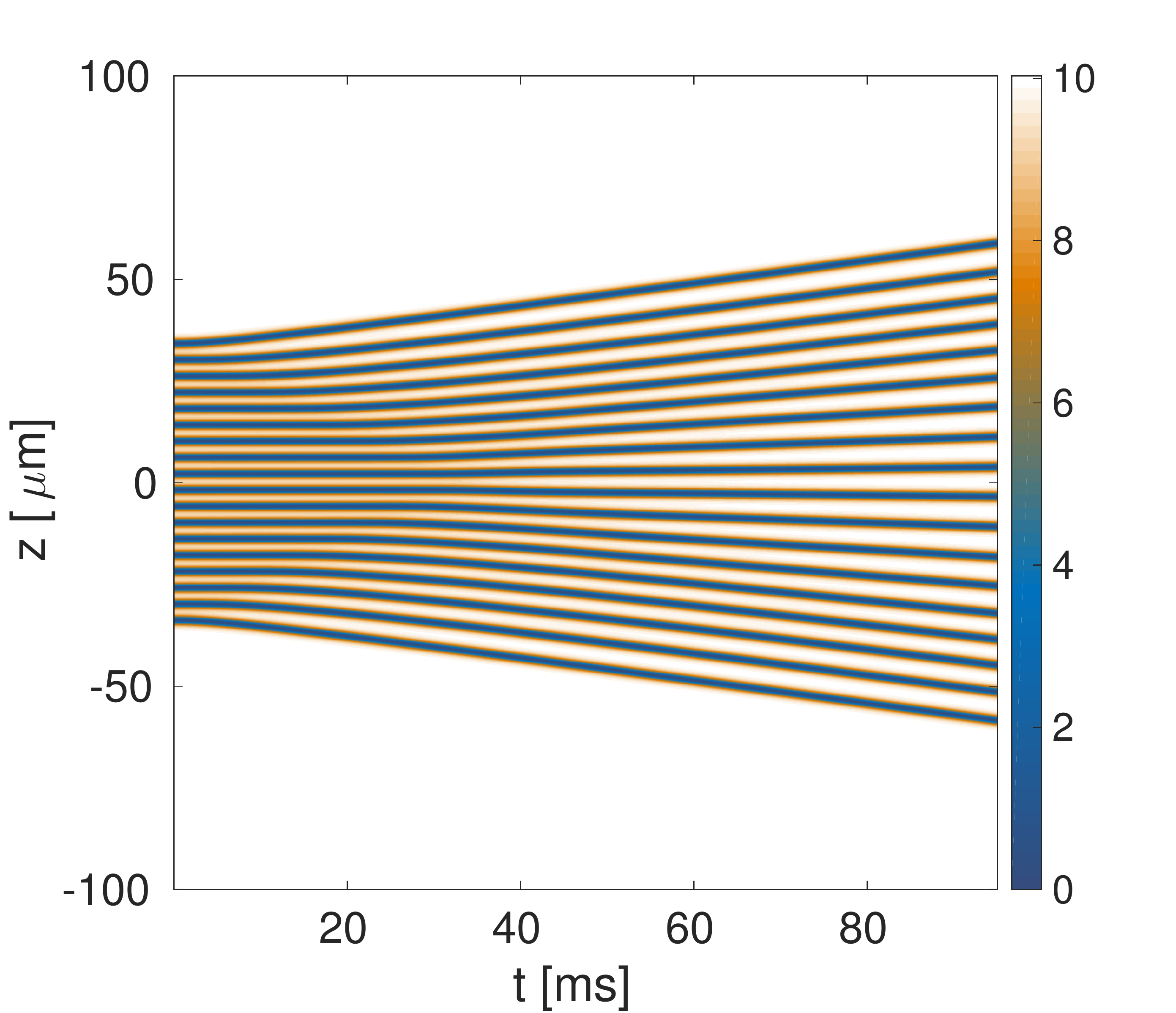} &  \\
&  &  \\
\end{tabular}%
\caption{The spatiotemporal diagram for the density, $n_{1D}$ for
different numbers of dark solitons: (a) $N_s=6$, (b) $N_s=10$, (c) $N_s=14$
and (d) $N_s=18$. In all the cases the initial distance between the solitons
is $d=4$ $\mu$m. The other fixed parameters are the same as in Fig. \ref{FIG3R}.}
\label{FIG5R}
\end{centering}
\end{figure}

In the case of two solitons, we have $z_{i}=z_{-i}=d/2$, where $d$ is the
initial inter-core separation. Frame (a) of Fig. \ref{FIG4R} shows the
spatiotemporal diagram for two solitons at $d=4\mu $m. One clearly observes
that both solitons separate in the course of the evolution. Frame (b) of
Fig. \ref{FIG4R} shows the speed taken after $90$ms of the evaluation as a
function of different initial inter-core separations between the dark
solitons. naturally, smaller initial separations generate higher speeds. In
fact, at this fix time the speed follows the law $v_{s}\sim d^{\alpha }$,
with $\alpha =-3.49$. Other features of the two-soliton interaction can be
found in Ref. \cite{Alphen18}.

Figure \ref{FIG5R} shows the spatiotemporal diagrams for the 1D density, $%
n_{1\mathrm{D}}$, for different numbers of dark solitons $N_{s}=(6,10,14,18)$%
. Similar to the case of two solitons, we observe that the solitons interact
repulsively. To measure the strength of the interaction is provided by the
distance between the central part and the positive-side border of the
soliton gas, $\delta z_{e}=|z_{\mathrm{central}}-z_{\mathrm{bond}}|$.

\begin{figure}[tbp]
\begin{centering}
\includegraphics[width=0.4\textwidth]{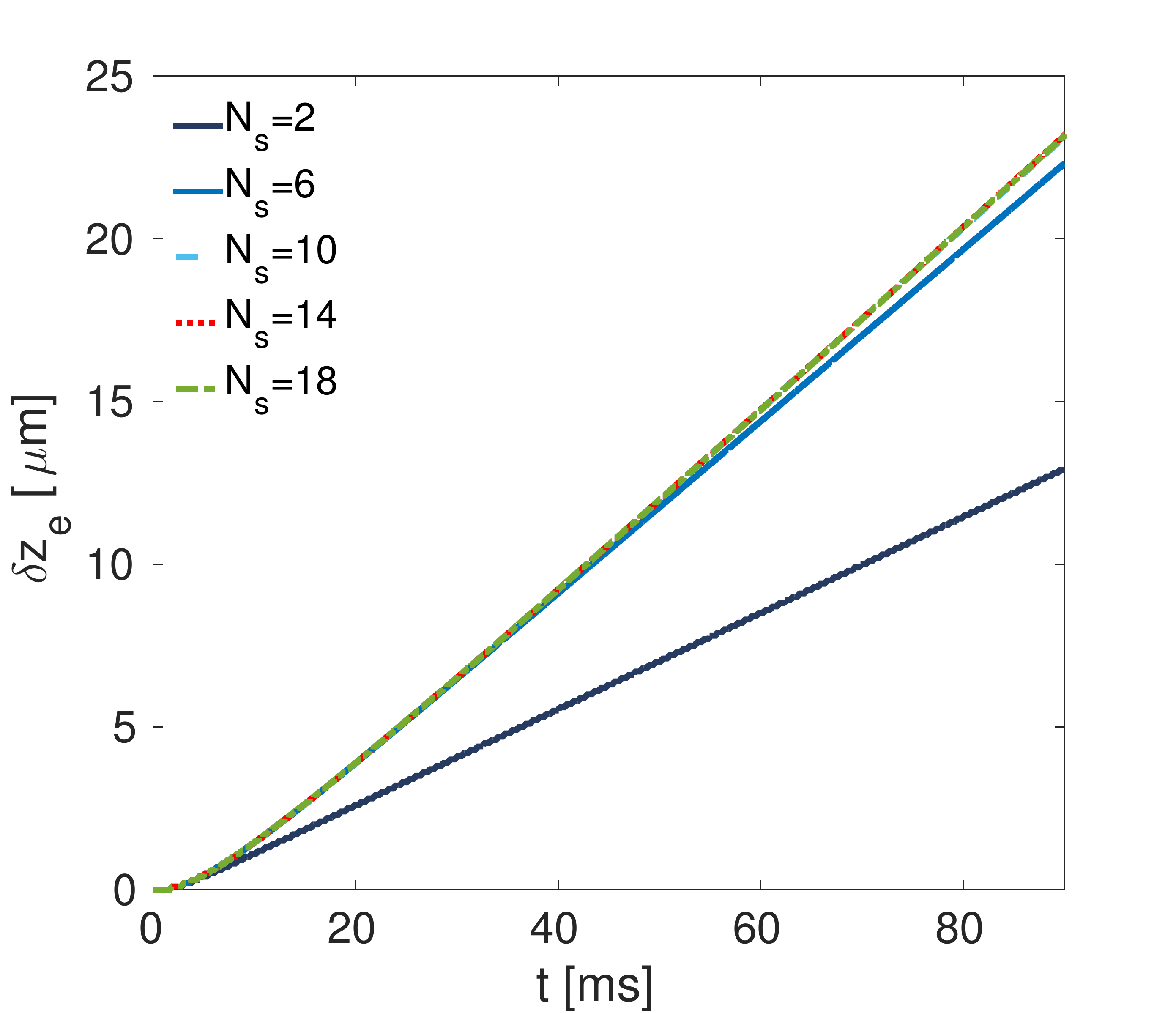} %
\caption{The distance between the central part and the edge at $z>0$ of the
dark-soliton gas, $\delta z_e$, as a function of time for different numbers
of the dark solitons, $N_s$. The other fixed parameters are the same as in Fig. \ref{FIG3R}.}
\label{FIG6R}
\end{centering}
\end{figure}

Figure \ref{FIG6R} shows $\delta z_{e}$ as a function of time for different
values of $N_{s}$. We obverse that it increases monotonously with time, and
its time derivative (speed) change as $N_{s}$ increases. Nevertheless, the
speed tends to a limit with the increase of the number of solitons, so that
there is no dramatic difference between $N_{s}=14$ and $N_{s}=18$. This
happens because the interaction between the solitons has an effective range,
as shown in frame (b) of Fig. \ref{FIG4R}, hence the solitons located near
the edges interact weaker with the central ones.

\begin{figure}[tbp]
\begin{centering}
\begin{tabular}{lll}
(a) & (b) &  \\
\includegraphics[width=0.3\textwidth]{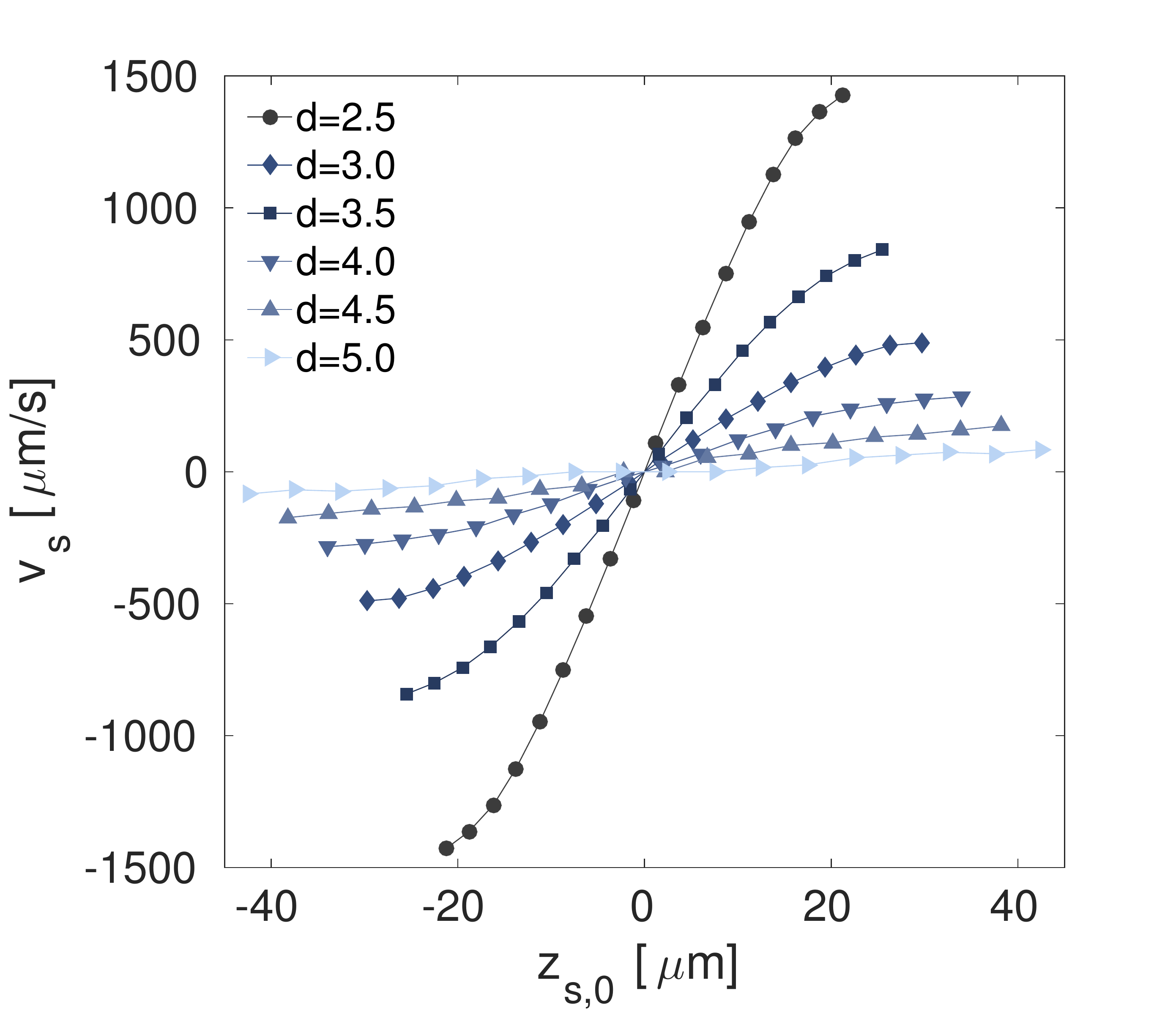} & %
\includegraphics[width=0.3\textwidth]{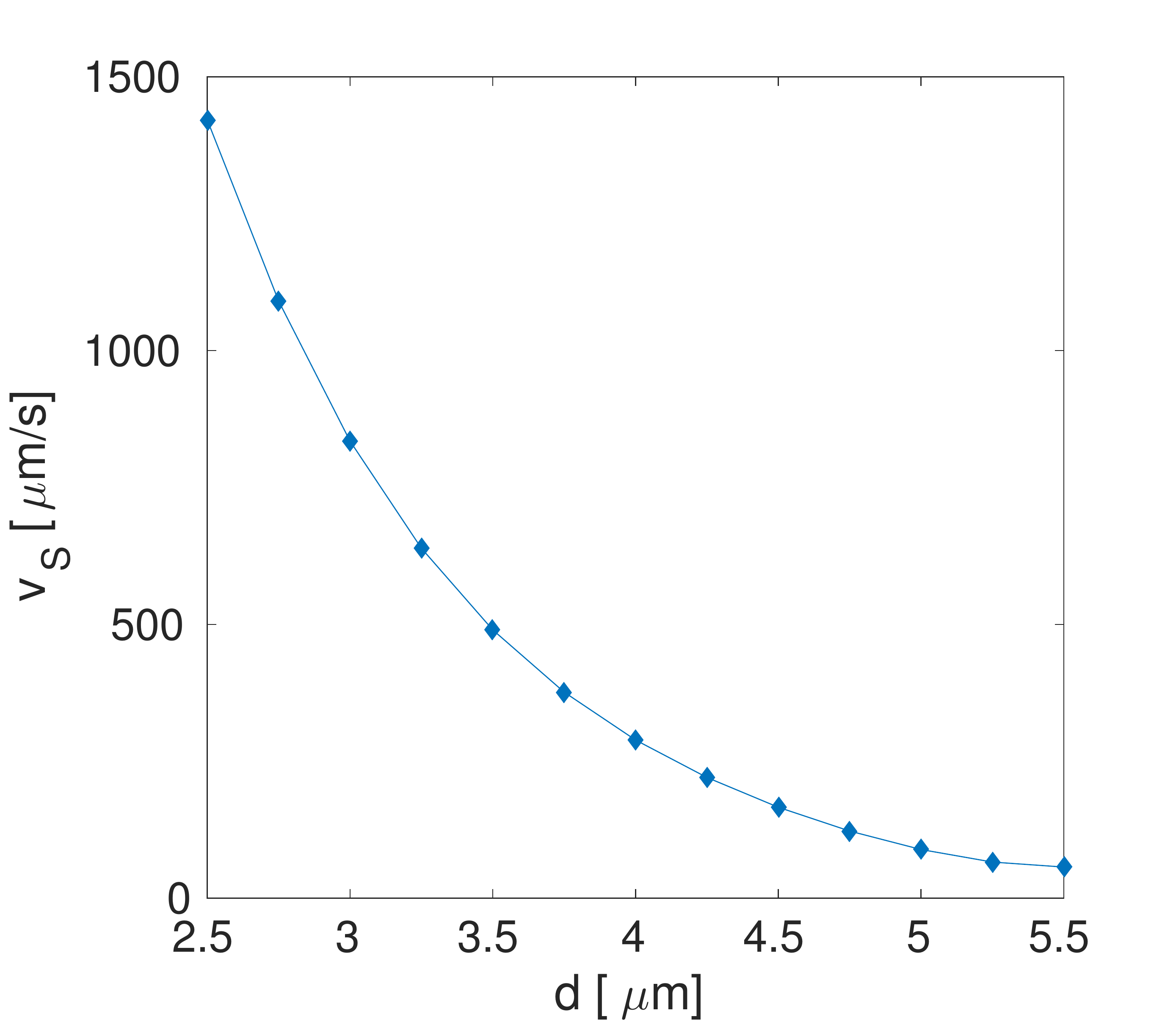} &  \\
\end{tabular}%
\caption{Dependence of the speed of the dark solitons for $N_s=18$.
(a) The distribution of the speed of each soliton at $t=90$ms for different initial distances $d$.
(b) The speed of the dark soliton near the edge at $t=90$ms as a function of $d$.
The other fixed parameters are the same as in Fig. \ref{FIG3R}.}
\label{FIG7R}
\end{centering}
\end{figure}

To analyze the case of a large number of solitons, it is enough to take $%
N_{s}=18$. Frame (a) of Fig. \ref{FIG7R} shows the speed at $t=90$ms of each
soliton as a function of its initial position for different initial
distances $d$. We observe that the central solitons have smaller speeds than
their counterparts placed near the edges, so that the speed is given by $%
v_{s}\simeq \tanh (\gamma _{d}z_{d})$ with $\gamma _{d}=-0.01d+0.077$ in the
range of Fig. \ref{FIG7R}. Frame (b) of Fig. \ref{FIG7R} shows the speed of
the soliton located near the positive edge at $t=90$ms. Similar to the
two-soliton case, the speed decays with the increase of the
initial distance, $v_{s}\sim d^{\alpha }$, with $\alpha =-3.385$.

\begin{figure}[tbp]
\begin{centering}
\begin{tabular}{llll}
(a) & (b) & (c) & \\
\includegraphics[width=0.28\textwidth]{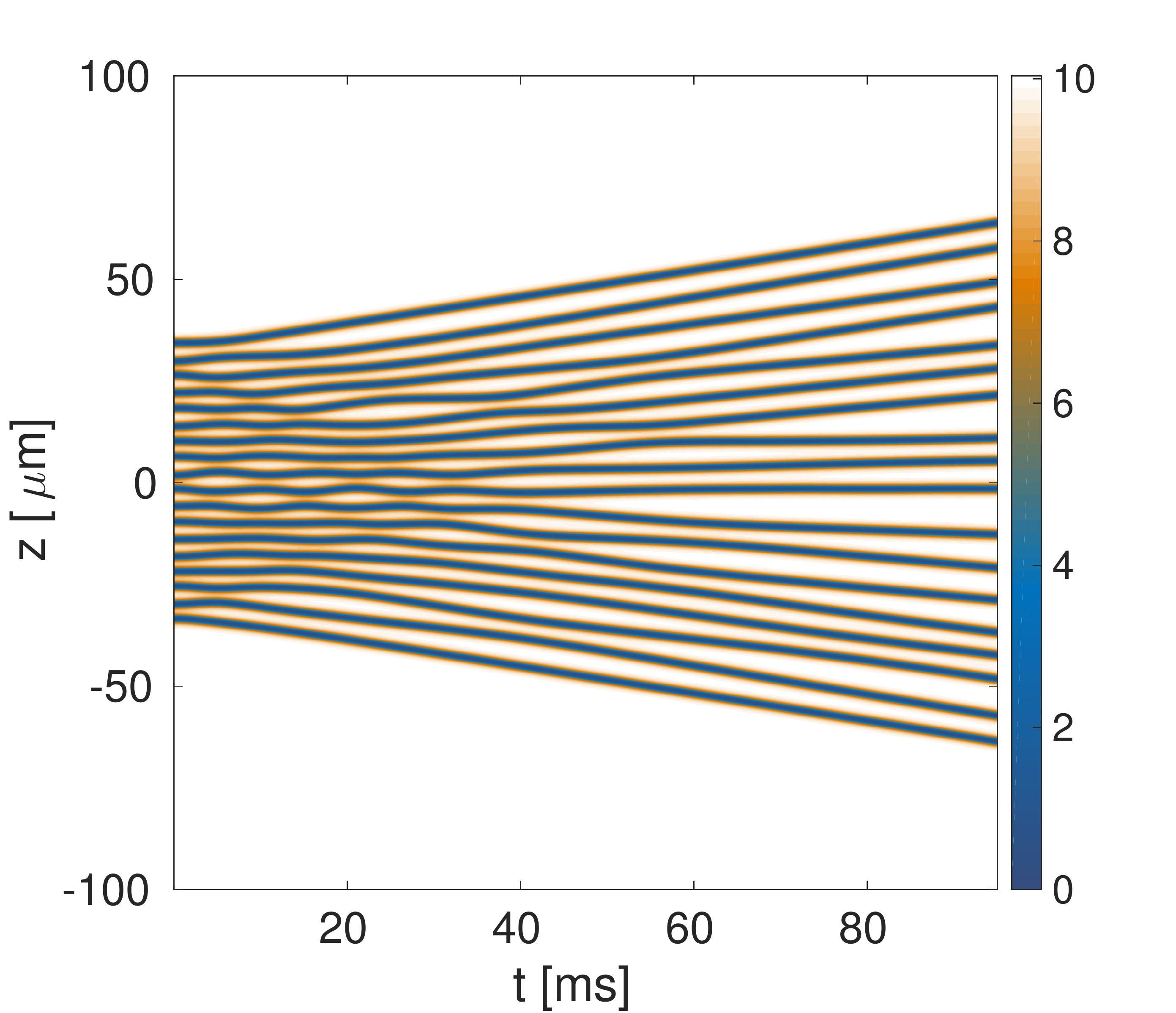} & %
\includegraphics[width=0.28\textwidth]{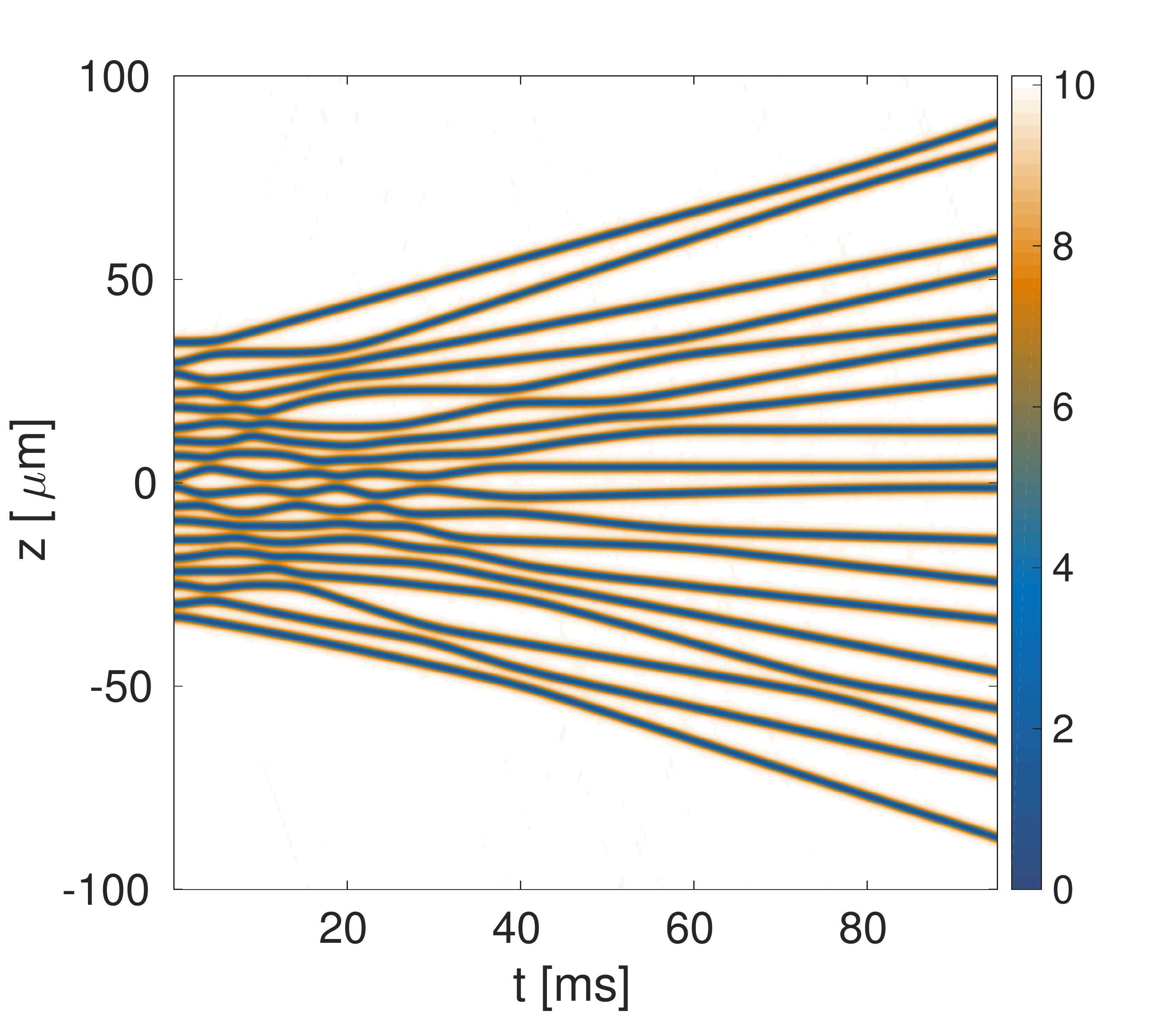} & %
\includegraphics[width=0.28\textwidth]{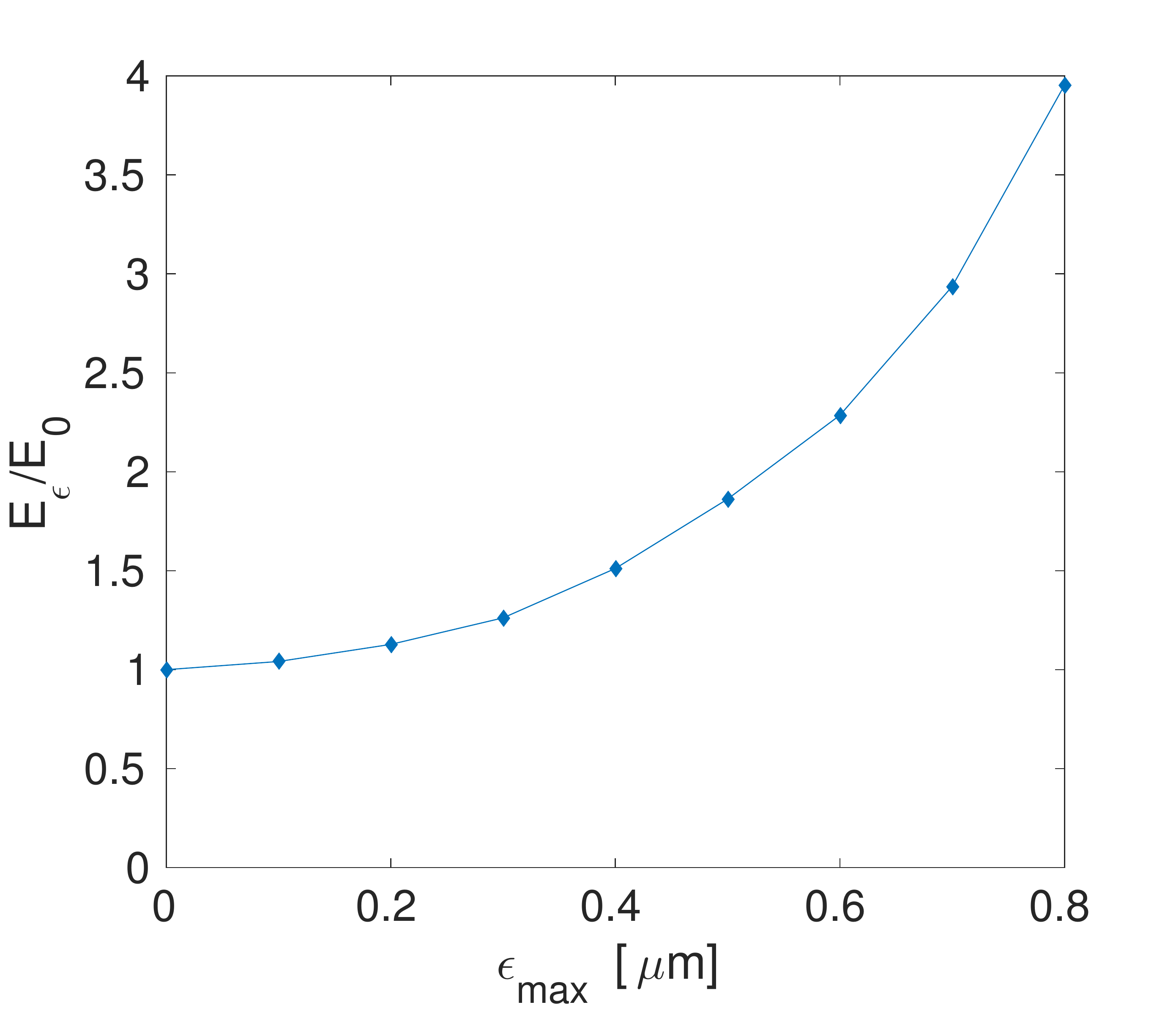} & \\
\end{tabular}%
\caption{Spatiotemporal diagrams of the density $n_{1D}$ for $N_s=18$ for
two different initial conditions in the presence of random perturbation $\epsilon$.
Panels (a) and (b) display the results for $\epsilon$ taking values in the ranges
of $[-\epsilon_{\max},\epsilon_{\max}]$, with $\epsilon_{\max}=0.4$ $\mu$m and
$0.8$ $\mu$m, respectively.
(c) $E_{\epsilon}$ normalized to $E_{0}$ as a function of $\epsilon_{\max}$,
i.e., the amplitude of the randomly varying variable.}
\label{FIG8R}
\end{centering}
\end{figure}

Finally, we consider random initial positions of the solitons, with $%
N_{s}=18 $. We define the initial positions as $z_{j,\epsilon
}=z_{j,0}+\epsilon $, where $\epsilon $ is a random fluctuation, and $%
z_{j,0} $ are the soliton positions, with the mean distance between them $%
d=4~\mu $m, like in the symmetric case. Figure \ref{FIG8R} shows the
spatiotemporal diagrams of the 1D density $n_{1\mathrm{D}}$ for two
different random realizations. In particular, we assume that $\epsilon $
takes random values in the ranges $[-\epsilon_{\max},\epsilon_{\max}]=
[-0.4,0.4]~\mu $m and $[-0.8,0.8]~\mu $m, in panels (a) and (b),
respectively. It is observed that the speed of the expansion is higher than
in the absence of the randomness, because the interaction energy generates
higher internal pressure in the gas of solitons. We analyze the influence of
the random-fluctuation magnitude, $\epsilon $ on the dynamics. In
particular, we calculate the sum of the squared velocities at the final
moment of time,
\begin{equation}
E_{\epsilon }=\sum\limits_{j=1}^{N_{s}}v_{j,\epsilon }^{2}.
\end{equation}%
Panel (c) of Fig. \ref{FIG8R} shows $E_{\epsilon }$ normalized to $E_{0}$
(the kinetic energy of the set of dark solitons with equidistant initial
positions) as a function of $\epsilon_{\max}$. We can observe that $%
E_{\epsilon}$ strongly increases with the growth of $%
\epsilon_{\max}$, which naturally means that the gas of solitons expands
faster when the fluctuations are stronger.

\section{The Bose - Fermi Mixture}

\label{S3}

In this section we consider a dilute superfluid mixture formed by $N_{%
\mathrm{B}}$ bosonic atoms of mass $m_{\mathrm{B}}$, and $N_{\mathrm{F}}$
fermionic atoms of mass $m_{\mathrm{F}}$ and spin $s_{\mathrm{F}}$. The
atoms interact through the pseudopotential, $\delta (\mathbf{r})$ \cite%
{Bloch08}. We assume that bosons form a BEC, described by the
Gross-Pitaevskii equation \cite{Bloch08}, while the local density
approximation \cite{Bloch08} applies to the description of the weakly
interacting fermionic component. Accordingly, the dynamical
equations can be derived from the functional,
\begin{equation}
\mathcal{S}=\int {dtd{\mathbf{r}}\left( \mathcal{L}_{\mathrm{B}}+\mathcal{L}%
_{\mathrm{F}}+\mathcal{L}_{\mathrm{BF}}\right) },  \label{Eq17}
\end{equation}
where $\mathcal{L}_{B}$ and $\mathcal{L}_{\mathrm{F}}$ are the Lagrangian
densities of the Bose and Fermi components, while $\mathcal{L}_{\mathrm{BF}}$
accounts for the interaction between them \cite{Diaz15}:
\begin{equation}
{\mathcal{L}_{B}}=\frac{{i\hbar }}{2}\left( {\Psi _{\mathrm{B}}^{\ast }\frac{%
{\partial {\Psi _{\mathrm{B}}}}}{{\partial t}}-{\Psi _{\mathrm{B}}}\frac{{%
\partial \Psi _{\mathrm{B}}^{\ast }}}{{\partial t}}}\right) -\frac{{{\hbar
^{2}}}}{{2{m_{\mathrm{B}}}}}{\left\vert {\nabla {\Psi _{\mathrm{B}}}}%
\right\vert ^{2}}-{U_{B}(\mathbf{r})\left\vert {{\Psi _{\mathrm{B}}}}%
\right\vert ^{2}}-\frac{1}{2}{g_{\mathrm{B}}}{\left\vert {{\Psi _{\mathrm{B}}%
}}\right\vert ^{4}},
\end{equation}%
\begin{equation}
{\mathcal{L}_{\mathrm{F}}}=\frac{{i\hbar }}{2\lambda _{1}}\left( {\Psi _{%
\mathrm{F}}^{\ast }\frac{{\partial {\Psi _{\mathrm{F}}}}}{{\partial t}}-{%
\Psi _{\mathrm{F}}}\frac{{\partial \Psi _{\mathrm{F}}^{\ast }}}{{\partial t}}%
}\right) -\frac{{{\hbar ^{2}}}}{{2\lambda _{2}{m_{\mathrm{F}}}}}{\left\vert {%
\nabla {\Psi _{\mathrm{F}}}}\right\vert ^{2}}-{U_{\mathrm{F}}(\mathbf{r})}{%
\left\vert {{\Psi _{\mathrm{F}}}}\right\vert ^{2}}-\frac{1}{2}{{g_{\mathrm{F}%
}}}{\left\vert {{\Psi _{\mathrm{F}}}}\right\vert ^{4}}-\frac{3\beta C_{%
\mathrm{F}}\hbar ^{2}}{10m_{\mathrm{F}}}{\left\vert {{\Psi _{\mathrm{F}}}}%
\right\vert ^{10/3}},
\end{equation}%
\begin{equation}
{\mathcal{L}_{B\mathrm{F}}}=-\frac{1}{2}{g_{\mathrm{BF}}}{\left\vert {{\Psi
_{B}}}\right\vert ^{2}}{\left\vert {{\Psi _{\mathrm{F}}}}\right\vert ^{2}}.
\label{Eq20}
\end{equation}%
Here ${g_{B}}\equiv 4\pi {\hbar ^{2}}{a_{B}}/${$m_{\mathrm{B}}$}, ${g_{%
\mathrm{F}}}\equiv 4\pi {\hbar ^{2}}({a_{\mathrm{F}}}/{m_{\mathrm{F}}})[2S_{%
\mathrm{F}}/(2S_{\mathrm{F}}+1)]$, and ${g_{\mathrm{BF}}}\equiv 4\pi {\hbar
^{2}}${$a_{\mathrm{BF}}$}$/${$m_{\mathrm{BF}}$} are three interaction
parameters of the mixture, with $a_{\mathrm{B}}$, $a_{\mathrm{F}}$ and $a_{%
\mathrm{BF}}$ being the respective scattering lengths; ${m_{\mathrm{BF}}}%
\equiv {m_{B}}{m_{\mathrm{F}}}/({m_{B}}+{m_{\mathrm{F}}})$ is the reduced
mass; and $U_{\mathrm{B}/\mathrm{F}}{(\mathbf{r})}$ are external potentials
acting on bosons/fermions. Complex wave functions $\Psi _{\mathrm{B}/\mathrm{%
F}}\left( \mathbf{r},t\right) $ are normalized to the respective numbers of
particles, $N_{\mathrm{B}/\mathrm{F}}$. The other parameters of the
fermionic Lagrangian density are the same as in Sec. \ref{S2}.

Varying action $\mathcal{S}$ with respect to $\Psi _{\mathrm{B}}^{\ast }$
and to $\Psi _{\mathrm{F}}^{\ast }$, we derive the following system of
nonlinear Schr\"{o}dinger equations for bosons and fermions:
\begin{equation}
i\hbar {\partial _{t}}{\Psi _{B}}=\left[ {-\frac{{{\hbar ^{2}}}}{{2{m_{B}}}}{%
\nabla ^{2}}+{g_{B}}{{\left\vert {{\Psi _{B}}}\right\vert }^{2}}+{g_{\mathrm{%
BF}}}{{\left\vert {{\Psi _{\mathrm{F}}}}\right\vert }^{2}}+{U_{B}}}\right] {%
\Psi _{B},}  \label{Eq21}
\end{equation}
\begin{equation}
\frac{{i\hbar }}{{{\lambda _{1}}}}{\partial _{t}}{\Psi _{\mathrm{F}}}=\left[
-\frac{{{\hbar ^{2}}}}{{2{\lambda _{2}}{m_{\mathrm{F}}}}}{\nabla ^{2}}+{{g_{%
\mathrm{F}}}{{\left\vert {{\Psi _{\mathrm{F}}}}\right\vert }^{2}}+{g_{%
\mathrm{BF}}}{{\left\vert {{\Psi _{B}}}\right\vert }^{2}}}+{U_{\mathrm{F}}}+%
\frac{{{\beta C_{\mathrm{F}}\hbar ^{2}}}}{{2{m_{F}}}}{{\left\vert {{\Psi _{%
\mathrm{F}}}}\right\vert }^{4/3}}\right] {\Psi _{\mathrm{F}}}.  \label{Eq22}
\end{equation}
We apply the formalism developed below to the ${}^{7}$Li-${}^{6}$Li mixture,
with the same scattering parameter for both species, $a_{\mathrm{B}}=a_{%
\mathrm{F}}=5$nm. The use of isotopes of the same alkali element is
suggested by the similarity of their electric polarizability, thus implying
similar external potentials induced by an optical trap. Unless specified
otherwise, in what follows below we consider configurations with fully
polarized fermions. Note that the BCS and unitarity regimes involve more
than one spin state of fermions, hence the magnetic trap will split the
respective spin energy levels. For this reason, we assume the presence of
the optical trap, which supports equal energy levels for all the spin
states, making it possible to discriminate different regimes of the
interaction in the BFM. In the BCS and unitarity regimes, we assume balanced
populations of the two spin components.

Our analysis is first presented for the GS and dynamics of perturbations
around it. In particular, for the GS we focus on determining the spatial
correlation $C_{s}$ between the spatial particle densities in both species,
defined as
\begin{equation}
{C_{s}}\left( {{\bar{n}_{B}},{\bar{n}_{\mathrm{F}}}}\right) =\frac{{%
\left\langle {{\bar{n}_{B}}{\bar{n}_{\mathrm{F}}}}\right\rangle }}{\sqrt{%
\left\langle {\bar{n}_{B}^{2}}\right\rangle \left\langle {\bar{n}_{\mathrm{F}%
}^{2}}\right\rangle }},  \label{Eq23}
\end{equation}
where ${{\bar{n}}_{\mathrm{B}/\mathrm{F}}}={n_{\mathrm{B}/\mathrm{F}}}%
-\left\langle {{n_{\mathrm{B}/\mathrm{F}}}}\right\rangle $, $\left\langle
{}\right\rangle $ standing for the spatial average. For dynamical
perturbations around the GS, a spatiotemporal correlation, which is defined
by replacing the spatial average with the spatiotemporal average, is known
as the Pearson coefficient $C_{s-t}$ \cite{Bragard04}. We remark that when $%
C_s=1$ and $C_s=-1$ the mixture is fully synchronized and anti-synchronized,
respectively; whereas, the mixture is not synchronized at $C_s=0$.

While numerical integration of this system in the 3D form is very heavy
computationally, the effective dimension may be reduced to 1D or 2D when the
system is tightly confined by a trapping potential. To this end, the VA is
employed, making use, as above, of the factorization of the 3D wave
function, which includes the Gaussian ansatz in the tightly confined
transverse directions. As mentioned above too, the factorization has been
widely used for Bose and Fermi systems separately, as it shown in Refs. \cite%
{Salasnich02a} and \cite{Diaz12}, respectively. In the next two subsections
we reduce the full 3D system to the corresponding 2D and 1D effective
systems, using the VA proposed in Ref. \cite{Diaz15}.

\subsection{The two-dimensional reduction}

Similar to the case of the pure Fermi gas, we derive 2D equations for the
disc-shaped configuration. Accordingly, the structure of the confinement
potential is taken as
\begin{equation}
{U_{\mathrm{B}/\mathrm{F}}}\left( {\mathbf{r}}\right) =\frac{1}{2}{m_{%
\mathrm{B}/\mathrm{F}}}\omega _{z,\mathrm{B}/\mathrm{F}}^{2}{z^{2}}+{U_{2%
\mathrm{D},\mathrm{B}/\mathrm{F}}}\left( \mathbf{r}_{\bot }\right) ,
\label{Eq24}
\end{equation}
where the second term corresponds to the strong harmonic-oscillator trap
acting along the $z$ direction. The corresponding factorized ansatz is
adopted as
\begin{equation}
\Psi _{\mathrm{B}/\mathrm{F}}\left( {{\mathbf{r}},t}\right) =\frac{1}{{{\pi
^{1/4}}\sqrt{\xi _{\mathrm{B}/\mathrm{F}}(\mathbf{r}_{\bot },t)}}}{\exp }%
\left( -\frac{{{z^{2}}}}{{2\left( \xi _{\mathrm{B}/\mathrm{F}}(\mathbf{r}%
_{\bot },t)\right) ^{2}}}\right) \phi _{\mathrm{B}/\mathrm{F}}(\mathbf{r}%
_{\bot },t)  \label{Eq25}
\end{equation}
where $\phi _{\mathrm{B}/\mathrm{F}}$ is normalized to $N_{\mathrm{B}/%
\mathrm{F}}$, and ${\xi _{\mathrm{B}/\mathrm{F}}}\left( {x,y,t}\right) $ are
widths of the gas in the confined direction. Substituting the factorized
ansatz (\ref{Eq25}) in action (\ref{Eq17}) and integrating over $z$, we
arrive at the following expression for the effective 2D action:
\begin{equation}
\mathcal{S}=\int {dtdxdy\left( \mathcal{L}_{2D\mathrm{,B}}+\mathcal{L}_{2%
\mathrm{D},\mathrm{F}}+\mathcal{L}_{2\mathrm{D,BF}}\right) },  \label{Eq26}
\end{equation}
where
\begin{equation}
\mathcal{L}_{2\mathrm{D,B}}=i\frac{\hbar }{2}\left( {\phi _{\mathrm{B}%
}^{\ast }{\partial _{t}}{\phi _{\mathrm{B}B}}-{\phi _{\mathrm{B}}}{\partial
_{t}}\phi _{\mathrm{B}}^{\ast }}\right) -{U_{2D\mathrm{,B}}}{n_{2\mathrm{D,B}%
}}-{e_{2\mathrm{D,B}}},
\end{equation}
\begin{equation}
{\mathcal{L}_{2\mathrm{D},\mathrm{F}}}=i\frac{\hbar }{{2{\lambda _{1}}}}%
\left( {\phi _{\mathrm{F}}^{\ast }{\partial _{t}}{\phi _{\mathrm{F}}}-{\phi
_{\mathrm{F}}}{\partial _{t}}\phi _{\mathrm{F}}^{\ast }}\right) -{U_{2%
\mathrm{D},\mathrm{F}}}{n_{2\mathrm{D},\mathrm{F}}}-{e_{2\mathrm{D},\mathrm{F%
}}},
\end{equation}%
\begin{equation}
{\mathcal{L}_{2\mathrm{D,BF}}}=-\frac{1}{{{\pi ^{1/2}}}}\frac{{{g_{B\mathrm{F%
}}}}}{\sqrt{\xi _{\mathrm{B}}^{2}+\xi _{\mathrm{F}}^{2}}}{n_{2\mathrm{D},%
\mathrm{B}}}{n_{2\mathrm{D},\mathrm{F}}},
\end{equation}%
so that $n_{2\mathrm{D,B/F}}\equiv \left\vert {\phi _{\mathrm{B}/\mathrm{F}}}%
\left( {x,y}\right) \right\vert ^{2}$ are the 2D particle densities of the
boson and fermion species, and ${e_{2\mathrm{D,B}}}$ and ${e_{2\mathrm{D},%
\mathrm{F}}}$ are their energy densities:
\begin{equation}
{e_{2\mathrm{D,B}}}=\frac{{{\hbar ^{2}}}}{{2{m_{\mathrm{B}}}}}{\left\vert {{%
\nabla _{\bot }}{\phi _{\mathrm{B}}}}\right\vert ^{2}}+\left[ {\frac{{{g_{%
\mathrm{B}}}}}{{{\ \sqrt{8\pi }}{\xi _{\mathrm{B}}}}}{n}_{2\mathrm{D,B}}+%
\frac{{{\hbar ^{2}}}}{{4{m_{\mathrm{B}}}\xi _{\mathrm{B}}^{2}}}}\right.
+\left. {\frac{1}{4}{m_{\mathrm{B}}}\omega _{z,\mathrm{B}}^{2}\xi _{\mathrm{B%
}}^{2}}\right] {n_{2\mathrm{D},\mathrm{B}}},
\end{equation}
\begin{eqnarray}
{e_{2\mathrm{D},\mathrm{F}}} &=&\frac{{{\hbar ^{2}}}}{{2{\lambda _{2}}{m_{%
\mathrm{F}}}}}{\left\vert {{\nabla _{\bot }}{\phi _{\mathrm{F}}}}\right\vert
^{2}}+\left[ {\frac{{{g_{\mathrm{F}}}}}{{{\ \sqrt{8\pi }}{\xi _{\mathrm{F}}}}%
}{n_{2\mathrm{D},\mathrm{F}}}+\frac{{{\hbar ^{2}}}}{{4\lambda _{2}{m_{%
\mathrm{F}}}\xi _{\mathrm{F}}^{2}}}+\frac{1}{4}{m_{\mathrm{F}}}\omega _{z,%
\mathrm{F}}^{2}\xi _{\mathrm{F}}^{2}}\right.  \notag \\
&&+\left. {\frac{{{\hbar ^{2}}}}{{2{m_{\mathrm{F}}}}}\xi \frac{3}{{5\xi _{%
\mathrm{F}}^{2/3}}}{C_{2\mathrm{D},\mathrm{F}}}n_{2\mathrm{D},\mathrm{F}%
}^{2/3}}\right] {n_{2\mathrm{D},\mathrm{F}}},
\end{eqnarray}
with $C_{2\mathrm{D},\mathrm{F}}\equiv {(3/5)^{1/2}}{(6/(2s_{\mathrm{F}%
}+1))^{2/3}}\pi $. The field equations for the 2D system are obtained by the
variation of the action $S$ given by Eq. (\ref{Eq26}) with respect to
variables $\phi _{\mathrm{B}}$ and $\phi _{\mathrm{F}}$:
\begin{eqnarray}
i\hbar {\partial _{t}}{\phi _{\mathrm{B}}} &=&\left[ {-\frac{{{\hbar ^{2}}}}{%
{2{m_{B}}}}\nabla _{\bot }^{2}+{U_{2\mathrm{D,B}}}+\frac{1}{{{\pi ^{1/2}}}}%
\frac{{{g_{\mathrm{BF}}}}}{\sqrt{\xi _{\mathrm{B}}^{2}+\xi _{\mathrm{F}}^{2}}%
}{n_{2\mathrm{D},\mathrm{F}}}}+\frac{{{g_{\mathrm{B}}}}}{{\sqrt{2\pi }{\xi _{%
\mathrm{B}}}}}{{\left\vert {{\phi _{\mathrm{B}}}}\right\vert }^{2}}\right.
\notag \\
&&+\left. \frac{{{\hbar ^{2}}}}{{4{m_{\mathrm{B}}}\xi _{\mathrm{B}}^{2}}}+%
\frac{1}{4}{m_{\mathrm{B}B}}\omega _{z,\mathrm{B}}^{2}\xi _{\mathrm{B}}^{2}%
\right] {\phi _{\mathrm{B}},}  \label{Eq32}
\end{eqnarray}%
\begin{eqnarray}
i\frac{\hbar }{{{\lambda _{1}}}}{\partial _{t}}{\phi _{\mathrm{F}}} &=&\left[
{-\frac{{{\hbar ^{2}}}}{{2{\lambda _{2}}{m_{\mathrm{F}}}}}\nabla _{\bot
}^{2}+{U_{2D,\mathrm{F}}}+\frac{1}{{{\pi ^{1/2}}}}\frac{{{g_{\mathrm{BF}}}}}{%
\sqrt{\xi _{\mathrm{B}}^{2}+\xi _{\mathrm{F}}^{2}}}{n_{2\mathrm{D,B}}}}+%
\frac{{{g_{\mathrm{F}}}}}{{\sqrt{2\pi }{\xi _{\mathrm{F}}}}}{\left\vert {{%
\phi _{\mathrm{F}}}}\right\vert ^{2}}\right.  \notag \\
&&+\left. \frac{{{\hbar ^{2}}}}{{2{m_{\mathrm{F}}}}}\xi \frac{1}{{\xi _{%
\mathrm{F}}^{2/3}}}{C_{2\mathrm{D},\mathrm{F}}}{\left\vert {{\phi _{\mathrm{F%
}}}}\right\vert ^{4/3}}+{\frac{{{\hbar ^{2}}}}{{4\lambda _{2}{m_{\mathrm{F}}}%
\xi _{\mathrm{F}}^{2}}}+\frac{1}{4}{m_{\mathrm{F}}}\omega _{z,\mathrm{F}%
}^{2}\xi _{\mathrm{F}}^{2}}\right] {\phi _{\mathrm{F}}.}  \label{Eq33}
\end{eqnarray}
Relations between $\xi _{\mathrm{B}/\mathrm{F}}$ and $\phi _{\mathrm{B}/%
\mathrm{F}}$ are produced by the Euler-Lagrange equations associated to $\xi
_{\mathrm{B}/\mathrm{F}}$:
\begin{equation}
{\kappa _{I,B}}\xi _{\mathrm{B}}^{4}-\frac{{{g_{B}}}}{\sqrt{2\pi }}{n_{2%
\mathrm{D,B}}}{\xi _{\mathrm{B}}}-\frac{{{\hbar ^{2}}}}{{{m_{\mathrm{B}}}}}%
=0,  \label{Eq34}
\end{equation}%
\begin{equation}
{\kappa _{I,\mathrm{F}}}\xi _{\mathrm{F}}^{4}-\frac{{2{\hbar ^{2}}}}{{5{m_{%
\mathrm{F}}}}}\xi {C_{2\mathrm{D},\mathrm{F}}}n_{2\mathrm{D},\mathrm{F}%
}^{2/3}\xi _{\mathrm{F}}^{4/3}-\frac{g_{\mathrm{F}}}{\sqrt{2\pi }}{n_{2%
\mathrm{D},\mathrm{F}}}{\xi _{\mathrm{F}}}-\frac{{{\hbar ^{2}}}}{\lambda
_{2}m_{\mathrm{F}}}=0,  \label{Eq35}
\end{equation}%
where ${\kappa _{I,\mathrm{F}}}\equiv {m_{\mathrm{F}}}\omega _{z,\mathrm{F}%
}^{2}+2{g_{\mathrm{BF}}}{n_{2\mathrm{D,B}}}/[{{\pi ^{1/2}}{{({\xi _{\mathrm{B%
}}^{2}+\xi _{\mathrm{F}}^{2}})}^{3/2}}}]$. Thus, Eqs. (\ref{Eq32})-(\ref%
{Eq35}) constitute a system of four 2D coupled equations produced by the
reduction of the underlying 3D system (\ref{Eq21}). Note also
that when $g_{BF}=0$, the system is decoupled and Eq. (\ref{Eq32})
corresponds to the dimensional reduction of the Gross-Pitaevskii equation.
Equations (\ref{Eq34}) and (\ref{Eq35}) for $\xi _{\mathrm{B}/\mathrm{F}}$
can be solved numerically by dint of the Newton's method. The basic external
potential is taken as the harmonic-oscillator one: $U_{2\mathrm{D},\mathrm{B}%
/\mathrm{F}}=m_{\mathrm{B}/\mathrm{F}}\omega _{x,\mathrm{B}/\mathrm{F}}^{2}{%
x^{2}}/2+m_{\mathrm{B}/\mathrm{F}}\omega _{y,\mathrm{B}/\mathrm{F}}^{2}{y^{2}%
}/2$. The simulations were based on the fourth-order Runge-Kutta algorithm
with $\Delta t=4.77$ $\mu $s. The spatial discretizations was performed with
$\Delta x=1$ $\mu $m, $\Delta y=1$ $\mu $m and $\Delta z=0.05$ $\mu $m. The
GS was found by means of the imaginary-time integration. We here focus on
the case when the number of bosons is much greater than the number of
fermions, \textit{viz}., $N_{\mathrm{B}}=5\times 10^{4}$ and $N_{\mathrm{F}%
}=2.5\times 10^{3}$.

\begin{figure}[tbp]
\centering
\resizebox{0.8\textwidth}{!}{
\includegraphics{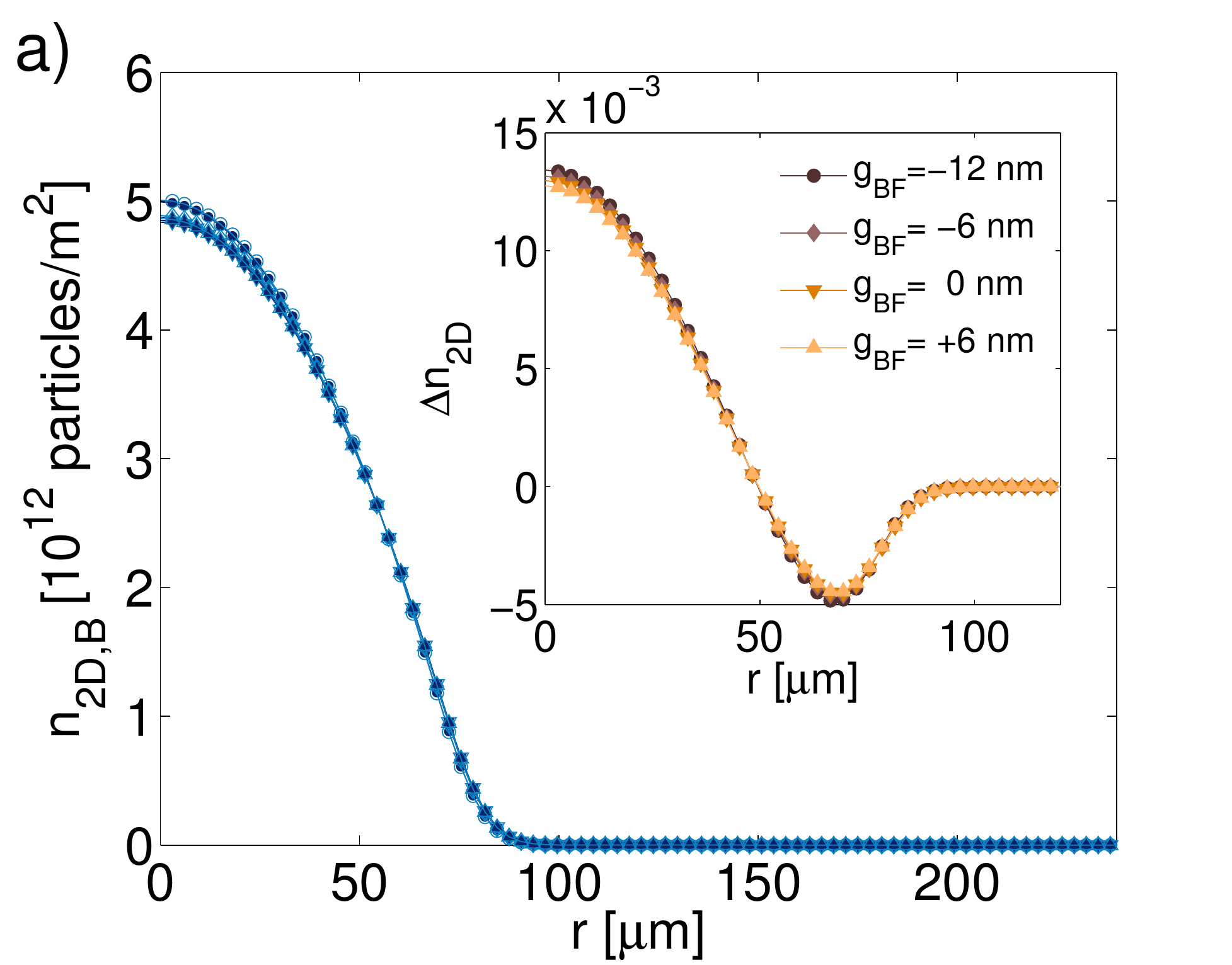}
\includegraphics{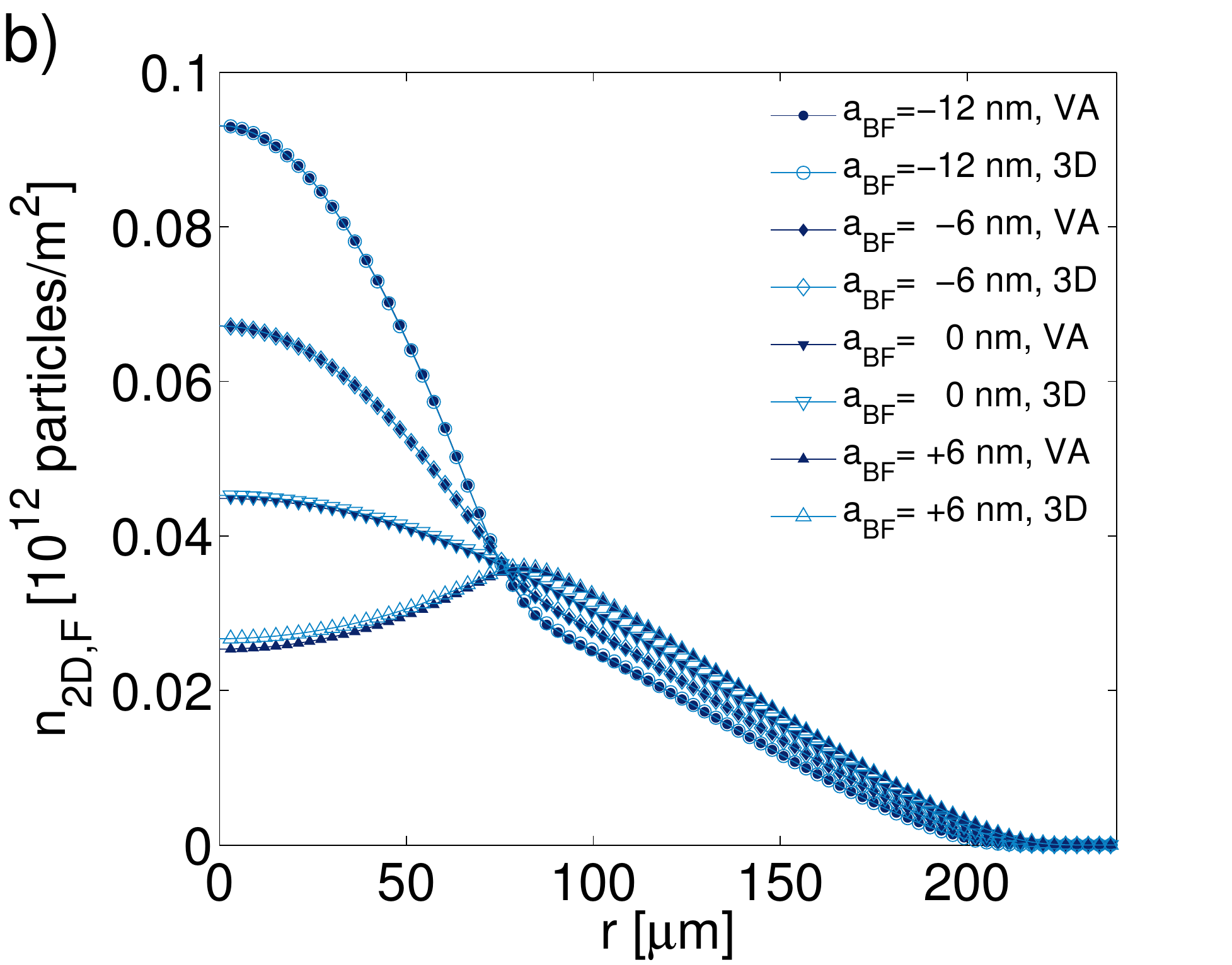}
}
\resizebox{0.8\textwidth}{!}{
\includegraphics{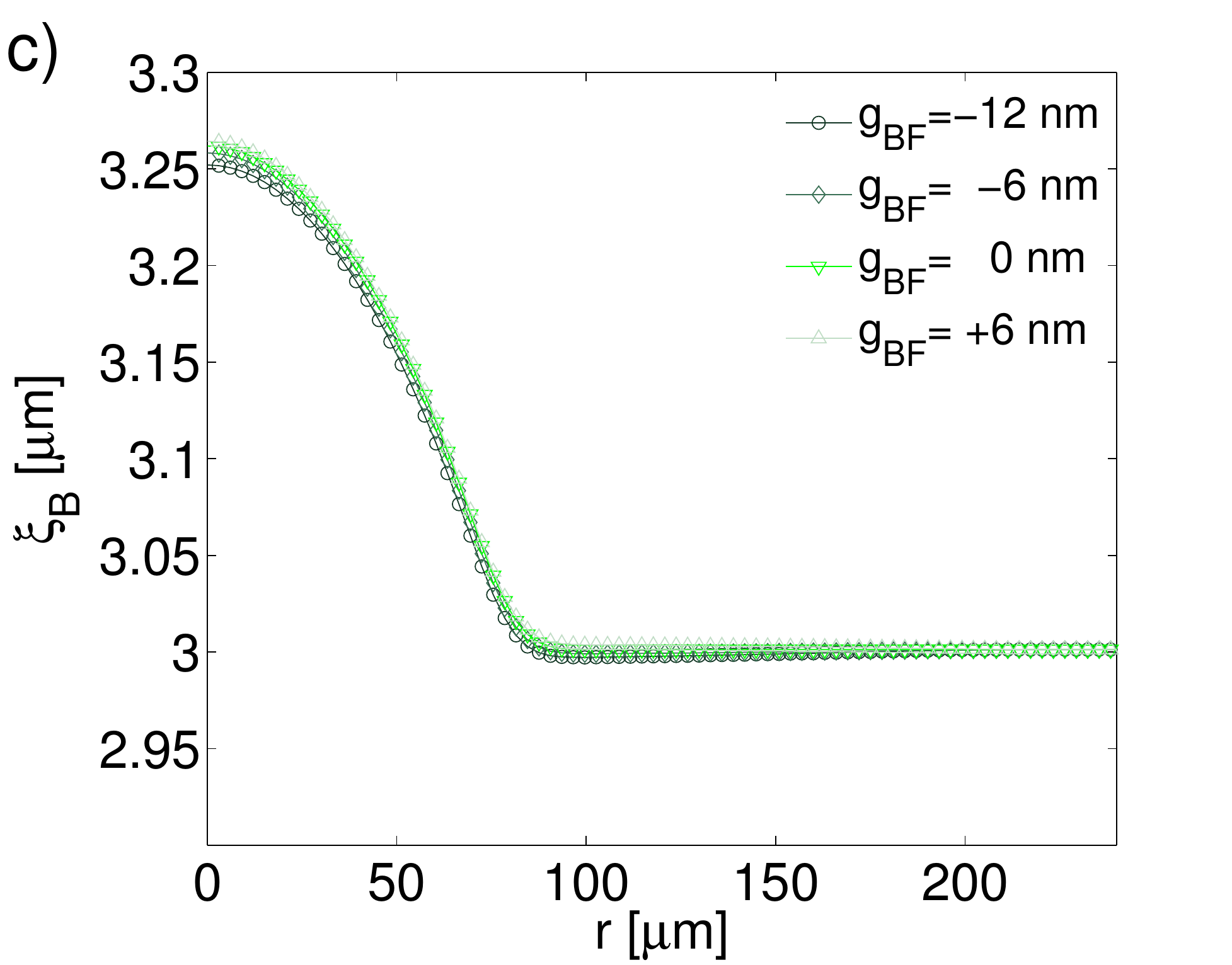}
\includegraphics{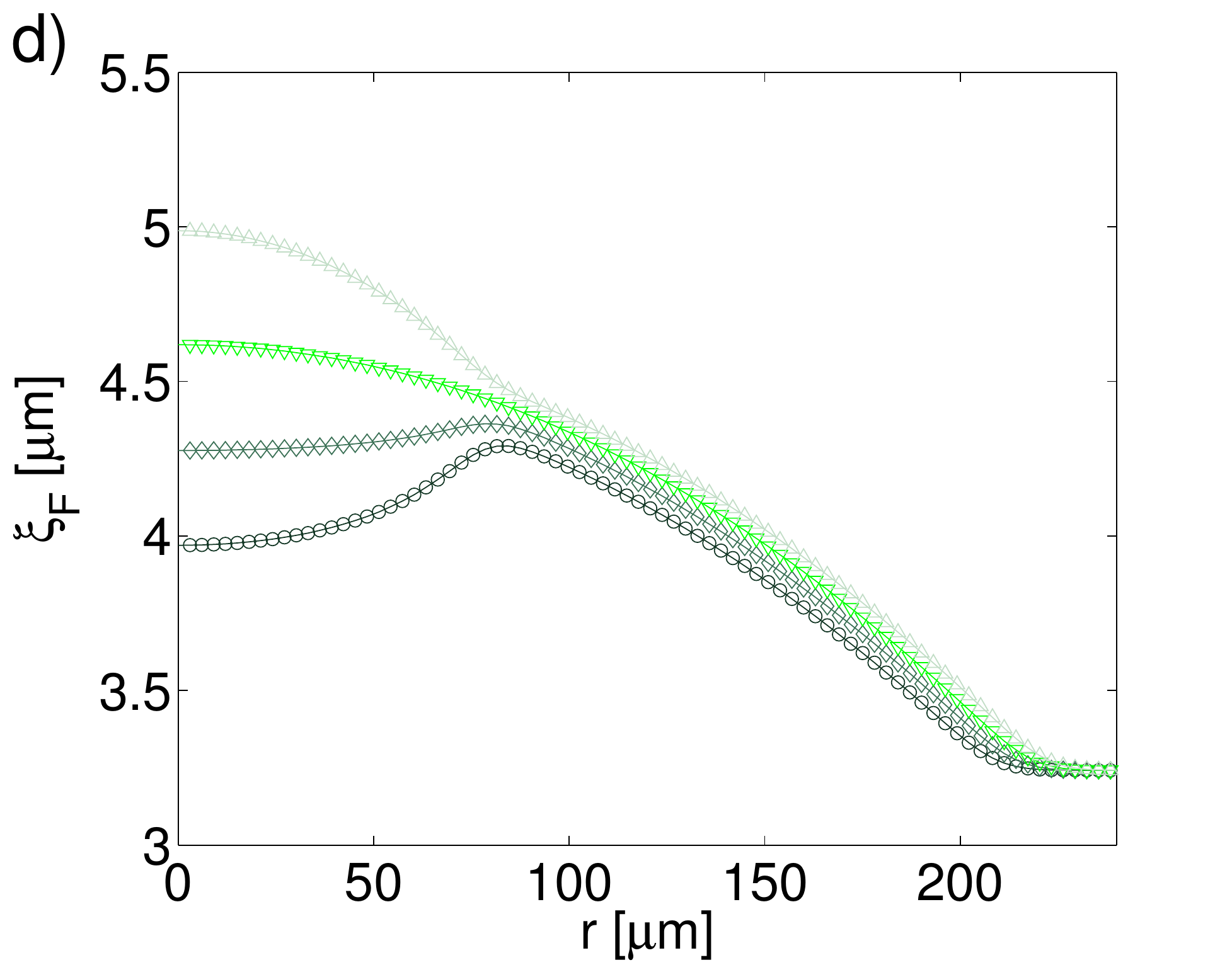}
}
\caption{The radial profile of the 2D particle density, and the respective
width for different values of interaction strength $a_{\mathrm{BF}}$. (a) $%
n_{2\mathrm{D},\mathrm{B}}$, (b) $n_{2\mathrm{D,F}}$, (c) $\protect\xi _{%
\mathrm{B}}$, and (d) $\protect\xi _{\mathrm{F}}$. The parameters are $N_{%
\mathrm{B}}=5\times 10^{4}$, $N_{\mathrm{F}}=2.5\times 10^{3}$, $a_{\mathrm{%
B/F}}=5$ nm, $\protect\omega _{z,\mathsf{B/F}}=1000$ Hz, and $\protect\omega %
_{x,\mathrm{B/F}}=\protect\omega _{y,\mathrm{B/F}}=30$ Hz. The inset in
panel (a) shows the difference between the VA and full 3D simulations, by
means of $\Delta n_{\mathrm{2D}}\equiv \bar{n}_{\mathrm{2D}}-n_{\mathrm{2D}}$%
. This figure is taken from Ref. \protect\cite{Diaz15}.}
\label{FIG9R}
\end{figure}

Frames (a) and (b) of Fig. \ref{FIG9R} show the radial profile of both 2D
bosonic and fermionic densities, $n_{2\mathrm{D},\mathrm{B}/\mathrm{F}}$,
respectively. The panels for the bosonic and fermionic components are the
left and right ones, respectively. Each density has been computed using the
VA and the full 3D system. To obtain the 2D profile from the 3D simulations,
Eqs. (\ref{Eq21}) and (\ref{Eq22}) were solved, and the 3D density was
integrated along the $z$ axis, $\bar{n}_{\mathrm{2D,B/F}}=\int_{-\infty
}^{+\infty }\left\vert \Psi _{\mathrm{2D,B/F}}(\mathbf{r})\right\vert dz$.
We infer that the repulsive mixture concentrates the bosons at the center,
while the attractive mixture concentrates both species at the center. Panels
(c) and (d) of Figure \ref{FIG9R} show the radial dependence of the width
for both bosonic and fermionic component, respectively. We observe that only
the width of the fermionic density profile varies significantly with the
change of the scattering length of the inter-species interaction, which is a
consequence of a greater number of bosons in comparison with fermions. It is
clearly seen that fermions are stronger confined when the interaction is
attractive, and their spatial distribution significantly expands when the
interaction is repulsive. Similar results have been reported in Refs. \cite%
{Adhikari08,Maruyama09,Salasnich07a}.

\begin{figure}[tbp]
\centering
\resizebox{0.5\textwidth}{!}{
\includegraphics{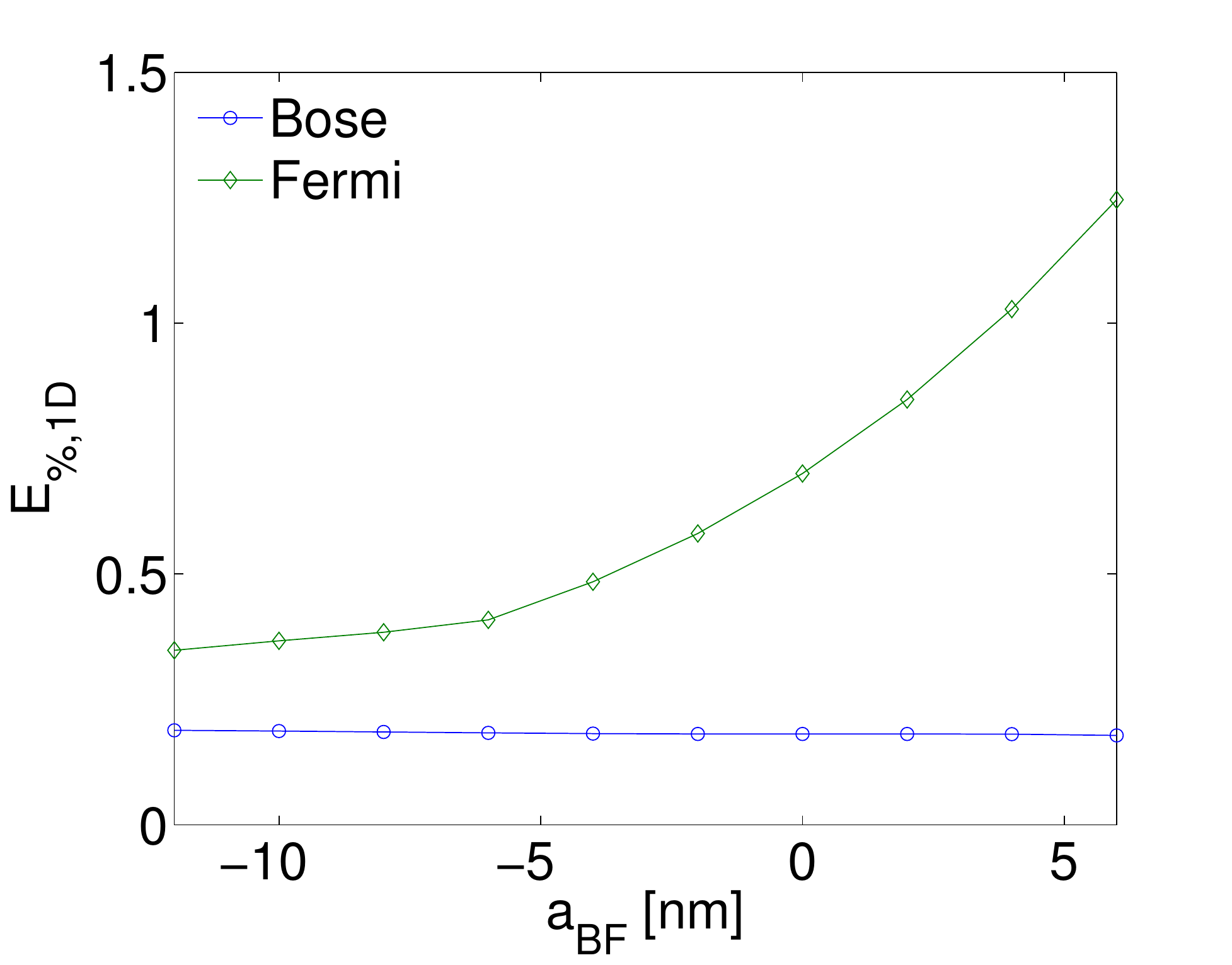}
}
\caption{The 2D overall percentage error of the VA versus the full 3D
system, as a functionof $a_{\mathrm{BF}}$ for both species. Parameters are
the same as in Fig.~\protect\ref{FIG9R}. This figure is taken from Ref.
\protect\cite{Diaz15}.}
\label{FIG10R}
\end{figure}

Now, to compare the results obtained from the VA with those produced by the
3D simulations, we note that both profiles are practically identical, except
for the repulsive case in which a discrepancy is observed. The inset in
panel (a) of Fig. \ref{FIG9R} shows that the difference between the two
results has a magnitude of nearly three orders of magnitude lower than the
density itself. We define the overall percentage error of the VA as $E_{\%,%
\mathrm{2D}}=\int \int \left\vert \rho _{\mathrm{2D}}-n_{\mathrm{2D}%
}\right\vert dxdy$ (for both species). Figure~\ref{FIG10R} shows the error
for both species as a function of interspecies scattering parameter, $a_{%
\mathrm{BF}}$. For bosons it takes values $\sim 0.2\%$, and does not change
much, as shown in the inset to panel (a) of Fig. \ref{FIG9R}. For fermions
the error is greater than for bosons throughout the observed range, but it
is quite small for the attractive mixture. Note that the error increases for
the repulsive mixture, but remains lower that $2\%$. Thus we conclude that
the 2D approximation is very accurate.

\begin{figure}[tbp]
\centering
\resizebox{0.5\textwidth}{!}{\includegraphics{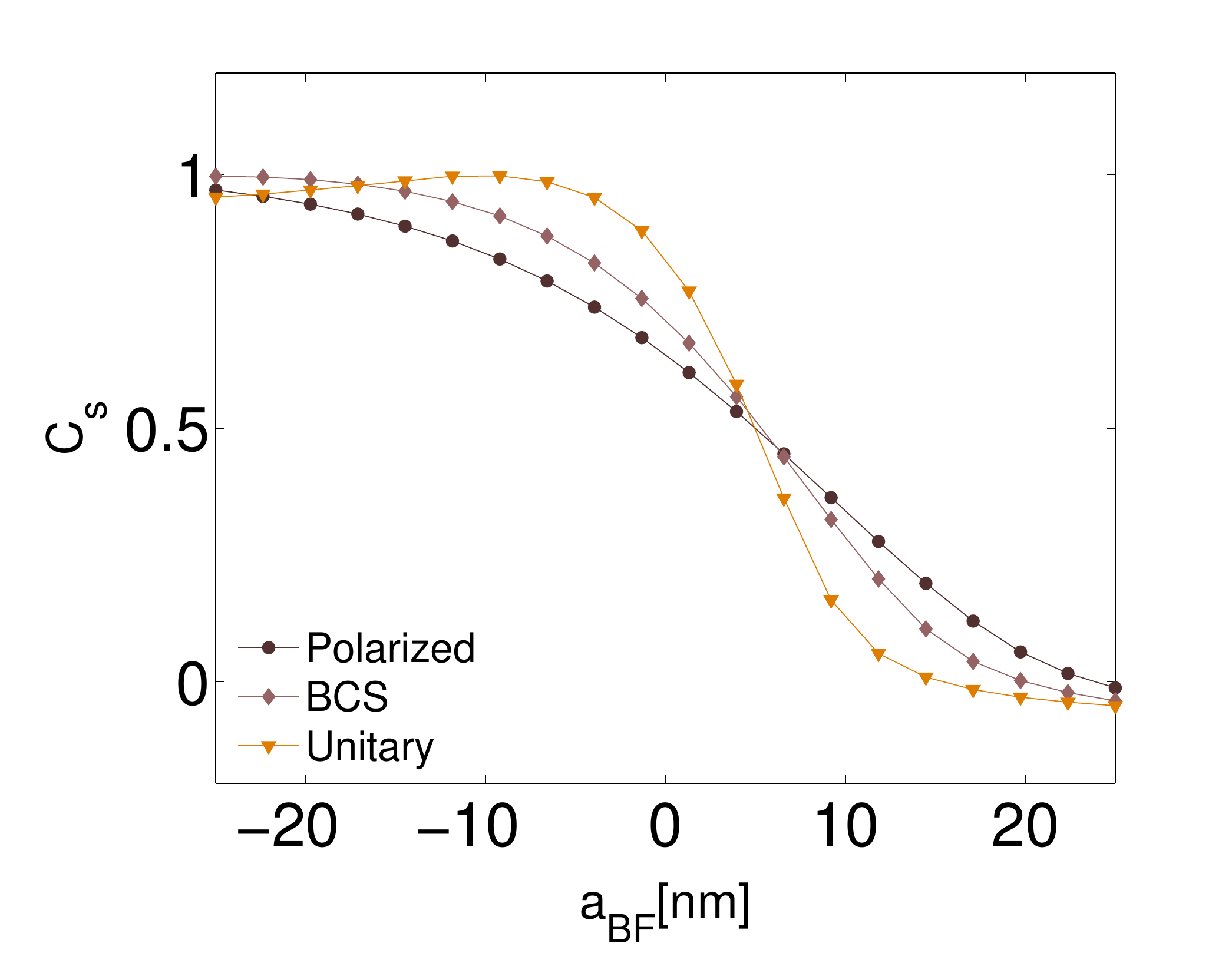}
}
\caption{(Color online) Spatial correlation $C_{s}$ of the GS of the 2D
mixture as a a function of $a_{\mathrm{BF}}$, for three fermionic regimes:
polarized, BCS, and unitarity. The fixed parameters are: $N_{\mathrm{B}%
}=5\times 10^{4}$, $N_{\mathrm{F}}=2.5\times 10^{3}$, $a_{\mathrm{B/F}}=5$
nm, $\protect\omega _{x,\mathrm{B/F}}=\protect\omega _{y,\mathrm{B/F}}=30$
Hz and $\protect\omega _{z,\mathrm{B/F}}=1000$ Hz. This figure is taken from
Ref. \protect\cite{Diaz15}.}
\label{FIG11R}
\end{figure}

Finally, we measure the correlations of the BFM states. To this end, the
spatial correlation, $C_{s}$, in the GS was calculated using the definition
given in Eq. (\ref{Eq23}). Figure \ref{FIG11R} presents the analysis of the
GS synchronization of the mixture as a function of $a_{\mathrm{BF}}$, where
three possible regimes are considered for the fermions: fully polarized,
BCS, and unitarity. Parameters of the Lagrangian density for each fermionic
regime are given in Table \ref{TT1}. When the interaction is attractive,
there is not a large discrepancy between the correlation curves. In fact,
for $a_{\mathrm{BF}} \in (-25, -15)$nm the values of $C_s \gtrsim 0.9$, and
therefore the GS states are synchronized. In the unitarity regime, it is
again observed that the correlation reaches a maximum close to $1$ at $a_{%
\mathrm{BF}}\approx -10$ nm, dropping to negative values when the mixture is
strongly repulsive. Also, we observe that the three curves demonstrate
stronger demixing when $a_{\mathrm{BF}}$ changes from negative to positive
values of $a_{\mathrm{BF}}$, and for $a_{\mathrm{BF}} \gtrsim 15$ the value
of $C_s$ tends to zero implying that the GS states are not synchronized.

\subsection{The one-dimensional reduction}

The 1D confinement means, as above, a cigar-shaped configuration elongated
in the direction of $z$. In this case, the corresponding confining
potentials trap is written as
\begin{equation}
U_{\mathrm{B}/\mathrm{F}}\left( {{\mathbf{r}},t}\right) =\frac{1}{2}m_{%
\mathrm{B}/\mathrm{F}}\omega _{t,{\mathrm{B}/\mathrm{F}}}^{2}r^{2}+U_{1%
\mathrm{D},{\mathrm{B}/\mathrm{F}}}\left( {z,t}\right) ,  \label{Eq36}
\end{equation}%
where $U_{1\mathrm{D},{\mathrm{B}/\mathrm{F}}}\left( {z,t}\right) $ are the
axial potentials. Assuming that the transverse trapping potential is strong
enough, the dimensional reduction is carried out by means of the usual
factorized ansatz for the wave functions,
\begin{equation}
\Psi _{\mathrm{B}/\mathrm{F}}\left( {{\mathbf{r}},t}\right) =\frac{1}{{{\pi
^{1/2}}{\sigma _{\mathrm{B}/\mathrm{F}}}\left( {z,t}\right) }}{\exp }\left( -%
\frac{r^{2}}{{2{\left( \sigma _{\mathrm{B}/\mathrm{F}}\left( {z,t}\right)
\right) ^{2}}}}\right) f_{\mathrm{B}/\mathrm{F}}\left( {z,t}\right) ,
\label{Eq37}
\end{equation}%
where $\sigma _{\mathrm{B}/\mathrm{F}}$ are the transverse GS Gaussians
widths. Here, the axial functions, $f_{\mathrm{B}/\mathrm{F}}$, are
normalized to $N_{\mathrm{B}/\mathrm{F}}$. For both species, we define the
axial density as ${n_{1D,{\mathrm{B}/\mathrm{F}}}}\equiv {\left\vert f_{%
\mathrm{B}/\mathrm{F}}\right\vert ^{2}}$. By means of a procedure similar to
the one outlined above for the 2D reduction, we derive the Euler-Lagrange
equations for the BFM in the 1D approximation:
\begin{eqnarray}
i\hbar {\partial _{t}}{f_{\mathrm{B}}} &=&\left[ {-\frac{{{\hbar ^{2}}}}{{2{%
m_{\mathrm{B}}}}}\partial _{Z}^{2}+{U{_{1\mathrm{D,B}}}}+\frac{1}{\pi }\frac{%
{{g_{\mathrm{BF}}}}}{{\sigma _{\mathrm{B}}^{2}+\sigma _{\mathrm{F}}^{2}}}{{%
\left\vert {{f_{\mathrm{F}}}}\right\vert }^{2}}}+{\frac{{{g_{\mathrm{B}}}}}{{%
2\pi \sigma _{\mathrm{B}}^{2}}}{{\left\vert {{f_{\mathrm{B}}}}\right\vert }%
^{2}}}\right.  \notag \\
&&+\left. \frac{{{\hbar ^{2}}}}{{2{m_{B}}\sigma _{\mathrm{B}}^{2}}}+\frac{1}{%
2}{m_{B}}\omega _{t,\mathrm{B}}^{2}\sigma _{\mathrm{B}}^{2}\right] {f_{%
\mathrm{B}},}  \label{Eq38}
\end{eqnarray}%
\begin{eqnarray}
i\frac{\hbar }{{{\lambda _{1}}}}{\partial _{t}}{f_{\mathrm{F}}} &=&\left[ {-%
\frac{{{\hbar ^{2}}}}{{2{\lambda _{2}}{m_{\mathrm{F}}}}}\partial _{Z}^{2}+{%
U_{1d,\mathrm{F}}}+\frac{1}{\pi }\frac{{{g_{\mathrm{BF}}}}}{{\sigma _{%
\mathrm{B}}^{2}+\sigma _{\mathrm{F}}^{2}}}{{\left\vert {{f_{B}}}\right\vert }%
^{2}}}+\frac{{{g_{\mathrm{F}}}}}{{2\pi \sigma _{\mathrm{F}}^{2}}}{{%
\left\vert {{f_{\mathrm{F}}}}\right\vert }^{2}}\right.  \notag \\
&&+\left. \frac{{{\hbar ^{2}}\xi }}{{2{m_{\mathrm{F}}}}}\frac{{{C_{\mathrm{F}%
,1\mathrm{D}}}}}{{\sigma _{\mathrm{F}}^{4/3}}}{{\left\vert {{f_{\mathrm{F}}}}%
\right\vert }^{4/3}}+{\frac{{{\hbar ^{2}}}}{{2{m_{\mathrm{F}}}\lambda
_{2}\sigma _{\mathrm{F}}^{2}}}+\frac{1}{2}{m_{\mathrm{F}}}\omega _{t,\mathrm{%
F}}^{2}\sigma _{\mathrm{F}}^{2}}\right] {f_{\mathrm{F}}}.  \label{Eq39}
\end{eqnarray}%
In addition, the algebraic relationships between $\sigma _{\mathrm{B}/%
\mathrm{F}}$ and $f_{\mathrm{B}/\mathrm{F}}$ are:
\begin{equation}
{\chi _{I,\mathrm{B}}}\sigma _{\mathrm{B}}^{4}-\frac{{{\hbar ^{2}}}}{{{m_{%
\mathrm{B}}}}}-\frac{{{g_{B}}}}{{2\pi }}{n_{1\mathrm{D,B}}}=0,  \label{Eq40}
\end{equation}%
\begin{equation}
{\chi _{I,\mathrm{F}}}\sigma _{B}^{4}-\frac{2}{5}\frac{{{\hbar ^{2}}}}{{{m_{%
\mathrm{F}}}}}\xi {C_{\mathrm{F},1\mathrm{D}}}n_{1\mathrm{D},\mathrm{F}%
}^{2/3}\sigma _{\mathrm{F}}^{2/3}-\frac{\hbar ^{2}}{\lambda _{2}m_{\mathrm{F}%
}}-\frac{g_{\mathrm{F}}}{{2\pi }}{n_{1\mathrm{D},\mathrm{F}}}=0,
\label{Eq41}
\end{equation}%
where $\chi _{I,\mathrm{B}/\mathrm{F}}\equiv m_{\mathrm{B}/\mathrm{F}}\omega
_{t,\mathrm{B}/\mathrm{F}}^{2}-2g_{\mathrm{BF}}n_{1\mathrm{D},\mathrm{F}/%
\mathrm{B}}/[\pi (\sigma _{\mathrm{B}}^{2}+\sigma _{\mathrm{F}}^{2})^{2}]$.
Thus, Eqs. (\ref{Eq38})-(\ref{Eq41}) constitute a system of four 1D coupled
equations produced by the reduction of the underlying 3D system (\ref{Eq21})
- (\ref{Eq22}). Simulations of the system were performed with mesh
parameters $\Delta t=0.5$ $\mu $s and $\Delta z=0.25$ $\mu $m. The external
potential is chosen here as the harmonic-oscillator one: $U_{1\mathrm{d},%
\mathrm{B/F}}=m_{\mathrm{B}/\mathrm{F}}\omega _{z,{\mathrm{B}/\mathrm{F}}%
}^{2}z^{2}/2$.

\begin{figure}[tbp]
\centering
\resizebox{0.6\textwidth}{!}{
\includegraphics{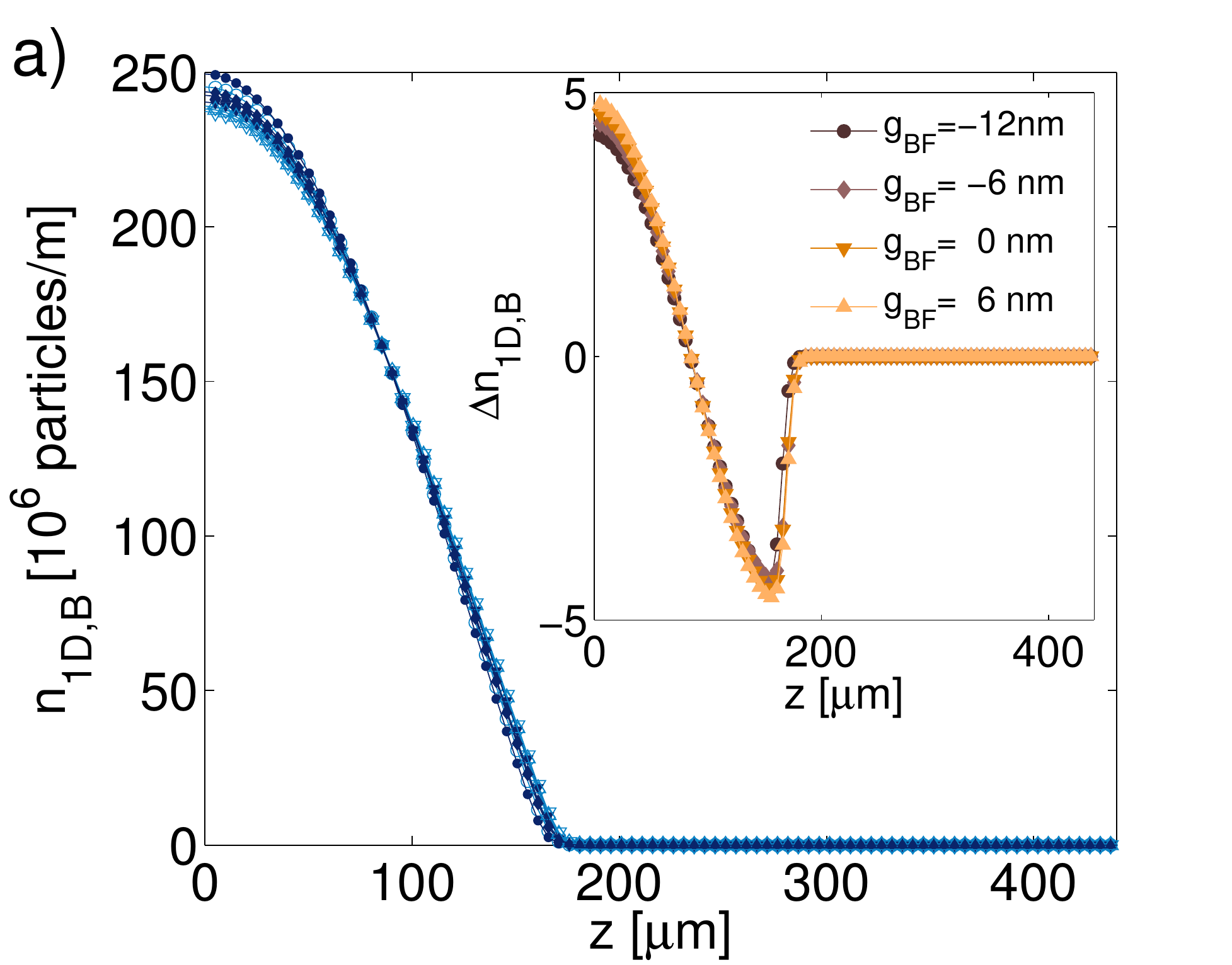}
\includegraphics{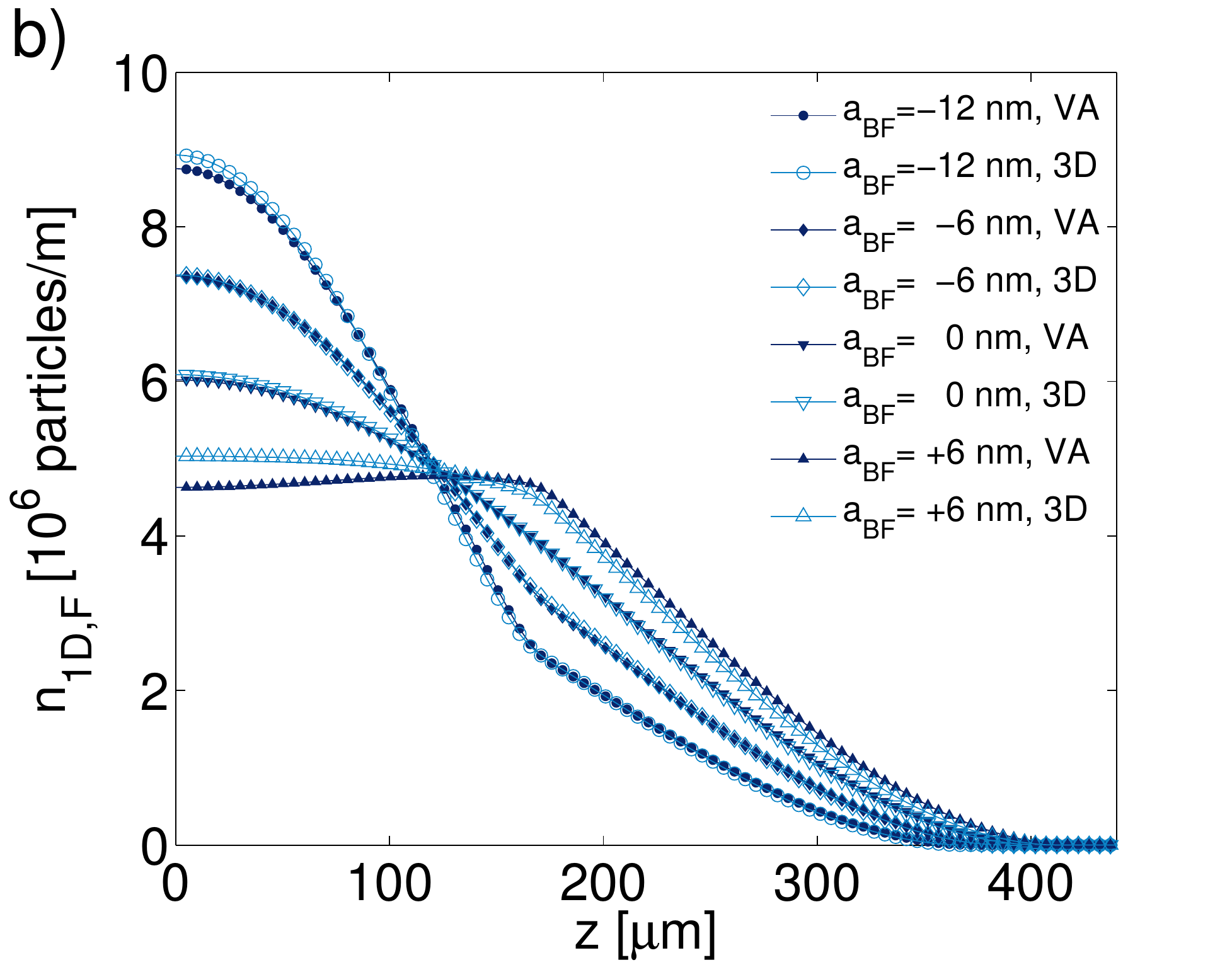}
}
\resizebox{0.6\textwidth}{!}{
\includegraphics{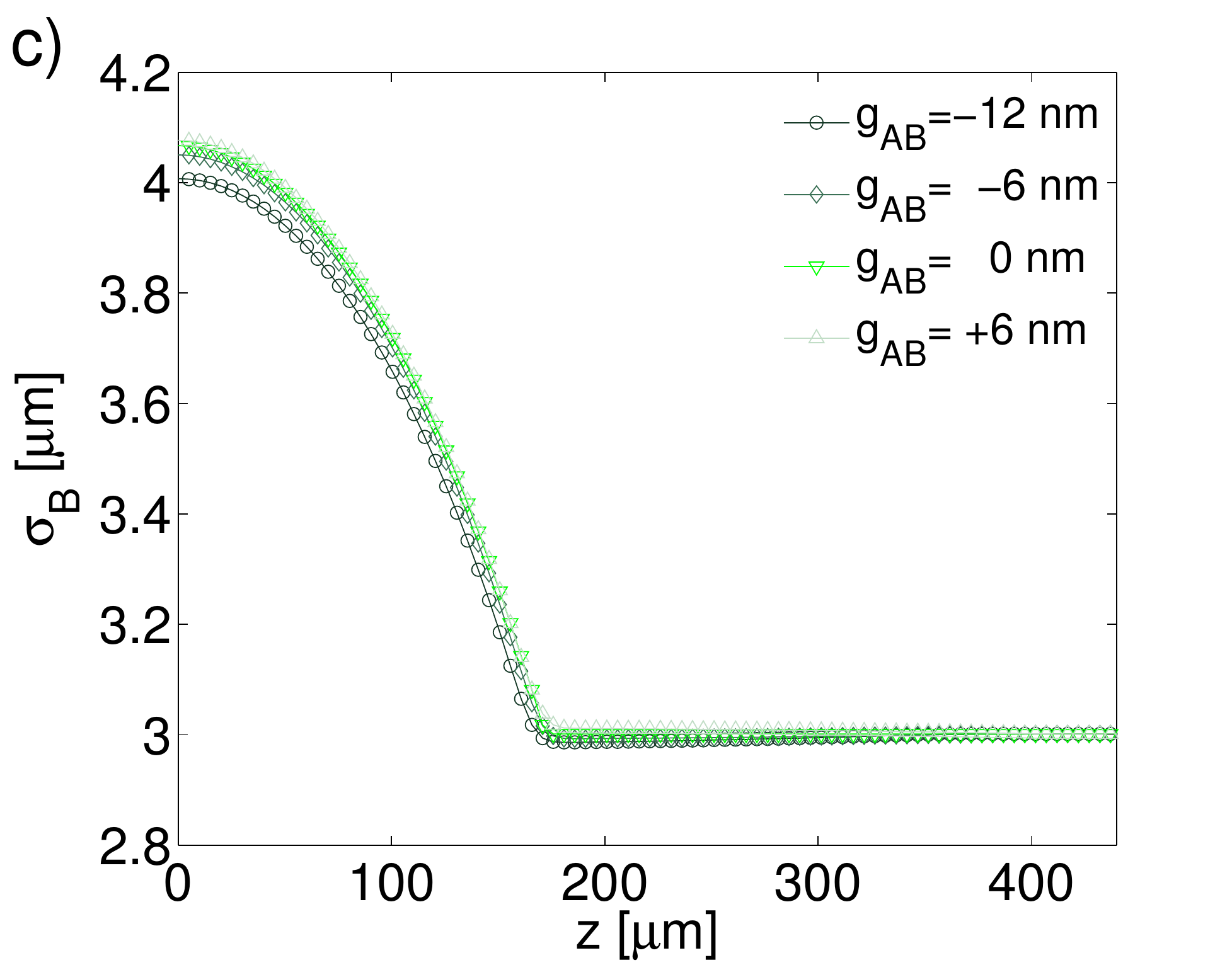}
\includegraphics{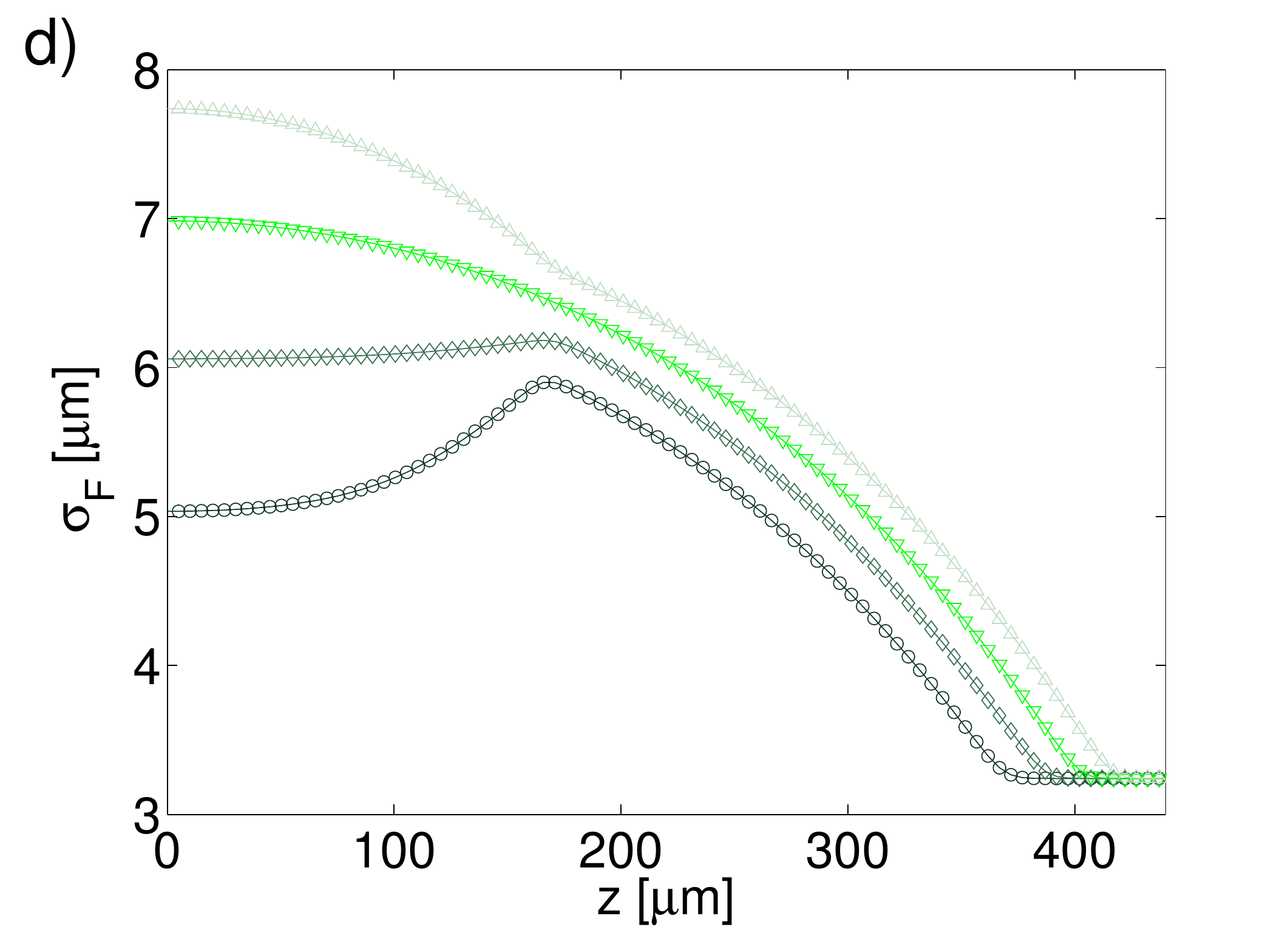}
}
\caption{Profiles of the particle density and the width in the confined
direction as a function the $z$-coordinate for different values of the
interaction strength $a_{\mathrm{BF}}$. (a) $n_{1\mathrm{D,B}}$, (b) $n_{1%
\mathrm{D,F}}$, (c) $\protect\sigma _{\mathrm{B}}$, and (d) $\protect\sigma %
_{\mathrm{F}}$. The parameters are $N_{\mathrm{B}}=5\times 10^{4}$, $N_{%
\mathrm{F}}=2.5\times 10^{3}$, $a_{\mathrm{B/F}}=5$ nm, $\protect\omega _{z,%
\mathrm{B/F}}=30$ Hz and $\protect\omega _{t,\mathrm{B/F}}=1000$ Hz. The
inset in panel (a) shows the difference between the VA and full 3D
simulations, by means of $\Delta n_{\mathrm{1D,B}}=\bar{n}_{\mathrm{1D,B}%
}-n_{\mathrm{1D,B}}$. This figure is taken from Ref. \protect\cite{Diaz15}.}
\label{FigBFM1D}
\end{figure}

The effect of the magnitude and sign of the interaction parameter on the
spatial profile of both species, and the accuracy of the VA compared to the
3D solution, can be analyzed by varying the scattering length, $a_{\mathrm{BF%
}}$. In particular, we consider a mixture with more bosons than fermions,
\textit{viz}., $N_{\mathrm{B}}=5\times 10^{4}$, $N_{\mathrm{F}}=2.5\times
10^{3}$. Because of this condition, the bosonic profile is mainly determined
by its self-interaction and the external potential. Frames (a) and (b) of
Fig. \ref{FigBFM1D} show the spatial dependence of $n_{1\mathrm{D,B}}$ and $%
n_{1\mathrm{D},\mathrm{F}}$, respectively. These densities are calculated
using both the reduced equations (\ref{Eq38}) - (\ref{Eq41}) and the full
numerical simulations of Eqs. (\ref{Eq21}) and (\ref{Eq22}). In the latter
case, the densities are calculated as $\bar{n}_{1\mathrm{D},\;\mathrm{j}%
}(z)=\int \int \left\vert \Psi _{\mathrm{j}}(\mathbf{r})\right\vert ^{2}dxdy$
with $j=(F,B)$. We observe that variations of the bosonic density profile
are very small in comparison to the significant changes of the inter-species
scattering length. The situation is opposite for the fermionic species. As
the repulsive scattering length increases, the fermions tend to be pushed to
the periphery of the bosonic-gas density profile. This phenomenon is known
as \textit{demixing} \cite{Adhikari2006,Adhikari08,Salasnich07a,Adhikari2007}%
. On the other hand, for the attraction case, fermions are, naturally,
concentrated in the same region where the bosons are located. Frames (c) and
(d) of Fig. \ref{FigBFM1D} correspond to the profiles of $\sigma _{\mathrm{B}%
}$ and $\sigma _{\mathrm{F}}$. We observe that the width of the bosonic
profile slightly increases while proceeding from the inter-species
attraction to repulsion. A similar trend is observed for fermions, as shown
in panel (d). However, the effect is amplified in the spatial zone of the
interaction with the bosons, where the gas is compressed in the case of the
attraction, and expands in the case of the repulsion. Note that the
fermionic component expands in the confined direction much more than its
bosonic counterpart, and that the fermionic width markedly varies, following
changes in the density.

Further, one can see in the inset of panel (a) of Fig. \ref{FigBFM1D} the
difference between the density calculated by means of the VA and the full 3D
simulation, $\Delta n_{\mathrm{1D,B}}=\bar{n}_{\mathrm{1D,B}}-n_{\mathrm{1D,B%
}}$ is really small. In fact, the difference between the bosonic profiles
obtained by both methods is $\sim 2\%$ of the maximum density for all cases
(the fact that the error changes very little with variations in $a_{\mathrm{%
BF}}$ is a consequence of the greater number of bosons). Frame (b) of Fig. %
\ref{FigBFM1D} shows that, for the attractive mixture, the variational
profile is very close to the 3D result, in particular for the case of $a_{%
\mathrm{BF}}=-6~$nm. For the repulsive mixture, it is observed that the
error increases, which is a consequence of the lower fermionic density at
the center of the 3D harmonic-oscillator potential, which plays the dominant
role for the bosons, hence a monotonously decreasing function in the
transverse direction, such as the Gaussian, is not a sufficiently good
approximation. We define the global error of the VA as $E_{\%,\mathrm{1D}%
}=\int_{-\infty }^{+\infty }\left\vert \bar{n}_{\mathrm{1D},\;j}-n_{\mathrm{%
1D}\;j}\right\vert dz$ (for both species). We have found that in the range
of $a_{\mathrm{BF}}\in \left( -6,6\right) $ nm the global error for the
bosonic species is around $2\%$ for all the values of $a_{\mathrm{BF}}$. For
the fermionic species, it goes from $0.5\%$ to $5\%$ depending on $a_{%
\mathrm{BF}}$, such that for positive value of $a_{\mathrm{BF}}$ the error
is higher than for negative ones, and the minimum error is attained at $a_{%
\mathrm{BF}}\approx -4$nm. This is a consequence of the fact that, for this
value of $a_{\mathrm{BF}}$, the interspecies interaction practically
compensates the Pauli repulsion, making the dynamics of the fully polarized
Fermi gas close to that governed by the linear Sch\"{o}dinger equation
(recall that the Gaussian is the solution for the ground state). When the
mixture becomes more attractive, the fermionic dynamics is dominated by the
bosons, producing a similar error for both species, while for the repulsive
mixture the Gaussian approximation is not appropriate. For the
non-interacting mixture, the error for the fermions is smaller than for the
bosons, because the fermionic density is very low, making the
self-interaction terms weak in comparison to the external potential,
therefore it is appropriate to use the Gaussian ansatz to approximate the 1D
dynamics. Finally, note that the error is lower in the 2D case in comparison
with 1D, because the reduction to 2D case is closer to the full 3D model.

\begin{figure}[tbp]
\centering
\resizebox{1.\textwidth}{!}{
\includegraphics{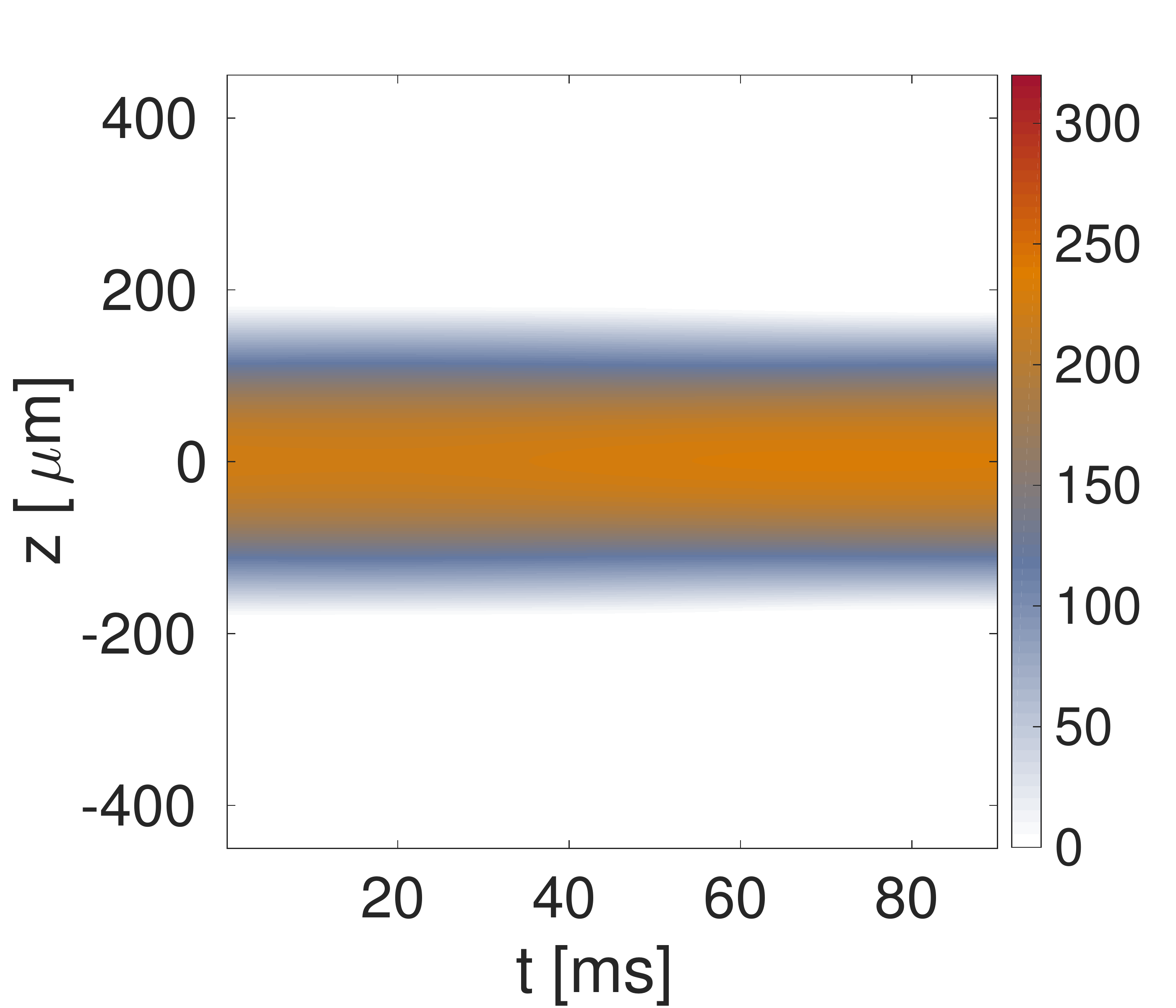}
\includegraphics{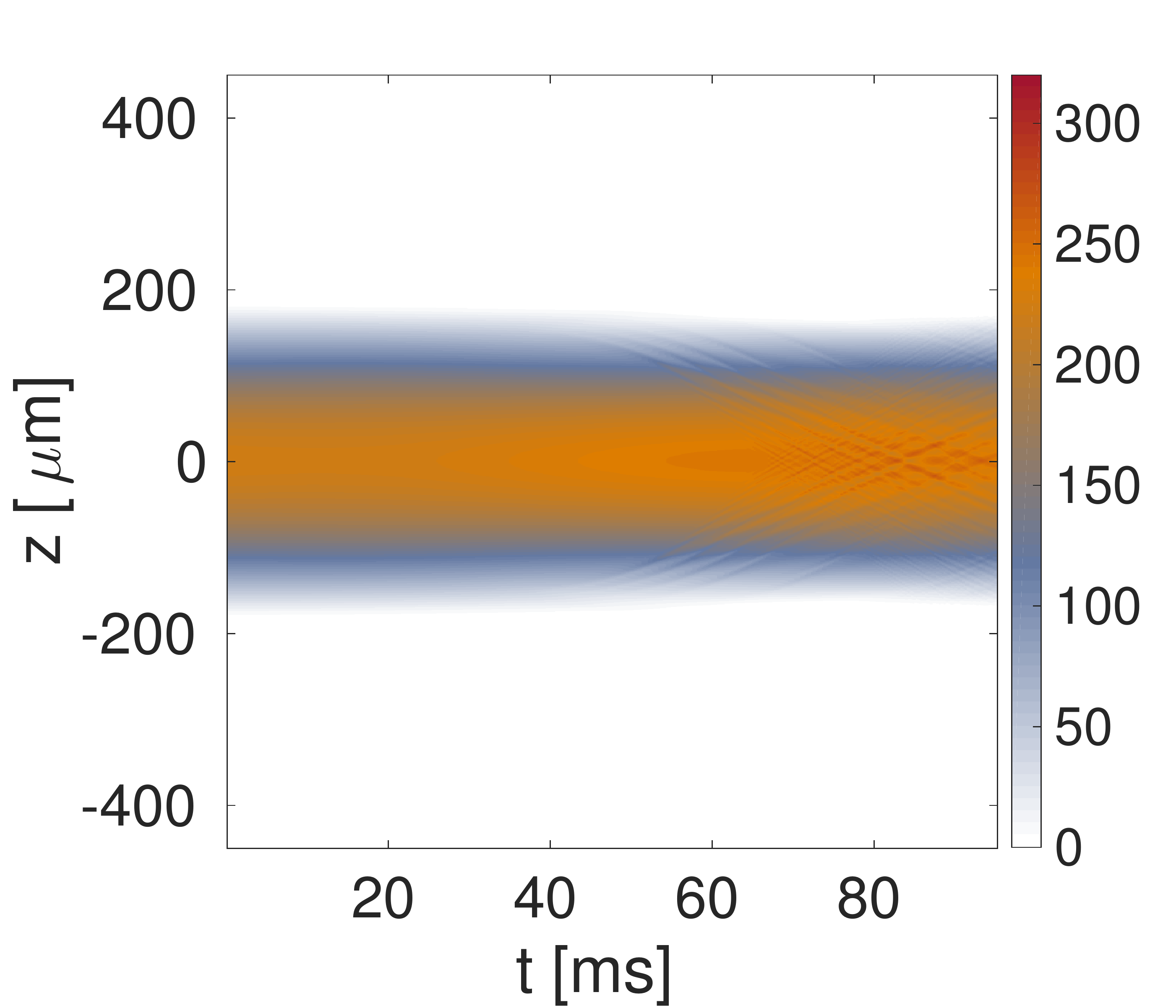}
\includegraphics{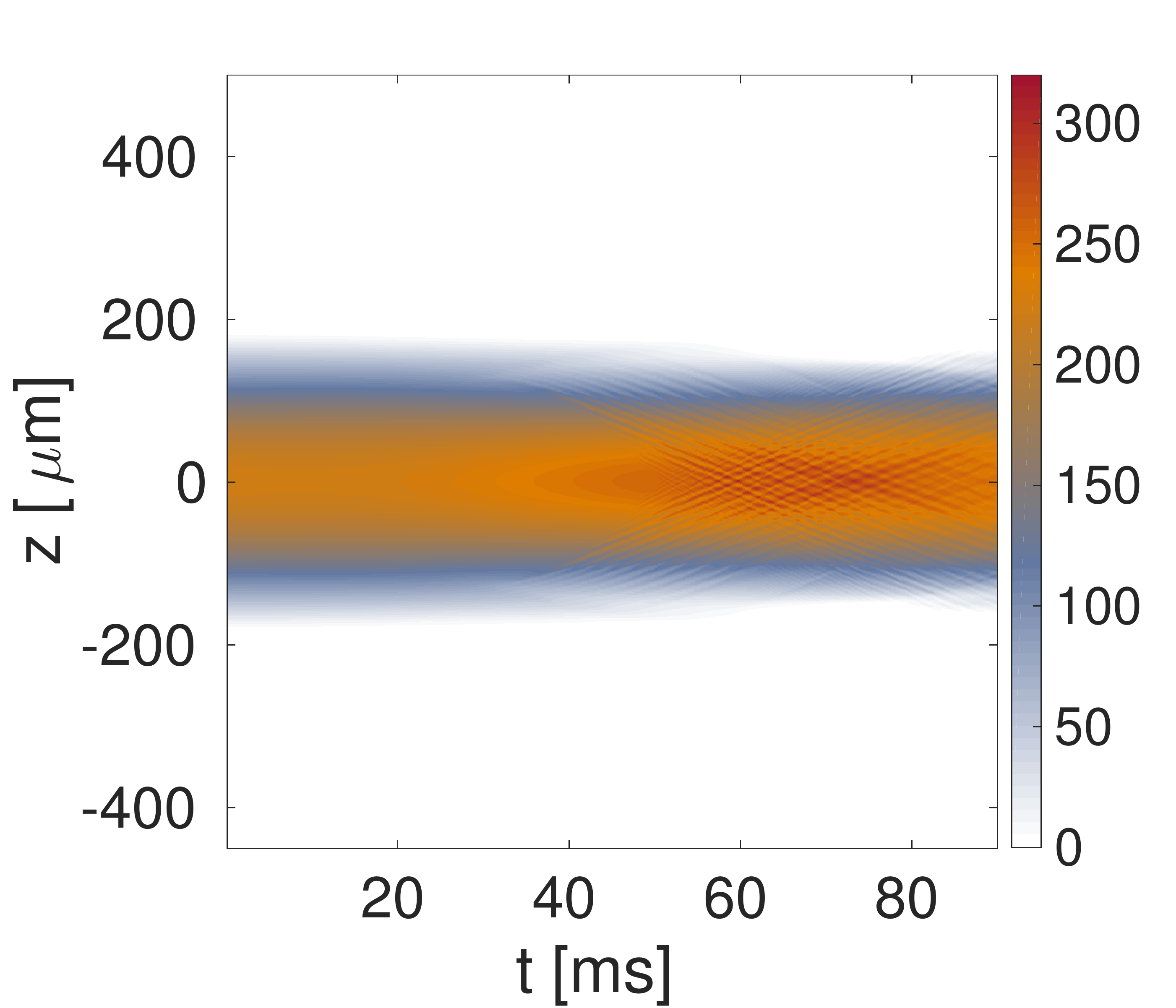}
}
\resizebox{1.\textwidth}{!}{
\includegraphics{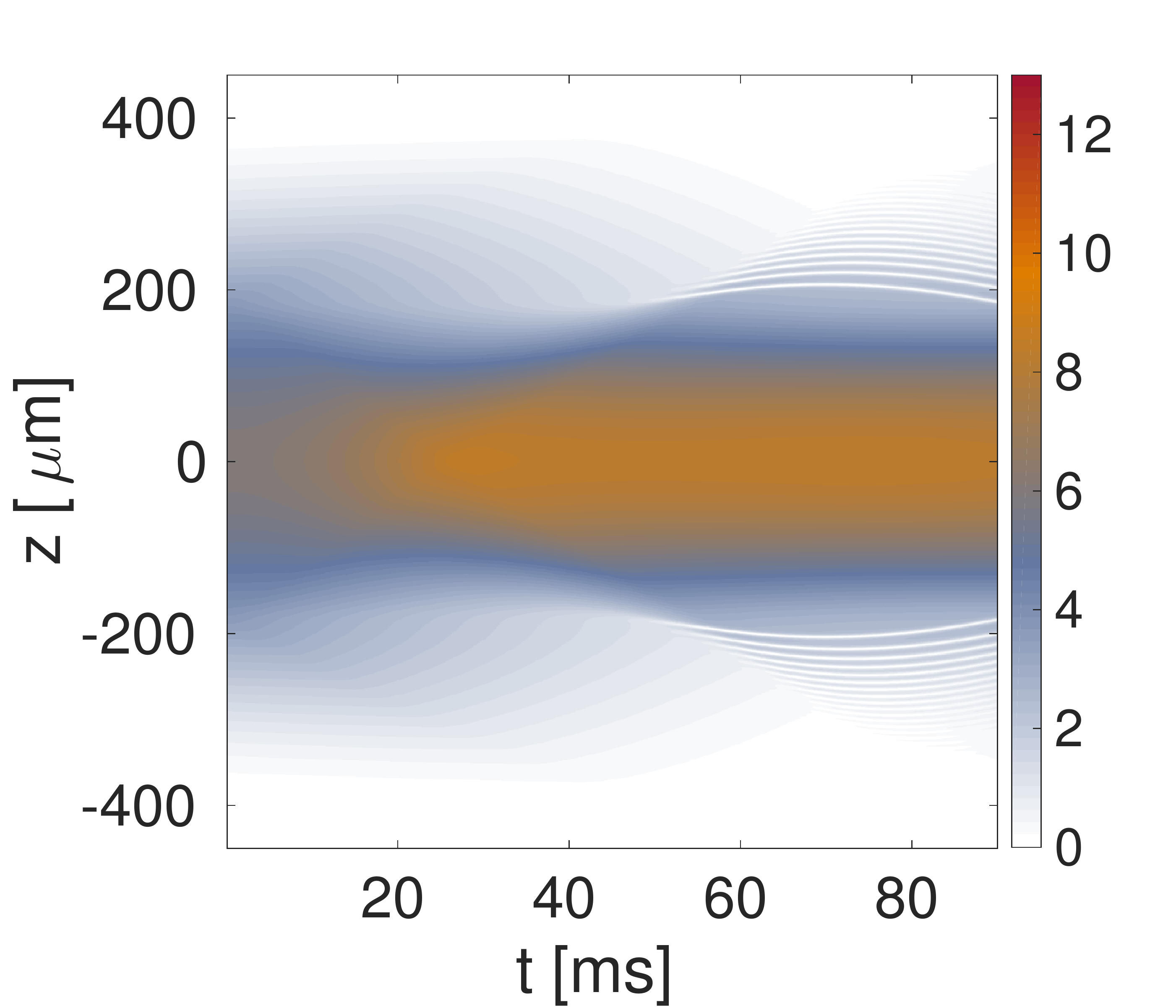}
\includegraphics{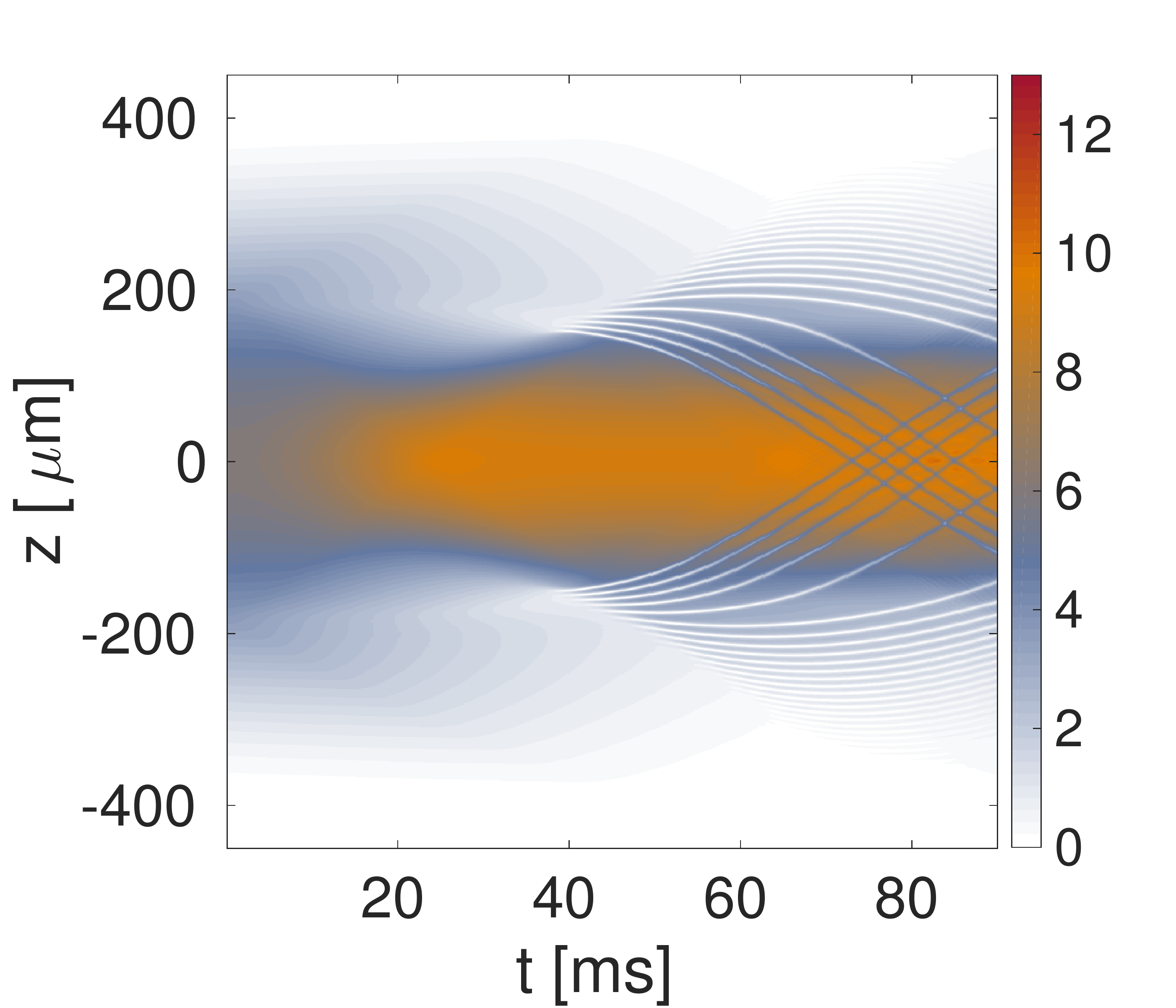}
\includegraphics{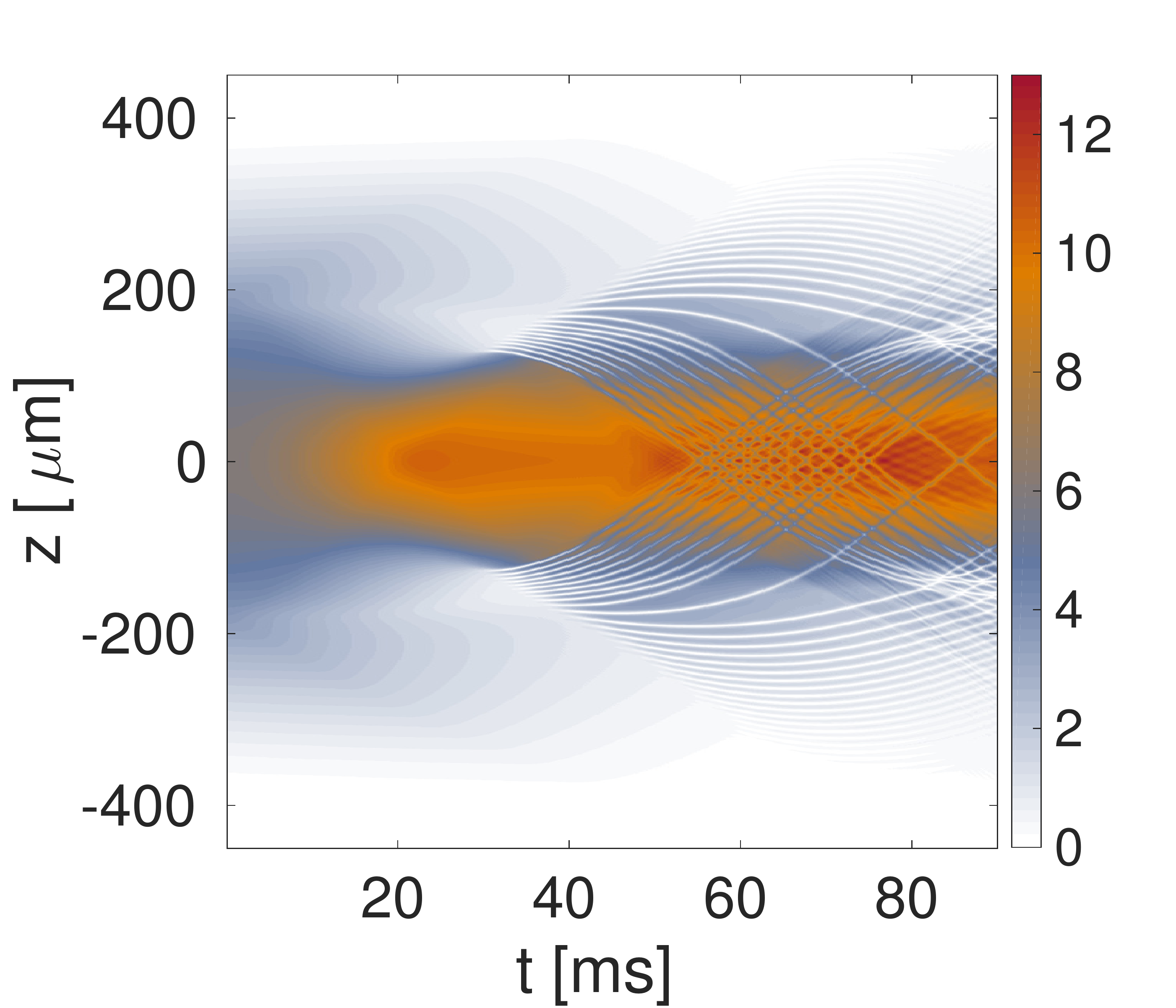}
}
\caption{(Color online) Space-time diagrams of the densities of bosons (top)
and fermions (bottom), for three different values of the interspecies
scattering parameter: (a,b) $a_{\mathrm{BF}}=-18nm$, (c,d) $a_{\mathrm{BF}%
}=-26nm$, and (e,f) $a_{\mathrm{BF}}=-34nm$. The initial conditions are the
same in all the cases, see the text. The other parameters are the same as in
Fig.~\protect\ref{FigBFM1D}}
\label{FigBFM1D2}
\end{figure}

Next, we address the BFM dynamics, considering a mixture with arbitrary
initial conditions for the 1D fields. To create the initial state, we start
with the GS found in the absence of the inter-species interaction ($a_{%
\mathrm{BF}}=0$). Then, at $t=0$, we switch the interaction on, which may
imply the application of the magnetic field, that gives rise to $a_{\mathrm{%
BF}}\neq 0$ via the FR. Figure~\ref{FigBFM1D2} shows three cases of the
temporal evolution with these initial conditions for $a_{\mathrm{BF}}=-18~%
\mathrm{nm}$, $a_{\mathrm{BF}}=-26~\mathrm{nm}$, and $a_{\mathrm{BF}}=-34~%
\mathrm{nm}$. In the first case (panels (a) and (b) of Fig. \ref{FigBFM1D2}%
), it is observed that the densities converge towards the center of the
potential, as may be expected, creating a pattern of oscillations around the
potential minimum; in addition, the fermions are affected by bosons, as
shown by the mark left by the bosons in the fermionic density. For the
second case (panels (c) and (d) of Fig. \ref{FigBFM1D2}), it is observed
that the increase in the strength of the attractive interaction generates
dark solitons in the fermionic density, some of which show oscillatory
dynamics very close to that observed in Refs. \cite%
{Yefsah13,Scott11,Shomroni09,Cardoso13,Donadello14}. The last case (panels
(e) and (f) of Fig. \ref{FigBFM1D2}) shows that the further increase of the
strength of the interspecies interaction generates a larger number of dark
solitons. In other words, we show that the attractive interaction of
fermions with bosons in a state different from the GS eventually generates a
gas of dark solitons.

\begin{figure}[tbp]
\centering{\
\resizebox{1.\textwidth}{!}{
\includegraphics{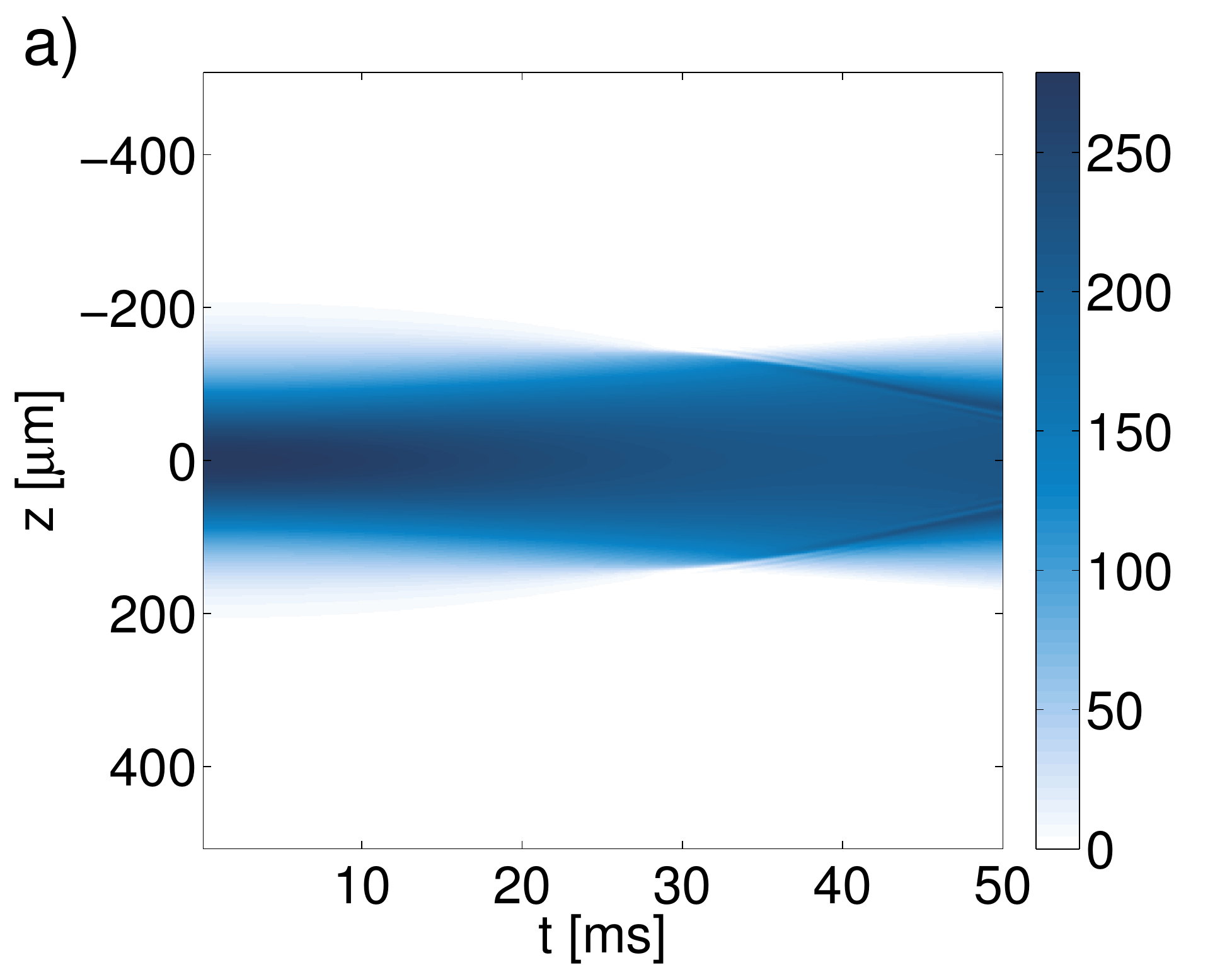}
\includegraphics{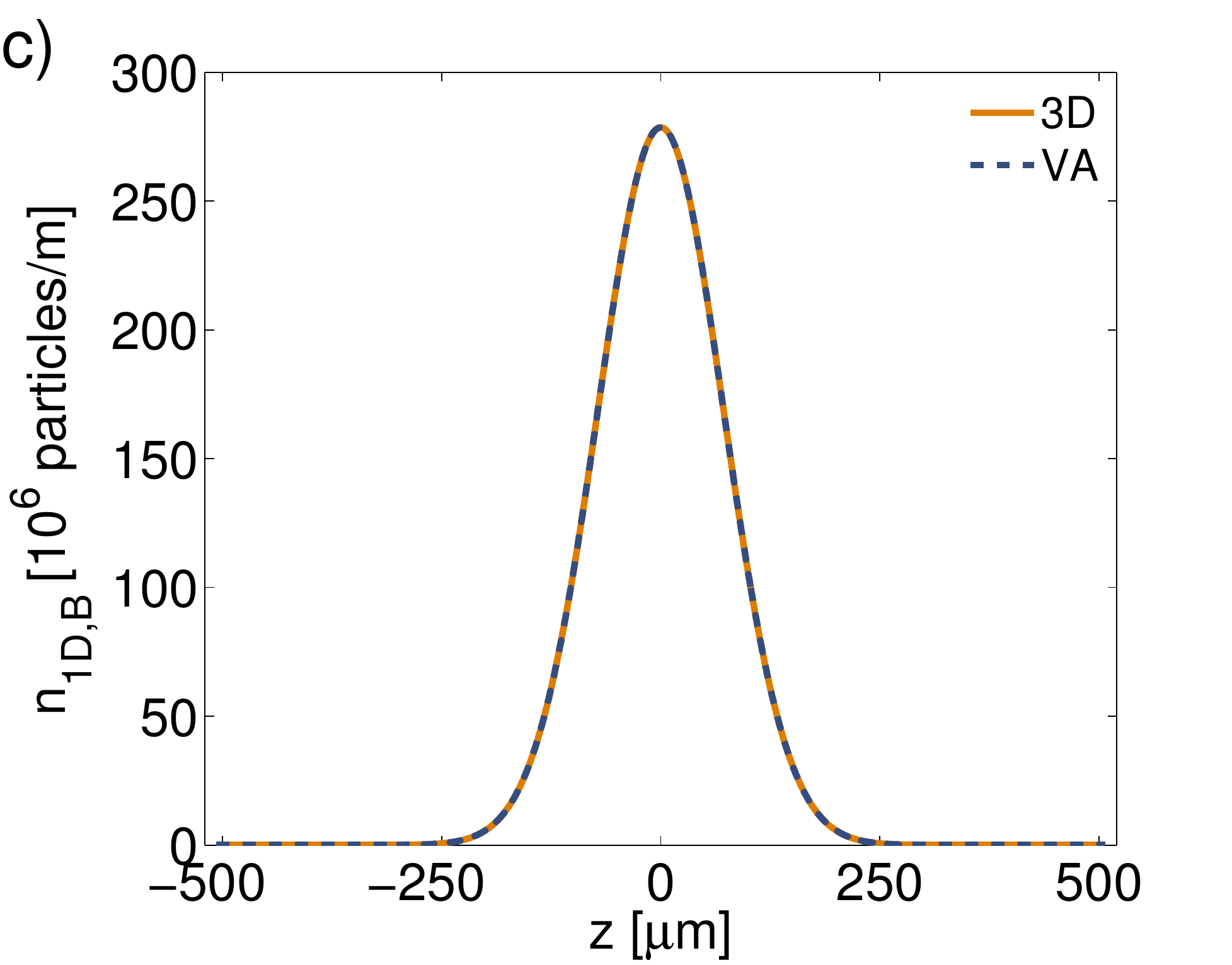}
\includegraphics{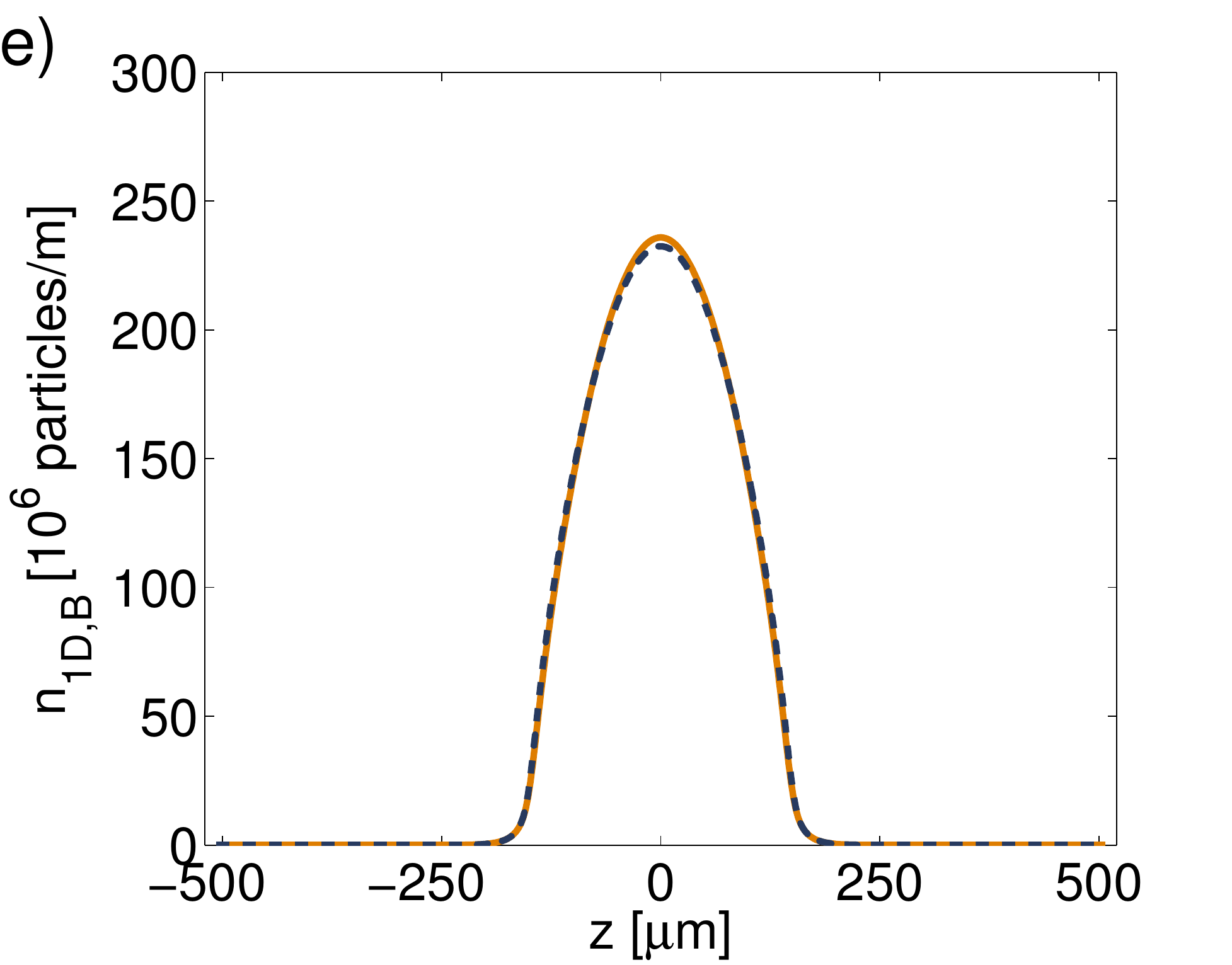}
\includegraphics{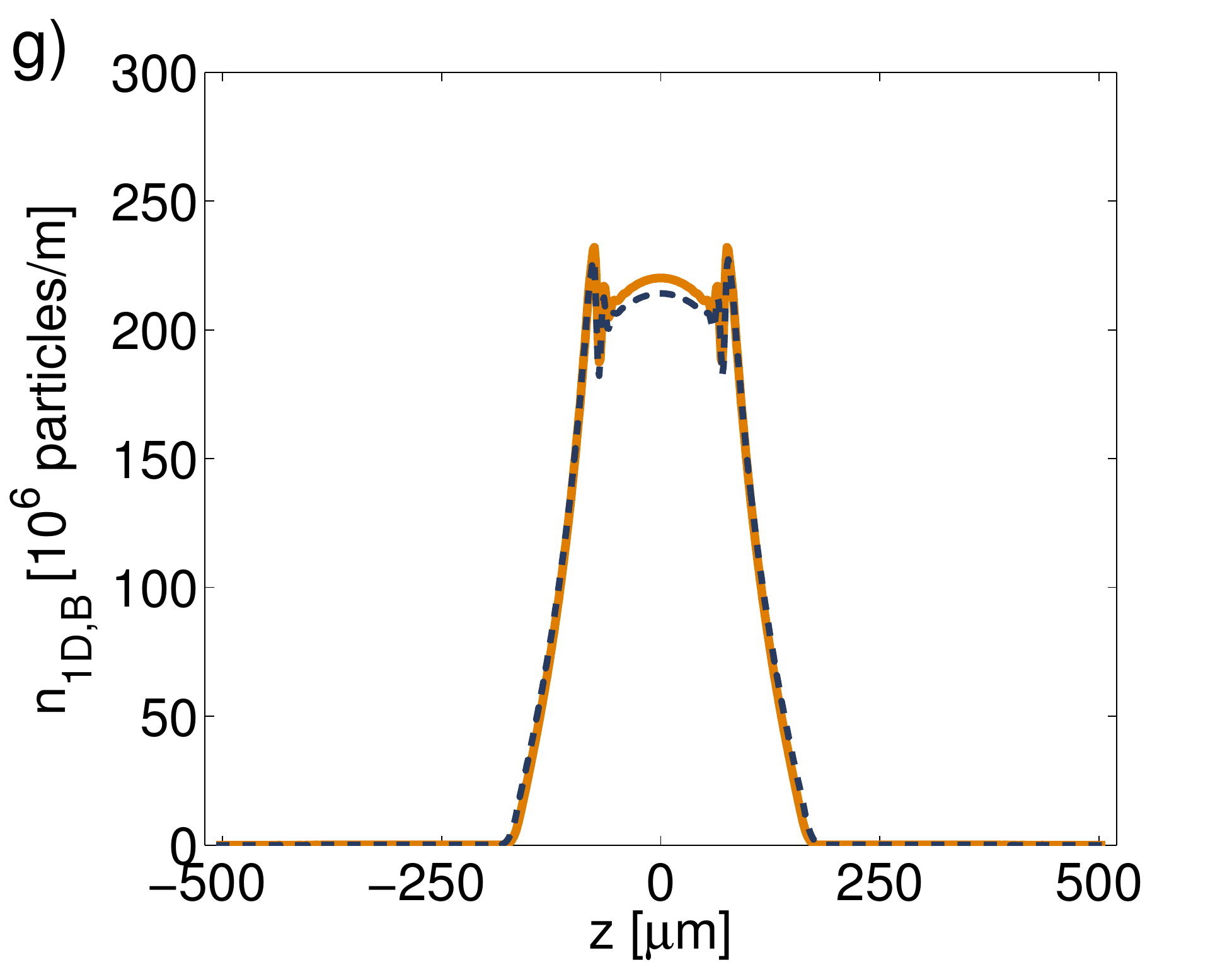}
}
\resizebox{1.\textwidth}{!}{
\includegraphics{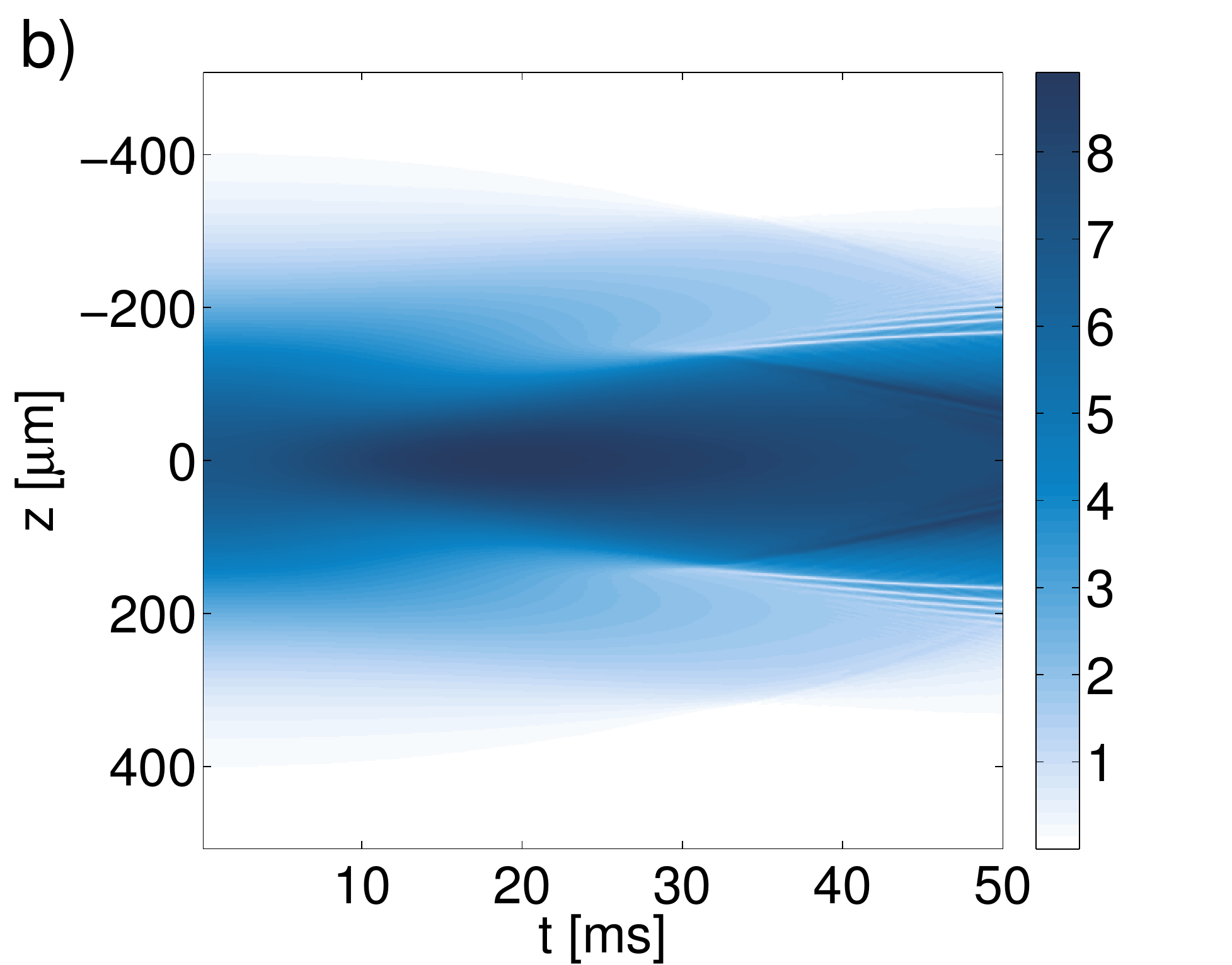}
\includegraphics{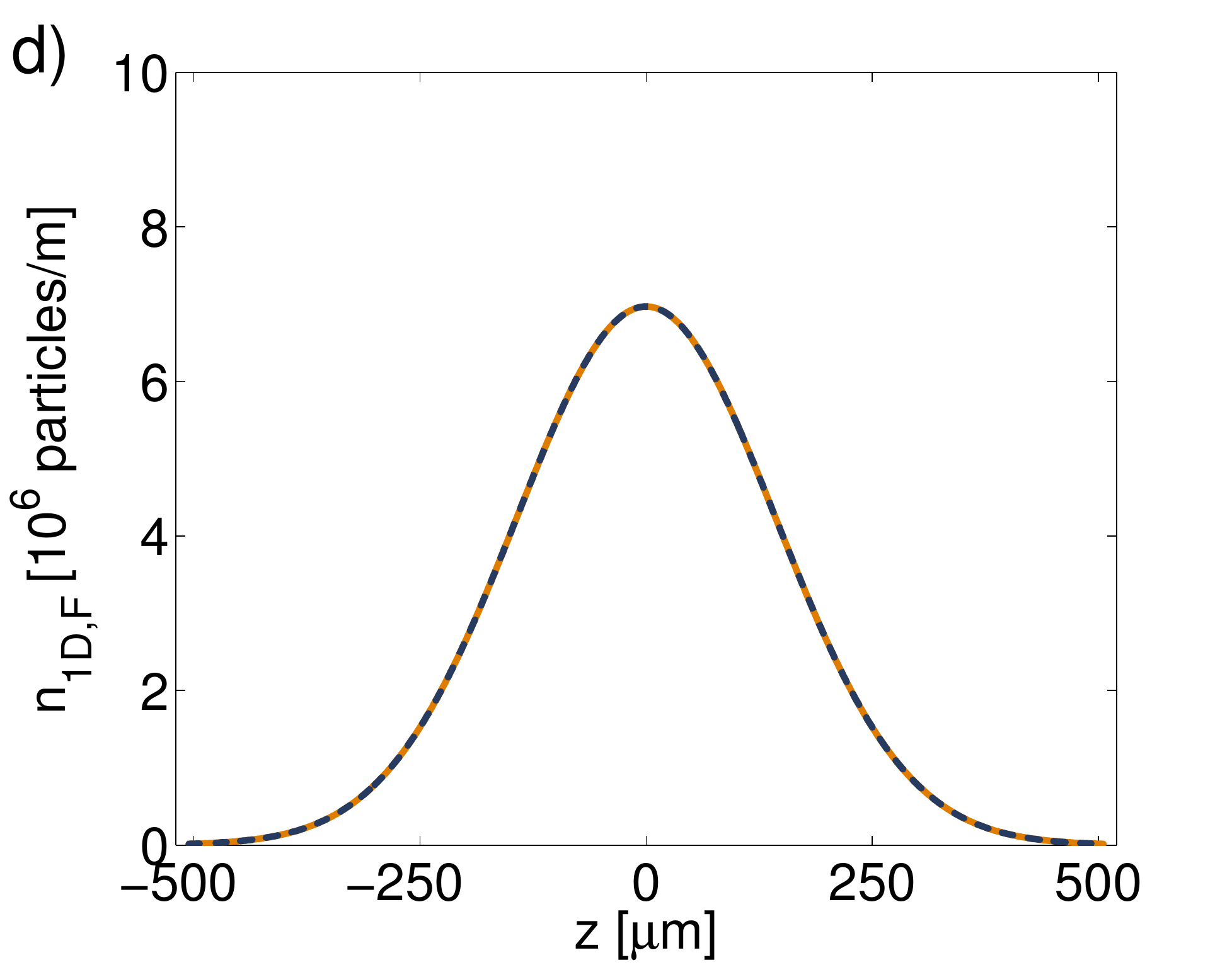}
\includegraphics{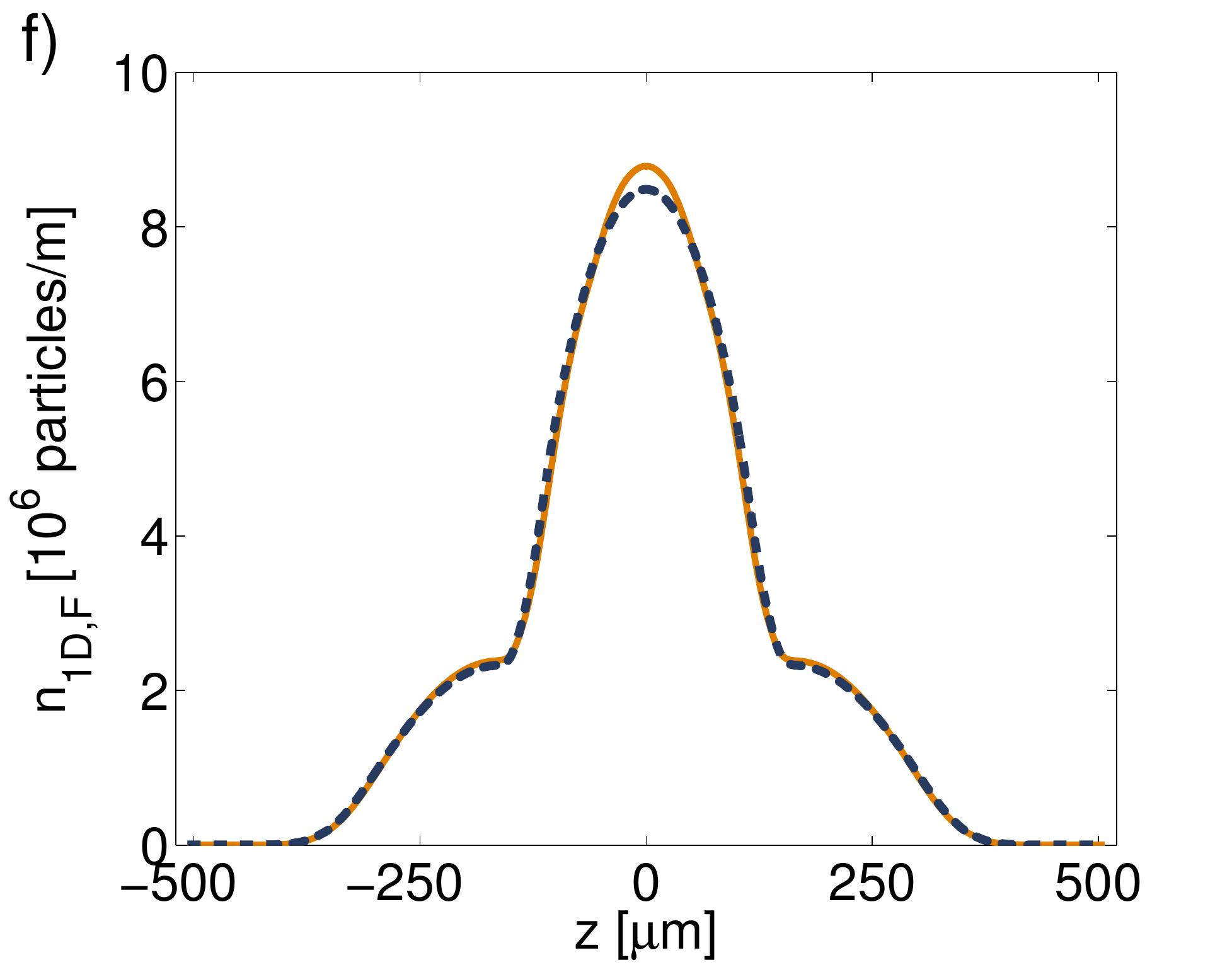}
\includegraphics{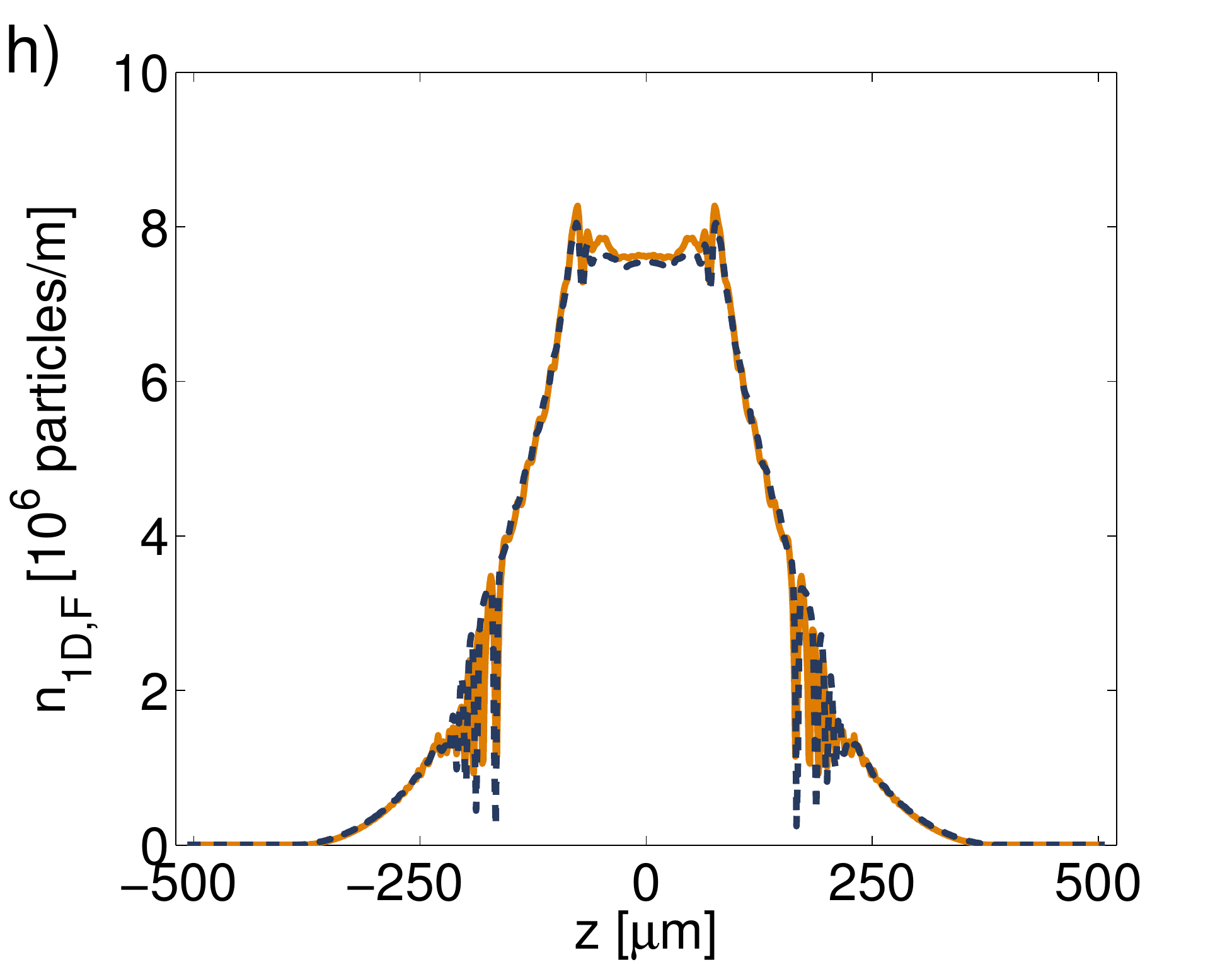}
}}
\caption{(Color online) Comparison of the dynamics, as produced by the 1D
VA, and from the 3D simulations. Spatiotemporal diagrams for bosons (a) and
fermions (b) are obtained from the 3D simulations. The other panels show
spatial profiles for: (c)$-$(d) $t=0~$ms, (e)$-$(f) $t=25$ ms, and (e)$-$(f)
$t=50$ ms. Here $a_{\mathrm{BF}}=-10$ nm, the initial conditions and other
fixed parameters being the same as in Fig.~\protect\ref{FigBFM1D}. This
figure is taken from Ref. \protect\cite{Diaz15}. }
\label{FigBFM1D3}
\end{figure}

Finally, we address the accuracy of the VA for the dynamical behavior near
the GS. Figure \ref{FigBFM1D3} displays the spatiotemporal dynamics of the
1D density, as produced by the solution of 3D equations (\ref{Eq21}) and (%
\ref{Eq22}) for $a_{\mathrm{BF}}=-10$ nm. The initial conditions for the 3D
dynamics are given by the ansatz based on Eq. (\ref{Eq25}), with the
Gaussian profile along the $z$ axes. Panels (a) and (b) of Fig. \ref%
{FigBFM1D3} show the spatiotemporal diagrams of the bosonic and fermionic
densities, making the emergence of dark solitons obvious. This result
corroborates that the dark solitons emerge too in the 3D dynamics, which is
approximated by the present 1D model. The other panels of Fig. \ref%
{FigBFM1D3} show a comparison of the 1D spatial profiles, as obtained from
the 3D simulations, and the 1D VA, for three instants of time: $t=0$ ms ((c)
and (d)), $t=25$ ms ((e) and (f)), and $t=50$ ms ((g) and (h)). The results
demonstrate that the VA profiles are very similar to their counterparts
produced by the 3D simulations, hence the present approximation provides
good accuracy and allows one to study dynamical features of the BFM in a
sufficiently simple form.

\section{Conclusion}

\label{S4}

In this brief review we have summarized results produced by the VA
(variational approximation) for reducing the 3D system to 1D and 2D forms
for the Fermi gas and BFM\ (Bose-Fermi mixture) in the framework of the
quasi-mean-field description \cite{Diaz12,Diaz15}. The method is based on
the Gaussian variational ansatz, which provides very accurate results for
the GSs (ground states) of the gases loaded in the disc- and cigar-shaped
traps. The reduced equations are useful, in particular, for modeling systems
with low atomic densities and large spatiotemporal variations of the
external potential. For the 1D case, the reduced equations provide results
by means of modest computational resources, allowing one to quickly explore
a vast volume of the parameter space. In the 2D case, the required
simulation time is still significantly lower than what is necessary for the
3D simulations. We have shown that, for the Fermi gases and BFMs alike, the
VA produces results with a very good accuracy, the error, in the comparison
to the 3D simulations, being lower than $5\%$, for both the GSs and
dynamical states.

For the Fermi gas case in the 2D approximation, we have considered the
example of the hexagonal superlattice, built as a superposition of two
triangular lattices with different angles among them. This possibility 
may be relevant for emulating condensed-matter settings, such
as graphene-like superlattices. In addition, we have presented results for
dark solitons, obtained in the framework of the 1D approximation. We have
verified that the interaction is repulsive \cite{Alphen18} and strongly
depends on the initial distance between the dark solitons.

Finally, for the BFM trapped in the harmonic-oscillator potential we have
shown that a change in the interaction strength can generate a gas of dark
solitons. The solitons oscillate under the action of the external potential.

\section*{Acknowledgents}

PD acknowledges partial financial support from DIUFRO project under grant
DI18-0066 and CMCC of the Universidad de La Frontera. DL acknowledge partial
financial support from Centers of Excellence with BASAL/CONICYT financing,
Grant FB0807, CEDENNA and CONICYT-ANILLO ACT 1410. PD and DL acknowledges
financial support from FONDECYT 1180905. The work of BAM is supported, in
part, by the joint program in physics between NSF and Binational (US-Israel)
Science Foundation through project No. 2015616, and by the Israel Science
Foundation, through grant No. 1287/17. The authors appreciate a support
provided by the PAI-CONICYT program (Chile), grant No. 80160086, and
hospitality of Instituto de Alta Investigaci\'{o}n, at Universidad de Tarapac%
\'{a} (Arica, Chile).

\section*{Appendix A: Nonlinear Schr\"{o}dinger equation for the fermionic
superfluid}

\label{secA}

Kim and Zubarev in Ref. \cite{Kim2004} proposed an effective hydrodynamic
equation for a Fermi gas, in the regime of the BCS-BEC crossover. The
equation was derived from the time-dependent density-functional theory and
has the form given by:
\begin{equation}
i\hbar {\partial _{t}}\Psi \left( {\mathbf{r},t}\right) =\left[ {\ -\frac{{%
\hbar ^{2}}}{{2m_\mathrm{F}}}{\nabla ^{2}}+U\left( {\mathbf{r}}\right) +\mu
\left( n\left( {\mathbf{r},t}\right) \right) }\right] \Psi \left( {\mathbf{r}%
,t}\right) ,  \label{E-b0}
\end{equation}
where $\Psi $ is a complex field that represent the superfluid wave
function, $n\left( {\mathbf{r},t}\right) ={\left\vert \Psi \left( {\mathbf{r}%
,t}\right) \right\vert ^{2}}$ is the particle density, and $\mu$ is the
chemical potential. In addition, the relationship between the chemical
potential and the energy density (energy per particle), $\varepsilon \left(
n\right)$, is given by:
\begin{equation}
\mu \left( n\right) =\frac{\partial }{\partial n}\left[ {n\varepsilon \left(
n\right) }\right]  \label{mu}
\end{equation}%
For the case of two spin states with balanced populations and a negative
scattering length, ${a_\mathrm{F}}<0$, the BCS limit corresponds to $k_{F}{%
\left\vert {a_\mathrm{F}}\right\vert }\ll 1$, where $k_{F}=(3\pi
^{2}n)^{1/3} $ is the Fermi wavenumber. In this limit $\varepsilon $ is
given by \cite{Huang1957}:
\begin{equation}
\varepsilon \left( n\right) =\frac{3}{5}{\varepsilon _{F}}\left[ {1+\frac{{10%
}}{{9\pi }}{k_{F}}{a_\mathrm{F}}+\frac{{4\left( {11-2\ln \left( 2\right) }%
\right) }}{{21{\pi ^{2}}}}{{\left( {{k_{F}}{a_\mathrm{F}}}\right) }^{2}}%
+\cdots }\right] ,  \label{E-b1}
\end{equation}%
where ${\varepsilon _{F}}={\hbar ^{2}}k_{F}^{2}/\left( {2m_\mathrm{F}}%
\right) $ is the Fermi energy. Taking the Eq. (\ref{E-b1}) into the Eq. (\ref%
{mu}) the chemical potential takes the form

\begin{equation}
\mu \left( n\right) =\frac{{\hbar ^{2}}}{{2m_\mathrm{F}}}{\left( {3{\pi ^{2}}%
}\right) ^{2/3}}{n^{2/3}}+\frac{{2{\hbar ^{2}}\pi {a_\mathrm{F}}}}{m_\mathrm{%
F}}n\left[ {1+{1.893 a_\mathrm{F}}{n^{1/3}}+\cdots }\right]  \label{E-b2}
\end{equation}%
where the first term corresponds to the effective Pauli repulsion, and the
following ones to the superfluidity due to collisions between the fermions
in different spin states. Substituting the latter expression in Eq.(\ref%
{E-b0}), and keeping only the first collisional term, we obtain the known
nonlinear Schr\"{o}dinger equation for the fermionic superfluid \cite%
{Kim2004,Adhikari2006b}
\begin{equation}
i\hbar {\partial _{t}}\Psi =\left[ {\ -\frac{{\hbar ^{2}}}{{2m_\mathrm{F}}}{%
\nabla ^{2}}+U\left( {\mathbf{r}}\right) +\frac{{\hbar ^{2}}}{{2m_\mathrm{F}}%
}{{\left( {3{\pi ^{2}}}\right) }^{2/3}}{n^{2/3}}+\frac{{2\pi {\hbar ^{2}}{a_%
\mathrm{F}}}}{m_\mathrm{F}}n}\right] \Psi ,  \label{E-b3}
\end{equation}%
where the last term is similar to one in the Gross-Pitaevskii equation for
bosons, but with an extra factor of $1/2$, as the Pauli exclusion principle
allows only atoms in different spin states interact via the scattering. We
remark that Eq. (\ref{E-b3}) implies equal particle densities and phases of
the wave functions associated with both spin states.

When we have a system with multiple atomic spin states, $\sigma _{j}$,
associated with vertical projection of the spin $s$ (with $2s_\mathrm{F}+1$
states), we treat the atoms per state as a fully polarized Fermi gas. The
term for the interactions by collisions between atoms in different spin
states, with the same scattering length ($a_\mathrm{F}$), correspond to the
scattering term in the Gross-Pitaevskii equation. The motion equation for
the atoms in spin states $j$ is given by:
\begin{eqnarray}
i\hbar {\partial _{t}}{\Psi _{j}}\left( {\mathbf{r},t}\right) &=&\left[ {\ -%
\frac{{\hbar ^{2}}}{{2m_\mathrm{F}}}{\nabla ^{2}}+U\left( {\mathbf{r}}%
\right) +\frac{{\hbar ^{2}}}{{2m_\mathrm{F}}}{{\left( {6{\pi ^{2}}}\right) }%
^{2/3}}{{n_{j}\left( {\mathbf{r},t}\right) }^{2/3}}}\right] {\Psi _{j}}%
\left( {\mathbf{r},t}\right)  \notag \\
&&+\frac{{4\pi {\hbar ^{2}}{a_\mathrm{F}}}}{m_\mathrm{F}}\sum\limits_{k\neq
j=-(s_\mathrm{F}+1/2)}^{s_\mathrm{F}+1/2}{{{n_{k}\left( {\mathbf{r},t}%
\right) }}{\Psi _{j}}}\left( {\mathbf{r},t}\right) ,  \label{E-1}
\end{eqnarray}%
where $\Psi _{j}$ is the wave function associated with spin projection $%
\sigma _{j}$, such that $n_{j}\left( {\mathbf{r},t}\right) ={\left\vert {{%
\Psi _{j}}\left( {\mathbf{r},t}\right) }\right\vert ^{2}}$ is the respective
particle density, and $V(\mathbf{r})$ an external potential, which is
assumed to be identical for all the spin states.

In the case of fully locally balanced populations, the density of particles
is the same in each component, $n_{1}=n_{2}=...=n_{2s_\mathrm{F}+1}$, hence
the total density is $n=n_{j}/(2s_\mathrm{F}+1)$. Assuming also equal phases
of the wave-function components, we define a single wave function, $\Psi =%
\sqrt{2s_\mathrm{F}+1}\Psi _{j}$, such that the Eq. (\ref{E-1}) take the
form
\begin{equation}
i\hbar {\partial _{t}}{\Psi }=\left[ -\frac{{{\hbar ^{2}}}}{{2{m_\mathrm{F}}}%
}{\nabla ^{2}}+{U\left( {\mathbf{r}}\right)}+\frac{{\hbar ^{2}}}{{2m_\mathrm{%
F}}}{{\left( {\frac{{6{\pi ^{2}}}}{{2s_\mathrm{F}+1}}}\right) }^{2/3}}
\left\vert {\Psi \left( {{\mathbf{r}},t}\right) }\right\vert ^{4/3}+g_%
\mathrm{F} \left\vert {\Psi \left( {{\mathbf{r}},t}\right) }\right\vert ^{2} %
\right] {\Psi },  \label{E-2}
\end{equation}%
where $g_\mathrm{F}\equiv 8s_\mathrm{F}\pi {\hbar ^{2}}{a_\mathrm{F}}/(2s_%
\mathrm{F}+1)m_\mathrm{F}$ is the scattering coefficient. This equation is
the same that Eq. \ref{Eq3} without considered the corrections of the first
principles calculations given by $\lambda_1$, $\lambda_2$ and $\beta$ \cite%
{Manini05,Salasnich09,Ancilotto09,Ancilotto12}. In particular, the fully
polarized gas, with the interactions between identical fermions suppressed
by the Pauli principle, formally corresponds to $s_\mathrm{F}=0$, hence $g_%
\mathrm{F}=0$, and the last term of Eq. \ref{E-2} vanishes.

Finally, the equation (\ref{E-2}) can be derived, as the Euler-Lagrange
equation,

\begin{equation}
\frac{{\delta \mathcal{L}}}{{\delta {\Psi ^{\ast }}}}=\frac{{\partial
\mathcal{L}}}{{\partial {\Psi ^{\ast }}}}-\frac{\partial }{{\partial t}}%
\frac{{\partial \mathcal{L}}}{{\partial \left( {{\partial _{t}}{\Psi ^{\ast }%
}}\right) }}-\nabla \frac{{\partial \mathcal{L}}}{{\partial \left( {\nabla {%
\Psi ^{\ast }}}\right) }}=0,  \label{E-5}
\end{equation}
from the corresponding action, $\mathcal{S}=\int {dtd{\mathbf{r}}\mathcal{L}}
$, with the Lagrangian density
\begin{eqnarray}
\mathcal{L} &=&i\frac{\hbar }{2}\left( {{\Psi ^{\ast }}\frac{{\partial \Psi }%
}{{\partial t}}-\Psi \frac{{\partial {\Psi ^{\ast }}}}{{\partial t}}}\right)
-\frac{{{\hbar ^{2}}}}{{2{m_\mathrm{F}}}}{\left\vert {\nabla {\Psi }}%
\right\vert ^{2}}-{U}(\mathbf{r}){\left\vert {\Psi \left( {{\mathbf{r}},t}%
\right) }\right\vert ^{4/3}}-  \notag \\
&&\frac{{{\hbar ^{2}}}}{{2{m_\mathrm{F}}}}\frac{3}{5}{\left( {\frac{{6{\pi
^{2}}}}{{2{s_\mathrm{F}}+1}}}\right) ^{2/3}}\left\vert {\Psi \left( {{%
\mathbf{r}},t}\right) }\right\vert ^{10/3}-\frac{1}{2}{g_\mathrm{F}}%
\left\vert {\Psi \left( {{\mathbf{r}},t}\right) }\right\vert ^{4},
\label{E-4}
\end{eqnarray}
where the asterisk stands for the complex conjugate. Similar Lagrangian
formalisms have been used, in the context of the density-functional theory,
in diverse settings \cite{Adhikari2006b,Adhikari2007,Kim2004a}.

\end{document}